\documentclass[longauth]{aa}

\usepackage{graphicx}
\usepackage{booktabs}
\usepackage{verbatim,multirow,multicol}
\usepackage{hhline}
\usepackage{soul}
\setstcolor{red}
\usepackage{txfonts}
\usepackage{color}
\usepackage{tablefootnote}

 \newcommand{\grau}{$^\circ$}
 \newcommand{\AS}{Alice Springs/AUS}
 \newcommand{\As}{Asiago/ITA}
 \newcommand{\Asv}{Asiago (0.9 m)/ITA}
 \newcommand{\Asl}{Asiago(1.8 m)/ITA}
 \newcommand{\Bor}{Borowiec/POL}
 \newcommand{\Coral}{Cairns (Coral Towers)/AUS}
 \newcommand{\Jewel}{Cairns (Jewel Box)/AUS}
 \newcommand{\CaiCT}{Cairns (0.4 m)/AUS}
 \newcommand{\CaiJB}{Cairns (0.5 m)/AUS}
 \newcommand{\Fly}{Flynn/AUS}
 \newcommand{\Gn}{Gnosca/CHE}
 \newcommand{\Kn}{Kninice/CZE}
 \newcommand{\Ko}{Konkoly/HUN}
 \newcommand{\LBJ}{La Bastide des Jourdans/FRA}
 \newcommand{\laHi}{La Hita/ESP}
 \newcommand{\LS}{La Silla/CHL}
 \newcommand{\LT}{Lake Tekapo/NZL}
 \newcommand{\Lu}{Lugano/CHE}
 \newcommand{\Mel}{Melbourne/AUS}
 \newcommand{\Me}{Mérida/VEN}
 \newcommand{\Mu}{Murrumbateman/AUS}
 \newcommand{\Ol}{Oliveira/BRA}
 \newcommand{\On}{Ondrejov/CZE}
 \newcommand{\Rec}{Reconquista/ARG}
 \newcommand{\Ro}{Rockhampton/AUS}
 \newcommand{\SV}{Samford Valley/AUS}
 \newcommand{\SPA}{San Pedro de Atacama/CHL}
 \newcommand{\Ya}{Yass/AUS}
 \newcommand{\Yi}{Yoron Island/JPN}
 \newcommand{\Ze}{Zeddam/NLD}

\newcommand{\AG}{Alan Gilmore}
\newcommand{\AM }{Alain Maury}
\newcommand{\AO}{Alberto Ossola}
\newcommand{\AP }{Andras Pál}

\newcommand{\as }{Ariel Stechina}
\newcommand{\CJ }{Cristóvão Jacques}
\newcommand{\CS }{Colin Snodgrass}
\newcommand{\Dh }{Dave Herald}
\newcommand{\DeH }{Dean Hooper}
\newcommand{\EJ }{Emmanuel Jehin}
\newcommand{\EP }{Eduardo Pimentel} 
\newcommand{\JMW}{Jan Marteen Winkel}
\newcommand{\JR }{João Ribeiro de Barros}
\newcommand{\JFP }{Joaquín Fabrega Polleri}
\newcommand{\JN }{John Newman}
\newcommand{\JB }{Jonathan Bradshaw}
\newcommand{\Jb }{Joseph Brimacombe}
\newcommand{\KH }{Kamil Hornoch}
\newcommand{\ON }{Orlando Naranjo}
\newcommand{\PS}{Patrick Sogorb}
\newcommand{\PK}{Pam Kilmartin}
\newcommand{\PP }{Petr Pravec}
\newcommand{\Ss }{Stefano Sposetti} 
\newcommand{\SK }{Stephen Kerr}
\newcommand{\Tc }{Tim Carruthers}
\newcommand{\TJ }{Tomas Janik}
\newcommand{\VN }{Valerio Nascimbeni}
\newcommand{\WH }{Willian Hanna}
\newcommand{\YUS}{Yuuji Ueno}

\begin{document} 

\title{Stellar occultations enable milliarcsecond astrometry for Trans-Neptunian objects and Centaurs}
\titlerunning{Stellar occultations enable milliarcsecond astrometry for TNOs and Centaurs}

\authorrunning{Rommel, F. L., Braga-Ribas, F. et al.}

\author{F. L. Rommel\inst{1,2,3} % email: flaviarommel@.on.br,
\and
F. Braga-Ribas\inst{2,1,3} % email: fribas@utfpr.edu.br,
\and
J. Desmars\inst{4,5} % email: josselin.desmars@obspm.fr,
\and
J. I. B. Camargo\inst{1,3} % email: camargo@linea.gov.br, camargo.jib@gmail.com,
\and
J. L. Ortiz\inst{6} % email: ortiz@iaa.es,
\and
B. Sicardy\inst{7} % email: Bruno.Sicardy@obspm.fr,
\and
R. Vieira-Martins\inst{1,3} % email: rvm@on.br,
\and
M. Assafin\inst{8,3} % email: massaf@astro.ufrj.br,
\and
P. Santos-Sanz\inst{6} % email: psantos@iaa.es,
\and
R. Duffard\inst{6} % email: duffard@iaa.es,
\and
E. Fernández-Valenzuela\inst{9} % email: estela@ucf.edu,
\and
J. Lecacheux\inst{7} % email: jl-x@laposte.net,
\and
B. E. Morgado\inst{7,1,3} % email: morgado.fis@gmail.com,
\and
G. Benedetti-Rossi\inst{7,3} % email: gugabrossi@gmail.com,
\and
A. R. Gomes-Júnior\inst{10} % email: altairgomesjr@gmail.com,
\and
C. L. Pereira\inst{2,3} % email: chryslp.fis@gmail.com,
\and
D. Herald\inst{11,12,13} % email: DRHerald@bigpond.net.au,
\and
W. Hanna\inst{11,12} % email: whhanna@me.com,
\and
J. Bradshaw\inst{14} % email: jonathan@the2bradshaws.com,
\and
N. Morales\inst{6} % email: nicolas@iaa.es,
\and
J. Brimacombe\inst{15} % email: jbrimaco@bigpond.net.au,
\and
A. Burtovoi\inst{16,17} % 
\and
T. Carruthers\inst{18} % email: tim@astrophoto.com.au,
\and
J. R. de Barros\inst{19} % email: joribar@gmail.com,
\and
M. Fiori\inst{20,17} % 
\and
A. Gilmore\inst{21} % email: alan.gilmore@canterbury.ac.nz,
\and
D. Hooper\inst{11,12} % email: dean@hooper.net.au,
\and
K. Hornoch\inst{22} % email: k.hornoch@centrum.cz,
\and
C. Jacques\inst{19} % email: cjacqueslf@yahoo.com.br,
\and
T. Janik\inst{11} % email: jazzer@centrum.cz,
\and
S. Kerr\inst{12,23} % email: Steve.Kerr@outlook.com.au,
\and
P. Kilmartin\inst{21} % email: pam.kilmartin@canterbury.ac.nz,
\and
Jan Maarten Winkel\inst{11} % email: jmwinkel@hetnet.nl,
\and
G. Naletto\inst{20,17} % 
\and
D. Nardiello\inst{24,17} % email: domenico.nardiello@lam.fr,
\and
V. Nascimbeni\inst{17,20} % email: valerio.nascimbeni@unipd.it,
\and
J. Newman\inst{11,13} % email: johncnewman@optusnet.com.au,
\and
A. Ossola\inst{25} % email: aossola@bluewin.ch,
\and
A. Pál\inst{26,27,28} % email: apal@szofi.net,
\and
E. Pimentel\inst{19} % email: pimentel.eduardo@gmail.com,
\and
P. Pravec\inst{22} % email: ppravec@astro.cz,
\and
S. Sposetti\inst{25} % email: stefanosposetti@ticino.com,
\and
A. Stechina\inst{29} % email: correo_ariel@hotmail.com,
\and
R. Szakats\inst{26} % email: szakats.robert@csfk.mta.hu,
\and
Y. Ueno\inst{30} % email: piscis-a@moon.gmobb.jp,
\and
L. Zampieri\inst{17} % email: luca.zampieri@inaf.it,
\and
J. Broughton\inst{31,12} % email: jb668587@gmail.com,
\and
J. B. Dunham\inst{11} % email: dunhamjoan@verizon.net,
\and
D. W. Dunham\inst{11} % email: dunham@starpower.net ,
\and
D. Gault\inst{12} % email: davegault@bigpond.com,
\and
T. Hayamizu\inst{30} % email: haya@po2.synapse.ne.jp,
\and
K. Hosoi\inst{30} % email: hosokatsuten@maroon.plala.or.jp,
\and
E. Jehin\inst{32} % email: ejehin@uliege.be,
\and
R. Jones\inst{11} % email: jones.robert@verizon.net,
\and
K. Kitazaki\inst{30} % email: tomi-01@sea.plala.or.jp,
\and
R. Komžík\inst{33} % email: rkomzik@ta3.sk,
\and
A. Marciniak\inst{34} % email: am@amu.edu.pl,
\and
A. Maury\inst{35} % email: maury2017@spaceobs.com,
\and
H. Mikuž\inst{36} % email: herman.mikuz@gmail.com,
\and
P. Nosworthy\inst{12} % email: peter@noswonky.com,
\and
J. Fabrega Polleri\inst{37} % email: fabrega@ae.com.pa,
\and
S. Rahvar\inst{38} % email: rahvar@sharif.edu,
\and
R. Sfair\inst{10} % email: rsfair@gmail.com,
\and
P. B. Siqueira\inst{10} % email: patricia.buzzatto@gmail.com,
\and
C. Snodgrass\inst{39} % email: colin.snodgrass@ed.ac.uk,
\and
P. Sogorb\inst{40} % email: patrick.sogorb@gmail.com,
\and
H. Tomioka\inst{30} % email: VYN03140@nifty.com,
\and
J. Tregloan-Reed\inst{41} % email: j.j.tregloan.reed@gmail.com,
\and
O. C. Winter\inst{10} % email: ocwinter@gmail.com,
}

\institute{Observatório Nacional/MCTIC, R. General José Cristino 77, Bairro Imperial de São Cristóvão, Rio de Janeiro (RJ), Brazil\\
\email{flaviarommel@.on.br}
\and
Federal University of Technology - Paraná (UTFPR / DAFIS), Rua Sete de Setembro, 3165, Curitiba (PR), Brazil
\and
Laboratório Interinstitucional de e-Astronomia - LIneA $\&$ INCT do e-Universo, Rua Gal. José Cristino 77, Bairro Imperial de São Cristóvão, Rio de Janeiro (RJ), Brazil
\and
Institut Polytechnique des Sciences Avancées IPSA, 63 boulevard de Brandebourg, F-94200 Ivry-sur-Seine, France
\and
Institut de Mécanique Céleste et de Calcul des Éphémérides, IMCCE, Observatoire de Paris, PSL Research University, CNRS,Sorbonne Universités, UPMC Univ Paris 06, Univ. Lille, 77, Av. Denfert-Rochereau, F-75014 Paris, France
\and
Instituto de Astrofísica de Andalucía, IAA-CSIC, Glorieta de la Astronomía s/n, 18008 Granada, Spain
\and
LESIA, Observatoire de Paris, Université PSL, CNRS, Sorbonne Université, Univ. Paris Diderot, Sorbonne Paris Cité, 5 place Jules Janssen, 92195 Meudon, France
\and
Observatório do Valongo/UFRJ, Ladeira Pedro Ant\^onio 43, Rio de Janeiro (RJ), Brazil
\and
Florida Space Institute, University of Central Florida, 12354 Research Parkway, Partnership 1, Orlando, FL, USA
\and
UNESP - São Paulo State University, Grupo de Dinâmica Orbital e Planetologia, CEP 12516-410, Guaratinguetá, SP, Brazil
\and
International Occultation Timing Association (IOTA), P.O. Box 423, Greenbelt, MD 20768, USA
\and
Trans-Tasman Occultation Alliance (TTOA), Wellington PO Box 3181, New Zealand
\and
Canberra Astronomical Society, Canberra, ACT, Australia
\and
Samford Valley Observatory (Q79), Queensland, Australia
\and
Coral Towers Observatory, Cairns, QLD 4870, Australia
\and
Centre of Studies and Activities for Space (CISAS) 'G. Colombo', University of Padova, Via Venezia 15, 35131
\and
INAF-Astronomical Observatory of Padova, Vicolo dell’Osservatorio 5, I-35122 Padova, Italy
\and
Jewel Box Observatory, 69 Falcon St, Bayview Heights QLD 4868, Australia
\and
SONEAR Observatory, Oliveira (MG), Brazil
\and
Department of Physics and Astronomy 'G. Galilei', University of Padova, Via F. Marzolo 8, 35131, Padova, Italy
\and
Mount John University Observatory, University of Canterbury, P.O. Box 56, Lake Tekapo 7945, New Zealand
\and
Astronomical Institute, Academy of Sciences of the Czech Republic, Fri\v{c}ova 298, 251 65 Ond\v{r}ejov, Czech Republic
\and
Astronomical Association of Queensland, 5 Curtis Street, Pimpama QLD 4209, Australia
\and
Aix Marseille Univ, CNRS, CNES, LAM, Marseille, France
\and
SOTAS - Stellar Occultation Timing Association Switzerland, Swiss Astronomical Society, Switzerland
\and
Konkoly Observatory, Research Centre for Astronomy and Earth Sciences, Konkoly-Thege Miklós út 15-17, 1121 Budapest, Hungary
\and
Eötvös  Loránd  University,  Department  of  Astronomy,  Pázmány Péter sétány 1/A, 1117 Budapest, Hungary
\and
ELTE Eötvös Loránd University, Institute of Physics, Pázmány Péter sétány 1/A, 1117 Budapest, Hungary
\and
Centro de Amigos de la Astronomia Reconquista - CAAR, Reconquista, Argentina
\and
Japan Occultation Information Network (JOIN), Japan
\and
Reedy Creek Observatory, Gold Coast, Queensland, Australia
\and
STAR Institute, Université de Liège, Allée du 6 août, 19C, 4000 Liège, Belgium
\and
Astronomical Institute, Slovak Academy of Sciences, 059 60 Tatranská Lomnica, Slovakia
\and
Astronomical Observatory Institute, Faculty of Physics, Adam Mickiewicz University, Poznan, Poland
\and
San Pedro de Atacama Celestial Explorations - SPACE, Chile
\and
University of Ljubljana, Faculty of Mathematics and Physics, Jadranska 19, 1000 Ljubljana, Slovenia
\and
Panamanian Observatory in San Pedro de Atacama - OPSPA
\and
Department of Physics, Sharif University of Technology, P.O. Box 11155-9161 Tehran, Iran
\and
Institute for Astronomy, University of Edinburgh, Royal Observatory, Edinburgh EH9 3HJ, UK
\and
Club d’astronomie Luberon Sud Astro, La Bastide des Jourdans, France
\and
Centro de Astronomía (CITEVA), Universidad de Antofagasta, Avenida U. de Antofagasta, 02800 Antofagasta, Chile}

\date{Received ; accepted }

\abstract
{Trans-Neptunian objects (TNOs) and Centaurs are remnants of our planetary system formation, and their physical properties have invaluable information for evolutionary theories. Stellar occultation is a ground-based method for studying these distant small bodies and has presented exciting results. These observations can provide precise profiles of the involved body, allowing an accurate determination of its size and shape.}   
% aims to head (mandatory)
{The goal is to show that even single-chord detections of TNOs allow us to measure their milliarcsecond astrometric positions in the reference frame of the \textit{Gaia} second data release (DR2). Accurated ephemerides can then be generated, allowing predictions of stellar occultations with much higher reliability.}
{We analyzed data from various stellar occultation detections to obtain astrometric positions of the involved bodies. The events published before the \textit{Gaia} era were updated so that the \textit{Gaia} DR2 stellar catalog is the reference, thus providing accurate positions. Events with detection from one or two different sites (single or double chord) were analyzed to determine the event duration.  Previously determined sizes were used to calculate the position of the object center and its corresponding error with respect to the detected chord and the International Celestial Reference System (ICRS) propagated \textit{Gaia} DR2 star position.}
{We derive 37 precise astrometric positions for 19 TNOs and four Centaurs. Twenty-one of these events are presented here for the first time. Although about 68\% of our results are based on single-chord detection, most have intrinsic precision at the submilliarcsecond level. Lower limits on the diameter of bodies such as Sedna, 2002 KX$_{14}$, and Echeclus, and also shape constraints on 2002 VE$_{95}$, 2003 FF$_{128}$ , and 2005 TV$_{189}$ are presented as valuable byproducts.}
{Using the \textit{Gaia} DR2 catalog, we show that even a single detection of a stellar occultation allows improving the object ephemeris significantly, which in turn enables predicting a future stellar occultation with high accuracy. Observational campaigns can be efficiently organized with this help, and may provide a full physical characterization of the involved object, or even the study of topographic features such as satellites or rings.}

\maketitle
%
%________________________________________________________________

\section{Introduction}
\label{Introduction}

Trans-Neptunian objects (TNOs) are a population of small bodies orbiting the Sun beyond Neptune with a semimajor axis greater than 30 au  \citep{Gladman08}. So far, a few thousand of them are known\footnote{https://www.minorplanetcenter.net/iau/lists/TNOs.html, https://www.minorplanetcenter.net/iau/lists/Centaurs.html}. Because TNOs are thought to be remnants of the primordial disk, they are an invaluable source of information about the primitive solar nebula as well as the history and evolution of the Solar System \citep{Nesvorny12}. Centaurs are a transient population between TNOs and Jupiter-family comets \citep{Fernandez02, Horner04}. Their orbits are between those of the giant planets. Almost one thousand of them are currently known. As they may have a common origin with the TNOs, they can also provide valuable information about the outer Solar System. 

Seen from Earth, these distant and small objects are faint sources (R $>$ 19 mag) and have small angular sizes on the sky (under 40 mas). Even though a large number of bodies have been discovered in the past 25 years, and NASA/New Horizons spacecraft flyby of (134340) Pluto and (486958) Arrokoth, also known as 2014 MU$_{69}$, have produced a plethora of data \citep{Stern19, Benecchi19, Buie2020, Earle20, Spencer20, Umurhan20},  our knowledge about orbital parameters and physical properties of  TNOs and Centaurs is still poor and fragmented. Radiometric measurements, which can provide sizes, albedos, and thermal properties, are available for about one hundred of them. Radiometry can achieve a precision for the sizes of 10-20\% \citep{Fornasier13, Lellouch13, MM12, Pal12, Santos-Sanz12,  Vilenius12, Mueller20}. 

To unveil fundamental properties such as density and shapes, we used one of the most accurate Earth-based techniques: stellar occultation \citep{Ortiz20_book}. The detection of an occultation from different sites allows us to combine the data sets to determine the physical parameters of the object (\citeauthor{Elliot10} \citeyear{Elliot10}, \citeauthor{Ortiz15} \citeyear{Ortiz15}, \citeauthor{Benedetti-Rossi16} \citeyear{Benedetti-Rossi16}, \citeyear{Benedetti-Rossi19}, \citeauthor{Schindler17} \citeyear{Schindler17}, \citeauthor{Dias-Oliveira17} \citeyear{Dias-Oliveira17} and the references therein). Observations with a high cadence in time and high signal-to-noise ratio allow detecting or setting upper limits to the presence of atmospheres at the nano bar level \citep{Sicardy11, Ortiz12, Braga-Ribas13} and the same for small structures such as jets or rings \citep{Elliot95, Braga-Ribas14, Berard17, Ortiz17}. A complete list of detected stellar occultations by outer Solar System objects that we are aware of can be found at \url{http://occultations.ct.utfpr.edu.br/results} \citep{Braga-Ribas19}.

These invaluable results can only be achieved if precise predictions of stellar occultations are available. We need to know the position of the occulted star and the ephemeris of the occulting body to make a good prediction. The \textit{Gaia} DR2 astrometric catalog \citep{Gaia18} today provides accurate information about position, proper motion, and parallax, allowing us to determine the position of the occulted stars at the instant of the event with a precision of a fraction of one milliarcsecond (mas). The ephemeris can be improved by combining classical astrometric observations in a weighting procedure that takes the quality of the observation and the reference catalog into account, such as made by Numerical Integration of the Motion of an
Asteroid, NIMA \citep{Desmars15}.

In this work, we show that whenever a body is detected to occult a star, we can measure its astrometric position with higher accuracy than is possible with classical astrometry. This position allows calculating precise ephemerides and in turn accurately predicting future stellar occultations by the body. In contrast to the positions obtained from classical astrometry, the photocenter in these positions is not displaced, for instance, by the solar phase angle or a companion body.

We present a set of 37 accurate astrometric positions derived from stellar occultations by TNOs and Centaurs. These events are part of our program of physical characterizing TNOs and Centaurs in an international collaboration under the umbrella of the ERC Lucky Star project\footnote{https://lesia.obspm.fr/lucky-star/}. Most of the events presented here were predicted by \cite{Assafin12} and \cite{Camargo14} and updated using astrometric observations of the candidate star and the occulting object made a few days before an observational campaign, see \cite{Benedetti-Rossi19}, as an example. After \textit{Gaia} releases, the astrometric efforts are focused on improving the object ephemeris (e.g., \citeauthor{Ortiz17} \citeyear{Ortiz17} and \citeauthor{Ortiz20} \citeyear{Ortiz20}). Last-minute astrometric observations were essential for many of the successful detections we present, and they were made using telescopes at Pico dos Dias Observatory in Brazil, at Pic du Midi Observatory in France, and with the Sierra Nevada, Calar Alto, Roque de Los Muchachos, and La Hita observatories in Spain.

A summary of the \textit{Gaia} mission and how the DR2 additional information was used to set the stellar positions to the event epoch, as well as the adopted target star position, are given in Sec. \ref{Gaia}. In Sec. \ref{Data} we present the method we used in the analysis of the different data sets of stellar occultations. Section \ref{Astrometric} describes the astrometric positions of four Centaurs and 19 TNOs. In Sec. \ref{Physical} we present some physical constraints for 3 TNOs we obtained: 2002 VE$_{95}$, 2003 FF$_{128}$, and 2005 TV$_{189}$. Discussion and conclusions are presented in Sec. \ref{Discussion}.

%__________________________________________________________________

\section{\textit{Gaia} catalog: target stars}
\label{Gaia}

\textit{Gaia} \citep{Gaia16} is an astrometric, photometric, and spectrometric mission of the European Space Agency (ESA). Its main goal is to measure and obtain a precise three-dimensional map of the spatial and velocity distribution of stars in our Galaxy. Published in April 2018, the second release \citep{Gaia18}, \textit{Gaia} DR2 for short, lists astrometric stellar positions with uncertainties below the mas level, as well as proper motion and parallaxes (for G mag. < 20 stars), and the median radial velocity (for G < 13) with error bars up to 1.2 mas/yr, 0.7 mas and 1.2 km/s.

The International Astronomical Union Standards of Fundamental Astronomy (SOFA) service\footnote{Software routines from the IAU SOFA collection were used. Copyright  International Astronomical Union Standards of Fundamental Astronomy (\url{http://www.iausofa.org}).} provides algorithms and procedures to implement standard models used in fundamental astronomy. In addition, the Navigation and Ancillary  Information Facility (NAIF) offers an extensive collection of application program interfaces, subroutines, and functions in the package called SPICE Toolkit\footnote{The Navigation and Ancillary Information Facility (\url{https://naif.jpl.nasa.gov/naif/toolkit.html}).}. Both tools were used in a \textsc{fortran 77} code that we wrote to propagate the stellar positions from the \textit{Gaia} DR2 reference epoch to the occultation epoch. The input data include correlations, magnitudes ($G$, $G_{\rm RP}$, and $G_{\rm BP}$), proper motions, parallaxes, radial velocity (when available), right ascension, and declination. The stellar positions are given in the International Celestial Reference System (ICRS) \citep{Arias95}, and their respective uncertainties are presented in Table \ref{table:stars_DR2}, along with the name of the occulting body as well as the date and time of the given position. The positions from the DR2 listed in Table \ref{table:stars_DR2}  were corrected for their space motion, and the origin was moved from the barycenter of the Solar System to the geocenter with the help of routines from the SOFA library. We used an unreleased counterpart of the SOFA routine PMPX\footnote{Additional information about this routine can be found on the appendix of documentation available here: \url{http://www.iausofa.org/2020_0721_F/sofa/sofa_ast_f.pdf.}} to propagate the stellar position in time from its proper motion and parallax. This routine is compliant with special relativity and presented occasional and negligible differences with respect to the released PMPX routine for the stars we considered. The parallax between barycentric and geocentric stellar positions for our sample can reach a non-negligible value of 3.36 mas. In brief, these are astrometric positions. Uncertainties were also propagated in time.

These accurate stellar positions were used as a reference to determine the celestial coordinates of the respective object. It is important to highlight that most of the uncertainties given in Table \ref{table:stars_DR2} are smaller than 1 mas. For reference, 1 mas corresponds to about 10 km at the average distance of our Centaur sample and a range of 20 to 70 km for our set of TNOs. 

%__________________________________________________________________

\section{Data reduction}
\label{Data}

Observation circumstances for each stellar occultation event we analyzed are presented in Table \ref{table:circumstances}. When available, the raw images of the 21 unpublished events were corrected for bias and flat field using standard procedures with the Image Reduction and Analysis Facility, IRAF \citep{Butcher}. The images were measured using differential photometry made with PRAIA (Package for Automatic Reduction of Astronomical Images, \cite{Assafin11}). Ingress and egress were obtained by modeling the occultation light curves, taking into account the finite exposure time, stellar diameter, and Fresnel diffraction, which were compared to the data using the minimum quadratic method ($\chi^2$ test). The observed and modeled light curves are presented in Appendix \ref{LC}. More information on the data reduction and modeling of the light curves can be found in \cite{Braga-Ribas13}, \cite{Benedetti-Rossi16}, \cite{Dias-Oliveira17}, and references therein. The derived ingress and egress times of unpublished events are presented in Table \ref{tab:times}, where the uncertainties are given at  the$1\sigma$ level ($\chi^2_{min}+1$). The table also presents the length of the corresponding chords in km and the expected equivalent diameter we used as reference when available in the literature.

The data sets we analyzed came from a great diversity of telescope sizes and detectors (video cameras and CCDs) and the image quality varies accordingly. The 37 events in this paper were divided into three different classes that we describe below, according to the different available information: events with positions already given in the literature, double-chord detection, and single-chord events.

%---------
\subsection{Updating object positions from the literature} 

A few stellar occultations by Centaurs and TNOs (mostly multichords) were published before the \textit{Gaia} releases. These works provided positions for the involved bodies, but the available stellar catalogs partly had important zonal errors (see the discussion in Sect. \ref{Discussion}). With \textit{Gaia} DR2, these positions can be updated now to provide accurate astrometry to be used in ephemeris calculations. We calculated the difference between stellar positions given in Table \ref{table:stars_DR2} and those provided by the respective publications (Table \ref{table:stars_published}). The resulting $\Delta \alpha\cdot\cos\delta$ and $\Delta \delta$ were added to the published astrometric position of the object. The precision of the new object coordinates (shown in Table \ref{table:objects_DR2}) depends on the uncertainties of the published center of the body and on the \textit{Gaia} DR2 star position propagated at the event epoch. They represent $\approx 24\%$ of all astrometric positions given in this work.

%---------
\subsection{Double-chord events} 

Two chords provide four points on the sky plane. They can be used to fit a circle, which is defined by three parameters (radius and center), representing the object limb. The error of the object position will depend only on the error of the obtained chords (i.e., the fit of a circle to the chords, considering their errors), and the assumption of a circular limb, while some ellipticity may be expected for these objects.

Three double-chord events are presented here for the first time:  2003 FF$_{128}$ (24 May 2017); 2002 KX$_{14}$ (19 September 2018); and 2005 RM$_{43}$ (24 December 2018). The derived ingress and egress are given in Table \ref{tab:times}. The event by Sedna was observed from two different sites. Unfortunately, they were less than 10 km apart. For the astrometric purposes, we therefore have only one effective chord on the sky plane. The times of the other four double-chord events were available in the literature and were used to derive the astrometric position of the object. The final celestial coordinates are given in Table \ref{table:objects_DR2} with the error bars. 

%--------

\subsection{Single-chord events} 

About  68 \% of the astrometric positions presented in this work come from single-chord occultations (i.e., only one site was able to detect the event). For the unpublished events, the ingress and egress times are given in Table \ref{tab:times}. 

A single-chord event provides two points on the sky plane that can be used under some assumptions to calculate the astrometric position of the object. From the literature, we collected the equivalent diameter (or the elliptical limb, when available) and fit only the center to the chord extremities. To account for the dimension uncertainties of the object, we fit circles with varying radii within the range given by radiometric measurements (or the ellipse parameters) to obtain its center.

\begin{figure}[!h]
    \centering
    \includegraphics[scale=0.36]{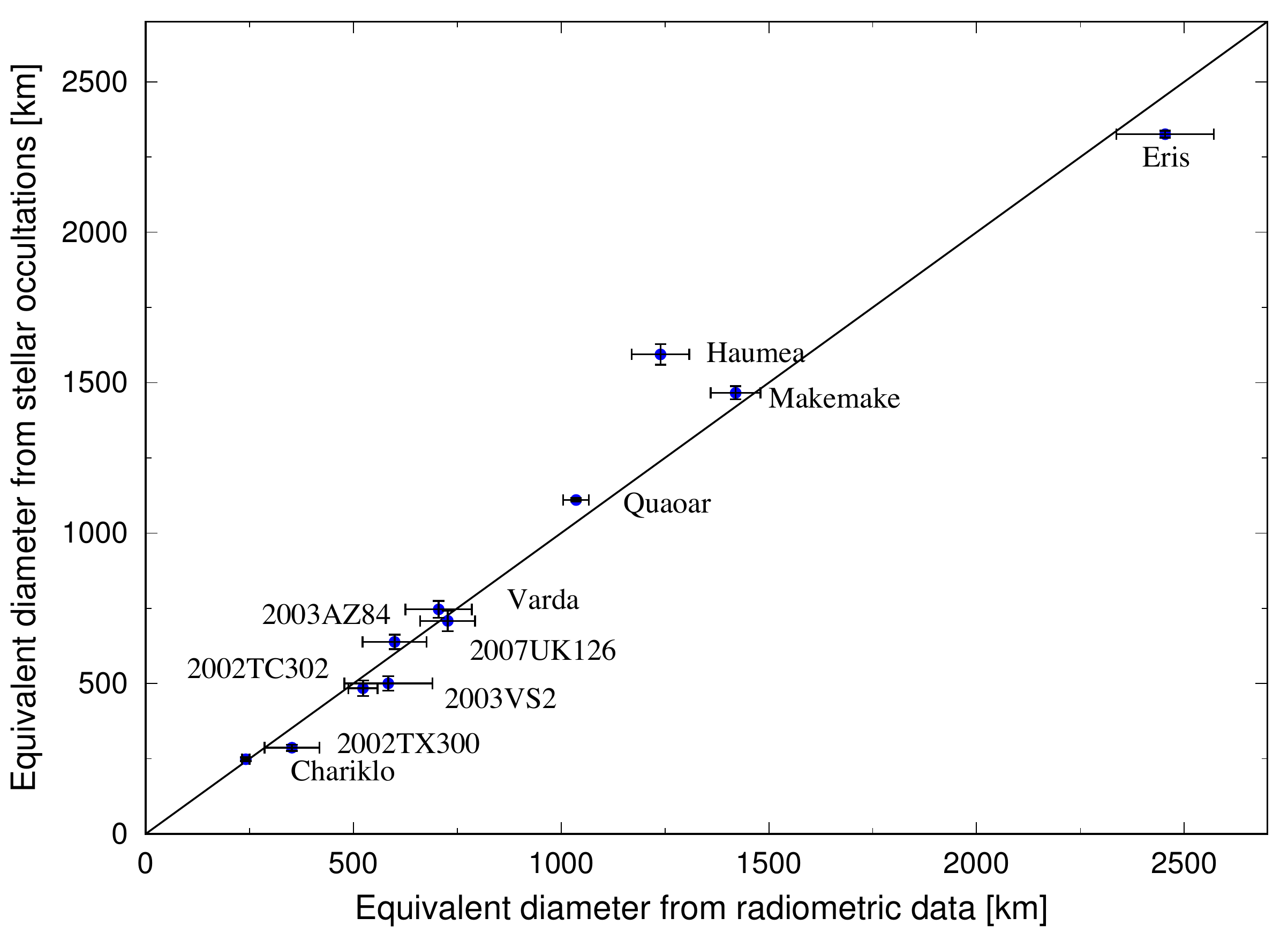}
    \caption{Direct comparison between equivalent diameters obtained using the radiometric technique (x-axis) and those obtained from stellar occultations (y-axis). The references for the object diameters from occultation and radiometric data are Eris \citep{Sicardy11,Santos-Sanz12}, Haumea  \citep{Ortiz17,Fornasier13}, Makemake  \citep{Ortiz12,Lim10},  Quaoar  \citep{Braga-Ribas13,Fornasier13}, Varda \citep{Souami, Vilenius14}), 2003 AZ$_{84}$  \citep{Dias-Oliveira17,MM12}, 2007 UK$_{126}$  \citep{Benedetti-Rossi16,Santos-Sanz12}, 2002 TC$_{302}$  \citep{Ortiz20,Fornasier13}, 2003 VS$_2$  \citep{Benedetti-Rossi19,MM12}, 2002 TX$_{300}$  \citep{Elliot10,Vilenius18}, and Chariklo  \citep{Leiva17,Lellouch17}.}
    \label{fig:radioXocc}
\end{figure}

To verify this assumption, we compared the sizes obtained from multichord stellar occultations to those from radiometric observations (Fig. \ref{fig:radioXocc}). On average, the two techniques agree well, mainly for the smaller bodies (< 1,000 km), although the error bars coming from stellar occultation determinations are considerably smaller. The only significant outlier is Haumea, a tri-axial body, known as a very complex system with moons and ring, which needs to be considered when individual results from the two techniques are compared. The direct comparison, as presented here, does not consider systematic effects that can affect radiometric data, such as the presence of satellites and observations in different aspect angles or at different heliocentric distances. Even so, radiometric sizes can be used as a good prior to obtain the object position from a single-chord event. If the object presents high oblateness, it will also affect the derived position by some value lower than half of its apparent size, that is, several mas. It could be considered as an additional error, but because of the lack of information, this is not considered in this work.

\begin{figure}[!ht]
\centering
    \includegraphics[scale=0.26]{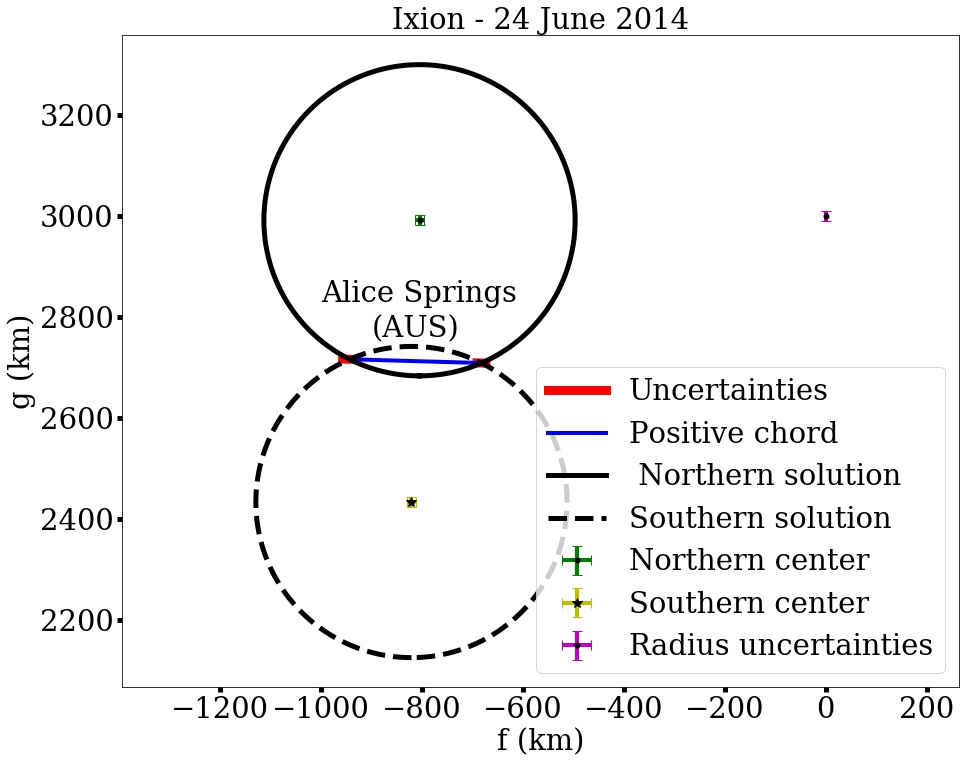}
    \caption{Example of two solutions for a single-chord stellar occultation (by the TNO Ixion on 24 June 2014). A circle with 308.25 km in radius was fit, which is the nominal value from \cite{Lellouch13}, and the given uncertainty is represented by the purple cross (in the upper right corner). The blue segment represents the observed chord. The chord length is defined by ingress and egress, which are the two points where the circles intercept each other. The red lines are the uncertainties of ingress and egress. The error bars at the center of the circles correspond to the uncertainties of the object position (fit center) relative to the \textit{Gaia} DR2 propagated star position. Each circle represents one of the two possible solutions, with the center to the north or south of the observed chord.}
    \label{fig:Ixion}
\end{figure}

A single-chord event has an ambiguity in the hemisphere of the body that occulted the star, leading to two different and equally valid solutions (Fig. \ref{fig:Ixion}). When this is the case, we list two astrometric positions in Table 3, one for each hemisphere. This ambiguity may be broken when a close negative chord\footnote{A negative chord means that data were acquired from a given site, and the event was not detected, for instance, the light curve does not show any clear flux drop close to the expected event time.} is available, but this is rare. Only 5 of the 23 single-chord events had close negative records: Eris (29 August 2013), 2007 JJ$_{43}$ (11 July 2019), 2017 OF$_{69}$ (27 August 2019), Bienor (02 April 2018), and Chiron  (29 November 2011). Chiron is described by \cite{Ruprecht15}.

In the stellar occultation by 2002 VE$_{95}$ detected on 3 December 2015, only one chord was detected. The length exceeded the upper limit of the best diameter available in the literature \citep{MM12}. An ellipse was fit to maintain the same equivalent area, with the chord as the equatorial diameter, and only one astrometric position is therefore provided in Table \ref{table:objects_DR2} (see the discussion in Sect. \ref{sec:VE95}). 

A single visual observation of a stellar occultation by the TNO 2005 TV$_{189}$ was reported on 13 November 2012. The observer reported two drops in the stellar flux, giving two consecutive chords with $\sim$50 km, separated by 403 km. The mean astrometric position between the two consecutive chords is provided in Table \ref{table:objects_DR2}. The given error corresponds to the difference between the first ingress and last egress times projected on the sky plane (for further information, see Sect. \ref{sec:TV189}). 

%__________________________________________________________________
\section{Occulting body positions}
\label{Astrometric}

As described in previous sections and considering the discussed circumstances, the positions of all objects at the occultation event times are given in Tables \ref{table:objects_DR2} and \ref{table:objects_DR2_N_S}. With the help of routines from the SPICE toolkit, these positions are also astrometric, that is, the origin is set at the observing site (the geocenter in the present case), the orientation of the coordinate axes is that of the ICRS, and the positions were corrected for the light-time delay.

Taking the \textit{Gaia} DR2 catalog into account, Table \ref{table:objects_DR2} presents updated positions of objects for which occultation events were available in the literature. It also presents accurate positions for events for which a single position could be obtained in this work (i.e., double- or single-chord detection, for which there is no ambiguity on the obtained position). 

Table \ref{table:objects_DR2_N_S} presents the positions of objects obtained from single-chord events, giving two possible solutions because of the ambiguity, as discussed in Sect. \ref{Data}. The correct solution may be retrieved when an accurate ephemeris is available, for instance, when previous astrometric positions from other occultation events, close in time, are available. As discussed below, this is the case for Quaoar, which has several occultation detections since 2011. In this case, the preferred position is highlighted in bold.

\begin{table}[!h]
\caption{Event times and chord lengths for all unpublished occultation detections. The equivalent diameters and respective references used to determine the astrometric positions are also provided. For three Plutinos no diameter was available in the literature, therefore we used its absolute magnitude given by MPC and a median albedo value of p = 0.089 $\pm$ 0.064 (\cite{Mueller20}, table 7.2).}
\label{tab:times}
\resizebox{\hsize}{!}{
\begin{tabular}{lllcc} 
\hline\hline
\multicolumn{1}{c}{\begin{tabular}[c]{@{}c@{}}\textbf{Occulting}\\ \textbf{Object}\end{tabular}}&
\multicolumn{1}{c}{\begin{tabular}[c]{@{}c@{}}\textbf{Sites}\\\textbf{Event Date}\\ (dd-mm-yyyy)\end{tabular}}&
\multicolumn{1}{c}{\begin{tabular}[c]{@{}c@{}}\textbf{Immersion}\\ \textbf{Emersion}\\(hh:mm:ss.ss $\pm$ ss.ss)\end{tabular}}&
\multicolumn{1}{c}{\begin{tabular}[c]{@{}c@{}} \textbf{Chord}\\ \textbf{Length}\\ (km)\end{tabular}} &
\multicolumn{1}{c}{\begin{tabular}[c]{@{}c@{}}\textbf{Equivalent}\\ \textbf{Diameter} \\ (km)  \end{tabular}} \\  \hline
(136199)            &\SV                &15:37:48.0   $\pm$  3.1         & \multirow{2}{*}{996 $\pm$ 120}          &\multirow{2}{*}{2326 $\pm$ 12\tablefootmark{a}}\\
Eris                &29-08-2013         &15:38:40.0   $\pm$  3.2         &                                         &\\ \hline                      
(50000)             &\Me                &02:40:12.4   $\pm$  1.5         &\multirow{2}{*}{1145 $\pm$ 67}           &\multirow{2}{*}{1110 $\pm$ 5\tablefootmark{b}}\\
Quaoar              &09-07-2013         &02:41:03.3   $\pm$  1.5         &                                         &\\ \hline
                    &\CaiCT             &12:24:12.4   $\pm$  5.6         &\multirow{2}{*}{1025 $\pm$ 135}          &\multirow{4}{*}{995 $\pm$ 80\tablefootmark{c}}\\
(90377)             &13-01-2013         &12:25:37.1   $\pm$  5.6         &                                         &\\ \cline{2-4}
Sedna               &\CaiJB             &12:23:52     $\pm$ 23.4         &\multirow{2}{*}{1305 $\pm$ 565}          &\\
                    &13-01-2013         &12:25:39     $\pm$ 23.4         &                                         &\\ \hline
                    & \Ol               &05:43:45.3   $\pm$ 0.23         &\multirow{2}{*}{126 $\pm$ 9}             &\multirow{5}{*}{610}\\
                    & 11-07-2019        &05:43:52.0   $\pm$ 0.23         &                                         &\\ \cline{2-4}
(278361)            & \Mu               & 13:05:31.4  $\pm$ 1.0          &\multirow{2}{*}{220.5  $\pm$ 20.1}       &\\
2007 JJ\textsubscript{43}& 07-08-2019   & 13:05:54.8  $\pm$ 1.1          &                                         &(+170 / -140)\tablefootmark{d}\\ \cline{2-4}
                    & \Mel              & 11:30:14.93 $\pm$ 0.48         &\multirow{2}{*}{278.8  $\pm$ 17.1}       &\\     
                    & 24-05-2020        & 11:30:26.66 $\pm$ 0.24         &                                         &\\ \hline   
(28978)             &\AS                &14:54:44.85  $\pm$ 0.55         &\multirow{2}{*}{257 $\pm$ 22}            &\multirow{1}{*}{617 }\\ 
Ixion               &24-06-2014         &14:54:55.35  $\pm$ 0.35         &                                         &(+19 / -20) \tablefootmark{e}\\ \hline
\multirow{2}{*}{2017 OF\textsubscript{69}} &\On& 21:19:33.82 $\pm$ 0.28  &\multirow{2}{*}{152.15 $\pm$ 12.0}       &\multirow{2}{*}{541.1 \tablefootmark{f}}\\
                    & 27-08-2019        & 21:19:40.90 $\pm$ 0.28         &                                         &\\ \hline
                    &\Gn                & 02:20:46.95 $\pm$ 0.42         &\multirow{2}{*}{455.11 $\pm$ 19.8}       &\multirow{5}{*}{524}\\
                    &24-12-2018         & 02:21:06.14 $\pm$ 0.42         &                                         &\\ \cline{2-4}
(145451)            &\Lu                & 02:20:46.42 $\pm$ 0.50         &\multirow{2}{*}{436.98 $\pm$ 18.3}       &\\
2005 RM\textsubscript{43}& 24-12-2018& 02:21:04.84 $\pm$ 0.28            &                                         &(+96 / -103)\tablefootmark{g}\\ \cline{2-4}
                    &\LT                & 10:56:29.10 $\pm$ 0.34         &\multirow{2}{*}{459.61 $\pm$ 8.6}        &\\
                    & 03-02-2019        & 10:57:10.84 $\pm$ 0.45         &                                         &\\ \hline   
                    &\Asv               &18:20:51.95  $\pm$ 0.45         &\multirow{2}{*}{310.26 $\pm$ 68}         &\multirow{3}{*}{423}\\      
(444030)            &16-11-2017         &18:21:11.8   $\pm$  3.9         &                                         &\\ \cline{2-4}
2004 NT\textsubscript{33}&\Asl          &18:20:53.073 $\pm$ 0.083        &\multirow{2}{*}{277.20 $\pm$ 3.5}        &(+87 / -80)\tablefootmark{h} \\         
                    &16-11-2017         &18:21:10.81  $\pm$ 0.14         &                                         &\\ \hline
                    &\Mu                &12:31:22.06  $\pm$ 0.13         &\multirow{2}{*}{421 $\pm$ 3}             &\multirow{4}{*}{455 $\pm$ 27\tablefootmark{i}}\\
(119951)            &19-09-2018         &12:32:05.11  $\pm$ 0.18         &                                         &\\ \cline{2-4}
2002 KX\textsubscript{14}&\Ya           &12:31:21.83  $\pm$ 0.28         &\multirow{2}{*}{431 $\pm$ 6}             &\\ 
                    &19-09-2018         &12:32:05.99  $\pm$  0.38        &                                         &\\ \hline 
(175113)            &\Rec               &01:35:55.97  $\pm$  0.63        &\multirow{2}{*}{322 $\pm$ 27}            &\multirow{1}{*}{468.2}\\
2004 PF\textsubscript{115}&28-09-2018&01:36:10.47  $\pm$ 0.58            &                                         &(+38.6/ -49.1) \tablefootmark{j} \\ \hline
(119979)            &\Ro                &11:57:50.65  $\pm$ 0.34         &\multirow{2}{*}{198 $\pm$ 17}            &\multirow{2}{*}{348 $\pm$ 45\tablefootmark{e}}\\
2002 WC\textsubscript{19}&30-12-2018&11:57:58.60  $\pm$ 0.34             &                                         &\\ \hline 
(55638)             &\SV                &12:30:04:660 $\pm$ 0.030        &\multirow{2}{*}{280.9 $\pm$ 1.8}         &\multirow{1}{*}{249.8}\\
2002 VE\textsubscript{95}&03-12-2015&12:30:16.600 $\pm$ 0.040            &                                         &(+13.5 / -13.1) \tablefootmark{j} \\ \hline
                    &\Mu                &13:41:42.2282 $\pm$ 0.006       &\multirow{2}{*}{111.90 $\pm$ 0.29}       &\multirow{4}{*}{205.8\tablefootmark{l}}\\
(469506)            &24-05-2017         &13:41:46.9760 $\pm$ 0.006       &                                         & \\ \cline{2-4}
2003 FF\textsubscript{128}&\Fly         &13:41:41.6305 $\pm$ 0.006       &\multirow{2}{*}{125.69 $\pm$ 0.34}       &\\
                    &24-05-2017         &13:41:46.9630 $\pm$ 0.009       &                                         &\\ \hline
                    &\Yi                &16:24:22.682 $\pm$ 0.038        &\multirow{2}{*}{53.9 $\pm$ 1.2}          &\multirow{3}{*}{199}\\
(54598)             &29-12-2017         &16:24:26.292 $\pm$ 0.042        &                                         &\\ \cline{2-4}
Bienor              &\Ko                &19:50:23.586 $\pm$ 0.032        &\multirow{2}{*}{92.43 $\pm$ 1.5}         &(+9 / -12) \tablefootmark{m} \\
                    &02-04-2018         &19:50:26.608 $\pm$ 0.018        &                                         &\\ \hline
\multirow{4}{*}{2005 TV\textsubscript{189}}&\Kn &22:26:58.28  $\pm$ 0.20 &\multirow{2}{*}{44 $\pm$ 11}             &\multirow{4}{*}{142.2\tablefootmark{n}}\\
                    &13-11-2012         &22:27:00.25  $\pm$ 0.30         &                                         &\\ \cline{2-4} 
                    &\Kn                &22:27:18.40  $\pm$ 0.10         & \multirow{2}{*}{58 $\pm$ 4}             &\\                      
                    &13-11-2012         &22:27:21.02  $\pm$ 0.10         &                                         &\\ \hline                    
(8405)              & S. P. de Atacama/CHL&05:23:18.72  $\pm$ 0.89       &\multirow{2}{*}{73 $\pm$ 44}             &\multirow{1}{*}{85}\\
Asbolus             &24-11-2013         &05:23:22.28  $\pm$ 0.89         &                                         &(+8 / -9)\tablefootmark{o} \\ \hline
(60558)             &\Ze                &23:36:45.56  $\pm$ 0.18         &\multirow{2}{*}{59 $\pm$ 9}              &\multirow{2}{*}{64.6 $\pm$ 1.6\tablefootmark{o}}\\
Echeclus            &25-06-2012         &23:36:48.98  $\pm$ 0.36         &                                         &\\ \hline 
\end{tabular}
}
\tablefoot{All the times here reported, obtained with video cameras plus time inserter, were corrected for the respective delays \citep{Barry15}. The references for the equivalent diameters presented in this table are
\tablefoottext{a}{\cite{Sicardy11}},
\tablefoottext{b}{\cite{Braga-Ribas13}},
\tablefoottext{c}{\cite{Pal12}},
\tablefoottext{d}{\cite{Pal2015}},
\tablefoottext{e}{\cite{Lellouch13}},
\tablefoottext{f}{$^\dag~$H mag. = 4.6,}
\tablefoottext{g}{\cite{Farkas20}},
\tablefoottext{h}{\cite{Vilenius14}},
\tablefoottext{i}{\cite{Vilenius12}},
\tablefoottext{j}{\cite{MM12}},
\tablefoottext{l}{$^\dag~$H mag. = 6.7},
\tablefoottext{m}{\cite{Lellouch17},
\tablefoottext{n}{$^\dag~$H mag. = 7.5},
\tablefoottext{o}{\cite{Duffard14}}.}}
\end{table}

\begin{table}[!h]
\caption{Astrometric positions of the occulting object for occultation events with an unambiguous solution. Each event has a label corresponding to the kind of available information: U for those updated from the literature, D for double-chord events, and S  for single chord with close-by negative observations or central-chord detection. These positions are in the ICRS as represented by the Gaia Celestial Reference Frame (Gaia-CRF2) \citep{GaiaCRF2}.}
\label{table:objects_DR2}      
\resizebox{\hsize}{!}{        
\begin{tabular}{lcllc}      
\hline\hline
\multicolumn{1}{c}{\begin{tabular}[c]{@{}c@{}}\textbf{Occulting}\\ \textbf{Object}\end{tabular}}&
\multicolumn{1}{c}{\begin{tabular}[c]{@{}c@{}}\textbf{Event Date}\\ (dd-mm-yyyy) \\ \textbf{Instant - UT} \\ (hh:mm:ss)\end{tabular}}&
\multicolumn{1}{c}{\begin{tabular}[c]{@{}c@{}}\textbf{Right Ascension}\\ (hh mm ss.ss)  \\ \textbf{Declination} \\ ($^\circ$ ' '')\end{tabular}}&
\multicolumn{1}{c}{\begin{tabular}[c]{@{}c@{}} \textbf{Astrometric}\\ \textbf{Uncertainty} \\ (mas) \end{tabular}} &
\multicolumn{1}{c}{\begin{tabular}[c]{@{}c@{}}\textbf{Label} \end{tabular}} \\   \hline
\multirow{3}{*}{(136199)}       & 06-11-2010         &  01 39 09.92735     &$\pm$ 1.0      & \multirow{2}{*}{U} \\                
                                & 02:20:00.000           & -04 21 12.04118     &$\pm$ 0.74                         \\ \cline{2-5}   
Eris                            & 29-08-2013         &  01 42 58.50247     &$\pm$ 1.2      & \multirow{2}{*}{S} \\                
                                & 15:38:00.000           & -03 21 17.39482     &$\pm$ 0.85                         \\ \hline  
(136472)                        & 23-04-2011         &  12 36 11.39029     &$\pm$ 1.1      & \multirow{2}{*}{U} \\                
Makemake                        & 01:36:00.00           & +28 11 10.3245      &$\pm$ 1.1                         \\  \hline
(136108)                        & 21-01-2017         &  14 12 03.199770    &$\pm$ 0.68     & \multirow{2}{*}{U} \\                
Haumea                          & 03:09:00.000           & +16 33 58.73306     &$\pm$ 0.57                         \\  \hline 
\multirow{3}{*}{(50000)}        & 04-05-2011         &  17 28 50.796367    &$\pm$ 0.42     & \multirow{2}{*}{U} \\                
                                & 02:40:00.000           & -15 27 42.80143     &$\pm$ 0.36                         \\ \cline{2-5} 
Quaoar                          & 17-02-2012         &  17 34 21.842097    &$\pm$ 0.26     & \multirow{2}{*}{U} \\                
                                & 04:30:00.000           & -15 42 10.33151     &$\pm$ 0.22                         \\ \hline  
\multirow{3}{*}{(208996)}       & 03-02-2012         &  07 45 54.776855    &$\pm$ 0.69     & \multirow{2}{*}{U} \\                
                                & 19:45:00.000           & +11 12 43.11513     &$\pm$ 0.44                         \\ \cline{2-5}
2003 AZ$_{84}$                  & 15-11-2014         &  08 03 51.288161    &$\pm$ 0.10     & \multirow{2}{*}{U} \\                
                                & 17:55:00.000         & +09 57 18.76056     &$\pm$ 0.13                         \\ \hline
(229762)                        & 15-11-2014         &  04 29 30.622370    &$\pm$ 0.40     & \multirow{2}{*}{U} \\                
2007 UK$_{126}$                 & 10:19:00.000          & -00 28 20.78081     &$\pm$ 0.43                         \\  \hline 
(278361)                        & 11-07-2019         &16 46 47.319636&$\pm$ 0.43& \multirow{2}{*}{S} \\              
2007 JJ$_{43}$                  & 05:44:00.000         &-26 16 50.42621&$\pm$ 0.32                    \\  \hline
\multirow{2}{*}{2017 OF$_{69}$} & 27-08-2019         &  20 13 12.06424     &$\pm$ 1.1      & \multirow{2}{*}{S} \\                
                                & 21:20:00.000         & -05 39 42.21020     &$\pm$ 0.59                         \\  \hline
(145451)                        & 24-12-2018         &  04 52 51.248767    &$\pm$ 0.26     & \multirow{2}{*}{D} \\                
2005 RM$_{43}$                  & 02:21:00.000           & +16 45 24.3162      &$\pm$ 3.5                         \\  \hline
(119951)                        & 19-09-2018         &  17 05 27.177401    &$\pm$ 0.37     & \multirow{2}{*}{D} \\                
2002 KX$_{14}$                  & 12:31:00.000          & -23 02 07.90824     &$\pm$ 0.80                         \\ \hline
(55636)                         & 09-10-2009         &  00 37 13.625731    &$\pm$ 0.27     & \multirow{2}{*}{U} \\                
2002 TX$_{300}$                 & 10:30:00.000           & +28 22 22.95729     &$\pm$ 0.27                         \\  \hline
(55638)                         & 03-12-2015         & 05 37 38.5189210&$\pm$ 0.04    & \multirow{2}{*}{S} \\          
2002 VE$_{95}$                  & 12:30:00.000           &  +08 05 09.419350& $\pm$ 0.064                     \\  \hline
(469506)                        & 24-05-2017         & 16 03 25.160345& $\pm$ 0.13     & \multirow{2}{*}{D} \\                 
2003 FF$_{128}$                 & 13:41:00.000         & -19 45 49.0695& $\pm$ 1.4                       \\  \hline
(54598)                         & 02-04-2018         & 03 00 51.843141&$\pm$ 0.15     & \multirow{2}{*}{S} \\                 
Bienor                          & 19:50:00.000        & +37 21 59.80884&$\pm$ 0.56                    \\ \hline
\multirow{3}{*}{(2060)}         & 07-11-1993         &  10 28 08.00772     &$\pm$ 2.4      & \multirow{2}{*}{U} \\                
                                & 13:15:00.00         & +03 30 36.3356      &$\pm$ 2.2                   \\ \cline{2-5}
Chiron                          & 29-11-2011         &  22 02 44.331192    &$\pm$ 0.23     & \multirow{2}{*}{U} \\                
                                & 08:16:00.000        & -05 55 12.0556      &$\pm$ 1.4                   \\ \hline
\multirow{2}{*}{2005 TV$_{189}$}& 13-11-2012         &  05 27 31.4687      &$\pm$ 10     & \multirow{2}{*}{S} \\          
                                & 22:27:00.00         & +12 59 31.7590      &$\pm$ 3.4                   \\  \hline \hline
\end{tabular}}
\end{table}

\begin{table}[!h]
\centering
\caption{Astrometric positions for the occulting object for occultation events with an ambiguous solution, i.e., when the center may be north or south of the observed chord. Our preferred solutions for Quaoar are shown in bold letters (see the discussion in Sect. \ref{Discussion}). These positions are in the ICRS as represented by the Gaia Celestial Reference Frame (Gaia-CRF2) \citep{GaiaCRF2}.}           
\label{table:objects_DR2_N_S} 
\resizebox{0.92\hsize}{!}{        
\begin{tabular}{lcllc}     
\hline\hline
\multicolumn{1}{c}{\begin{tabular}[c]{@{}c@{}}\textbf{Occulting}\\ \textbf{Object}\end{tabular}}&
\multicolumn{1}{c}{\begin{tabular}[c]{@{}c@{}}\textbf{Event Date}\\ (dd-mm-yyyy) \\ \textbf{Instant - UT} \\ (hh:mm:ss)\end{tabular}}&
\multicolumn{1}{c}{\begin{tabular}[c]{@{}c@{}}\textbf{Right Ascension}\\ (hh mm ss.ss)  \\ \textbf{Declination} \\ ($^\circ$ ' '')\end{tabular}}&
\multicolumn{1}{c}{\begin{tabular}[c]{@{}c@{}} \textbf{Astrometric}\\ \textbf{Uncertainty} \\ (mas) \end{tabular}} &
\multicolumn{1}{c}{\begin{tabular}[c]{@{}c@{}}\textbf{Solution} \end{tabular}} \\   \hhline{=====}
\multirow{11}{*}{(50000)}     &                                     &  17 28 47.613559       &$\pm$ 0.40  & \textbf{\multirow{2}{*}{Northern}} \\      
                              & 11-02-2011                      & -15 41 58.91524        &$\pm$ 0.34     & \\ \cline{3-5}
                              & 10:04:00.000                    &  17 28 47.613773       &$\pm$ 0.40     & \multirow{2}{*}{Southern}\\           
                              &                                     & -15 41 58.94064        &$\pm$ 0.34  &\\ \hhline{~====}
                              &                                     &  17 28 10.137009       &$\pm$ 0.44  & \textbf{\multirow{2}{*}{Northern}} \\ 
                              & 15-10-2012                      & -15 36 23.32076        &$\pm$ 0.40     & \\ \cline{3-5}
Quaoar                        & 00:45:00.000                    &  17 28 10.136524       &$\pm$ 0.44     & \multirow{2}{*}{Southern}\\           
                              &                                     & -15 36 23.35283        &$\pm$ 0.40  &\\ \hhline{~====}
                              &                                     &  17 34 40.463793       &$\pm$ 0.17  & \textbf{\multirow{2}{*}{Northern}} \\   
                              & 09-07-2013                      & -15 23 37.48276        &$\pm$ 0.30     & \\ \cline{3-5}
                              & 02:40:00.000                    &  17 34 40.463800       &$\pm$ 0.17     & \multirow{2}{*}{Southern}\\           
                              &                                     & -15 23 37.48297        &$\pm$ 0.30  &\\  \hhline{=====}
\multirow{3}{*}{(90377)}      &                                     &  03 32 44.708068       &$\pm$ 0.85  & \multirow{2}{*}{Northern} \\            
                              & 13-01-2013                      & +06 47 18.9865         &$\pm$ 1.2      & \\ \cline{3-5}
Sedna                         & 12:25:00.000                    &  03 32 44.708049       &$\pm$ 0.85     & \multirow{2}{*}{Southern}\\           
                              &                                     & +06 47 18.9851         &$\pm$ 1.2   &\\  \hhline{=====}
\multirow{7}{*}{(208996)}     &                                     &  07 43 41.83465        &$\pm$ 1.9   & \multirow{2}{*}{Northern} \\               
                              & 08-01-2011                      & +11 30 23.4571         &$\pm$ 1.4      & \\ \cline{3-5}
                              & 06:29:00.00                     &  07 43 41.83452        &$\pm$ 1.9      & \multirow{2}{*}{Southern}\\           
                              &                                     & +11 30 23.4411         &$\pm$ 1.4   &\\  \hhline{~====}
2003 AZ$_{84}$                &                                     &  07 58 56.722173       &$\pm$ 0.89  & \multirow{2}{*}{Northern} \\               
                              & 02-12-2013                      & +10 19 04.88115        &$\pm$ 0.64     & \\ \cline{3-5}
                              & 14:53:00.000                    &  07 58 56.722237       &$\pm$ 0.89     & \multirow{2}{*}{Southern}\\           
                              &                                     & +10 19 04.86927        &$\pm$ 0.64  &\\  \hhline{=====}
\multirow{7}{*}{(278361)}     &                                     &  16 45 23.14122        &$\pm$ 1.6   & \multirow{2}{*}{Northern} \\               
                              & 07-08-2019                      & -26 11 07.4081         &$\pm$ 2.7      & \\ \cline{3-5}
                              & 13:06:00.00                     &  16 45 23.14062        &$\pm$ 1.6      & \multirow{2}{*}{Southern}\\                
                              &                                     & -26 11 07.4261         &$\pm$ 2.7   &\\  \hhline{~====}
2007 JJ$_{43}$                &                                     &  16 57 27.967999       &$\pm$ 0.47  & \multirow{2}{*}{Northern} \\               
                              & 24-05-2020                      & -26 18 43.95822        &$\pm$ 0.28     & \\ \cline{3-5}
                              & 11:30:00.000                    &  16 57 27.967851       &$\pm$ 0.56     & \multirow{2}{*}{Southern}\\                
                              &                                     & -26 18 43.9741         &$\pm$ 3.0   &\\  \hhline{=====}
\multirow{3}{*}{(28978)}      &                                     &17 17 54.235706&$\pm$ 0.11& \multirow{2}{*}{Northern} \\      
                              & 24-06-2014                      &-26 40 55.17494&$\pm$ 0.39& \\ \cline{3-5}
Ixion                         & 14:54:00.000                &17 17 54.235664&$\pm$ 0.11& \multirow{2}{*}{Southern}\\               
                              &                                     &\-26 40 55.19446&$\pm$ 0.39&\\  \hhline{=====}
\multirow{3}{*}{(145451)}     &                                     &  04 50 01.05514        &$\pm$ 1.8   & \multirow{2}{*}{Northern} \\               
                              & 03-02-2019                      & +16 51 18.7545         &$\pm$ 4.2      & \\ \cline{3-5}
 2005 RM$_{43}$               & 10:57:00.00                     &  04 50 01.05497        &$\pm$ 1.8      & \multirow{2}{*}{Southern}\\                
                              &                                     & +16 51 18.7484         &$\pm$ 4.2   &\\  \hhline{=====}
\multirow{3}{*}{(444030)}     &                                     &  21 32 10.91190        &$\pm$ 4.8   & \multirow{2}{*}{Northern} \\          
                              & 16-11-2017                      & +18 37 31.0549         &$\pm$ 2.2      &  \\ \cline{3-5}
2004 NT$_{33}$                & 18:20:00.00                     &  21 32 10.91133        &$\pm$ 4.8      & \multirow{2}{*}{Southern}\\           
                              &                                     & +18 37 31.0512         &$\pm$ 2.2   &  \\  \hhline{=====}
\multirow{3}{*}{(119951)}     &                                     &  16 35 04.26341        &$\pm$ 1.8   & \multirow{2}{*}{Northern} \\            
                              & 26-04-2012                      & -22 15 22.7292         &$\pm$ 1.8      & \\ \cline{3-5}
2002 KX$_{14}$                & 02:35:00.00                     &  16 35 04.26334        &$\pm$ 1.8      & \multirow{2}{*}{Southern}\\
                              &                                     & -22 15 22.7354         &$\pm$ 1.8   &\\  \hhline{=====}
\multirow{3}{*}{  (175113)}&                            &   23 01 47.574516&  $\pm$ 0.65     & \multirow{2}{*}{  Northern} \\          
                              &   28-09-2018            &   -20 31 01.2497 &  $\pm$ 1.1    & \\ \cline{3-5}
  2004 PF$_{115}$      &   01:36:00.000 &   23 01 47.574730&  $\pm$  0.65       & \multirow{2}{*}{  Southern}\\           
                              &                                     &   -20 31 01.2602 &  $\pm$ 1.1 &\\  \hhline{=====}
\multirow{3}{*}{(119979)}     &                                     &  05 50 53.889168&$\pm$ 0.41         & \multirow{2}{*}{Northern} \\            
                              & 30-12-2018                      &  +19 55 52.6368 &$\pm$ 1.1      & \\ \cline{3-5}
2002 WC$_{19}$                & 11:58:00.000                &  05 50 53.889148&$\pm$ 0.41    & \multirow{2}{*}{Southern}\\           
                              &                                     &   +19 55 52.6269&$\pm$ 1.1    &\\  \hhline{=====}
\multirow{3}{*}{(54598)}      &                                     &  02 49 15.578865&$\pm$ 0.42 & \multirow{2}{*}{  Northern} \\             
                              & 29-12-2017                      &  +38 29 52.39778&$\pm$ 0.40     & \\ \cline{3-5}
Bienor                        & 16:24:00.000                &  02 49 15.579994&$\pm$ 0.42    & \multirow{2}{*}{  Southern}\\         
                              &                                     &  +38 29 52.38545&$\pm$ 0.40  &\\  \hhline{=====} 
\multirow{3}{*}{(8405)}       &                                     &  03 47 11.771545  &  $\pm$ 0.16     & \multirow{2}{*}{  Northern} \\             
                              & 24-11-2013                      &   +36 02 10.07830&  $\pm$ 0.59   & \\ \cline{3-5}
Asbolus                       & 05:23:00.000                &  03 47 11.771563  &  $\pm$ 0.16  & \multirow{2}{*}{  Southern}\\         
                              &                                     &  +36 02 10.07711 &  $\pm$ 0.59 &\\  \hhline{=====}
\multirow{3}{*}{(60558)}      &                                     &  17 24 26.139113       &$\pm$ 0.18  & \multirow{2}{*}{Northern} \\               
                              & 25-06-2012                      & -18 10 19.80427        &$\pm$ 0.46     & \\ \cline{3-5}
Echeclus                          & 23:36:00.000                    &  17 24 26.139100       &$\pm$ 0.18  & \multirow{2}{*}{Southern}\\           
                              &                                     & -18 10 19.80965        &$\pm$ 0.46  &\\  \hhline{=====}
\end{tabular}}
\end{table}

%____________________________________________________________________
\section{Physical constraints}
\label{Physical}

Even single- or double-chord events may bring further information on the objects physical properties. The chord length of each single-chord detection can be used as a lower limit to the equatorial diameter of the object, and this can be found in Table \ref{tab:times}. Furthermore, these values can be compared to size measurements made with other techniques, such as radiometric observations. We present three events below for which information on the object shape could also be obtained.

%-------------

\subsection{2002 VE$_{95}$}
\label{sec:VE95}

The first occultation by this Plutino\footnote{TNOs classified as Plutinos are those with a Pluto-like orbit, that is, they are in 2:3 mean-motion resonances with Neptune \citep{Fernandez20}.} was observed on 3 December 2015. It was a single-chord detection presenting a chord with a length of 279.5 $\pm$ 2.5 km. The equivalent diameter of 249.8 $^{+13.5}_{-13.1}$ km, the most precise so far, was determined by \cite{MM12} using measurements from the Herschel Space Observatory and the Spitzer Space Telescope. As the observed chord is longer than the reported equivalent diameter, we assumed an object with an elliptical shape on the plane of the sky to preserve the equivalent area for this reported diameter. Then we calculated the flattening ($\epsilon$) of the body by using  Eq. \ref{eq:Requiv}, where $R_{equiv}$ is the reported equivalent radius, and $a$ is the equatorial radius or half of the maximum length of the chord,

\begin{equation}
    R_{equiv}=a\cdot\sqrt{1-\epsilon}
    \label{eq:Requiv}
.\end{equation}

\begin{figure}[!h]
    \centering
    \includegraphics[scale=0.6]{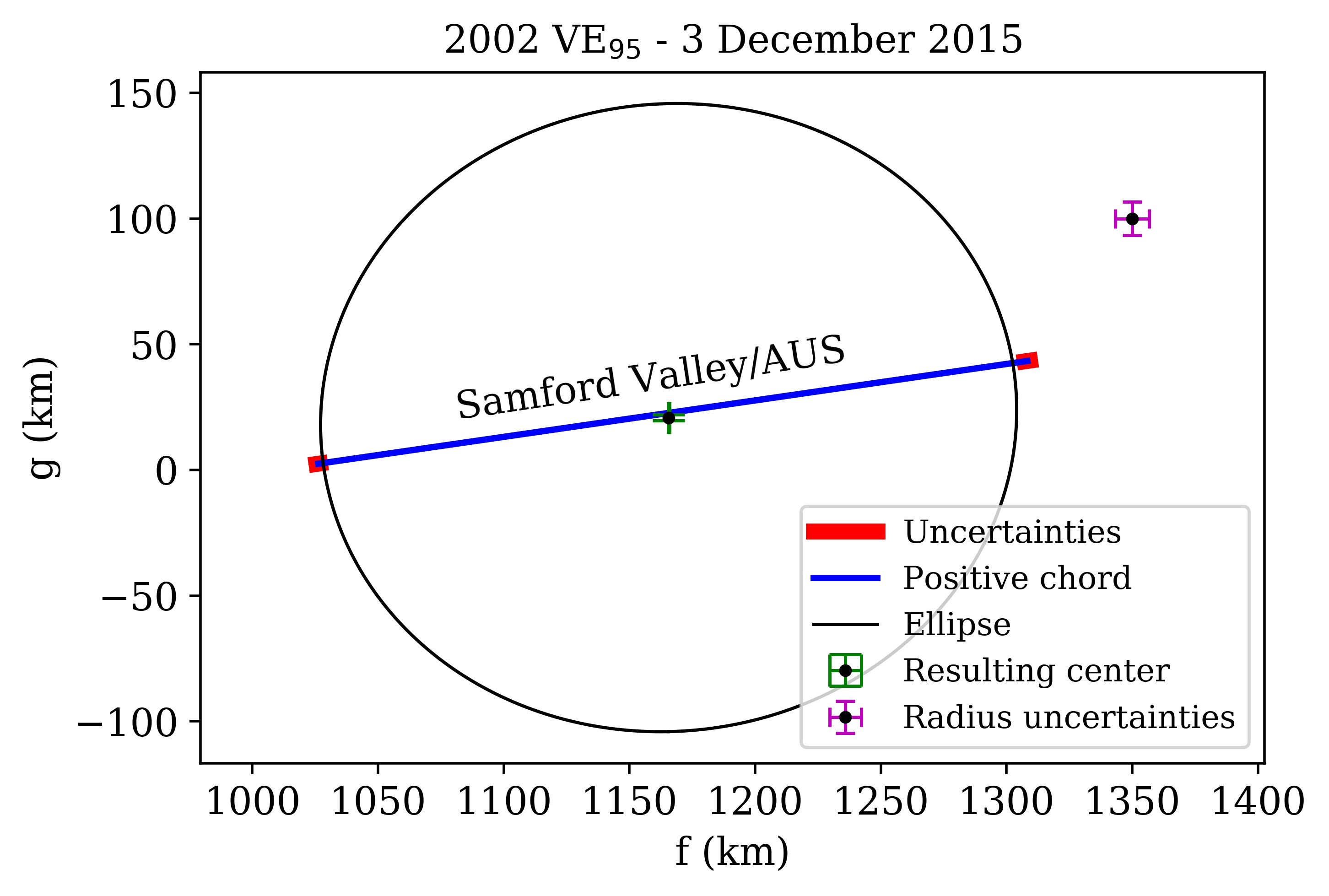}
    \caption{Occultation by 2002 VE$_{95}$ observed on 3 December 2015. The blue segment corresponds to the detected chord, with ingress and egress uncertainties in red. The green cross represents the center uncertainty. The purple bars are the radius uncertainties given by \cite{MM12}, and the black ellipse represents the object profile. }
    \label{fig:VE95}
\end{figure}

The result is an ellipse with a minimum flattening of 0.201, as presented in Fig. \ref{fig:VE95}. The propagated error can be calculated from Eq. \ref{eq:epsilon_err} and has a value of $\pm$ 0.084. In this case, considering the above assumptions, just one central chord allowed us to determine a minimum value for the object flattening. Only one astrometric position could then be derived for this single-chord detection,

\begin{equation}
    \begin{array}{lr}
      \centering \sigma\epsilon= \sqrt{\left( \frac{\partial \epsilon}{\partial a}\cdot \sigma a\right)^2 + \left( \frac{\partial \epsilon}{\partial {R_{equiv}}}\cdot \sigma{R_{equiv}} \right)^2 } = &  \\
      \\
         \multicolumn{2}{r}{ \sqrt{\left(\frac{2\cdot{R_{equiv}^2}}{a^3}\cdot \sigma a \right)^2 + \left(\frac{-2\cdot{R_{equiv}}}{a^2}\cdot \sigma{R_{equiv}} \right)^2}}
    \end{array}
    \label{eq:epsilon_err}
.\end{equation}   

% ------------

\subsection{2003 FF$_{128}$}
\label{sec:FF128}

\begin{figure}[!h]
    \centering
    \includegraphics[width=\linewidth]{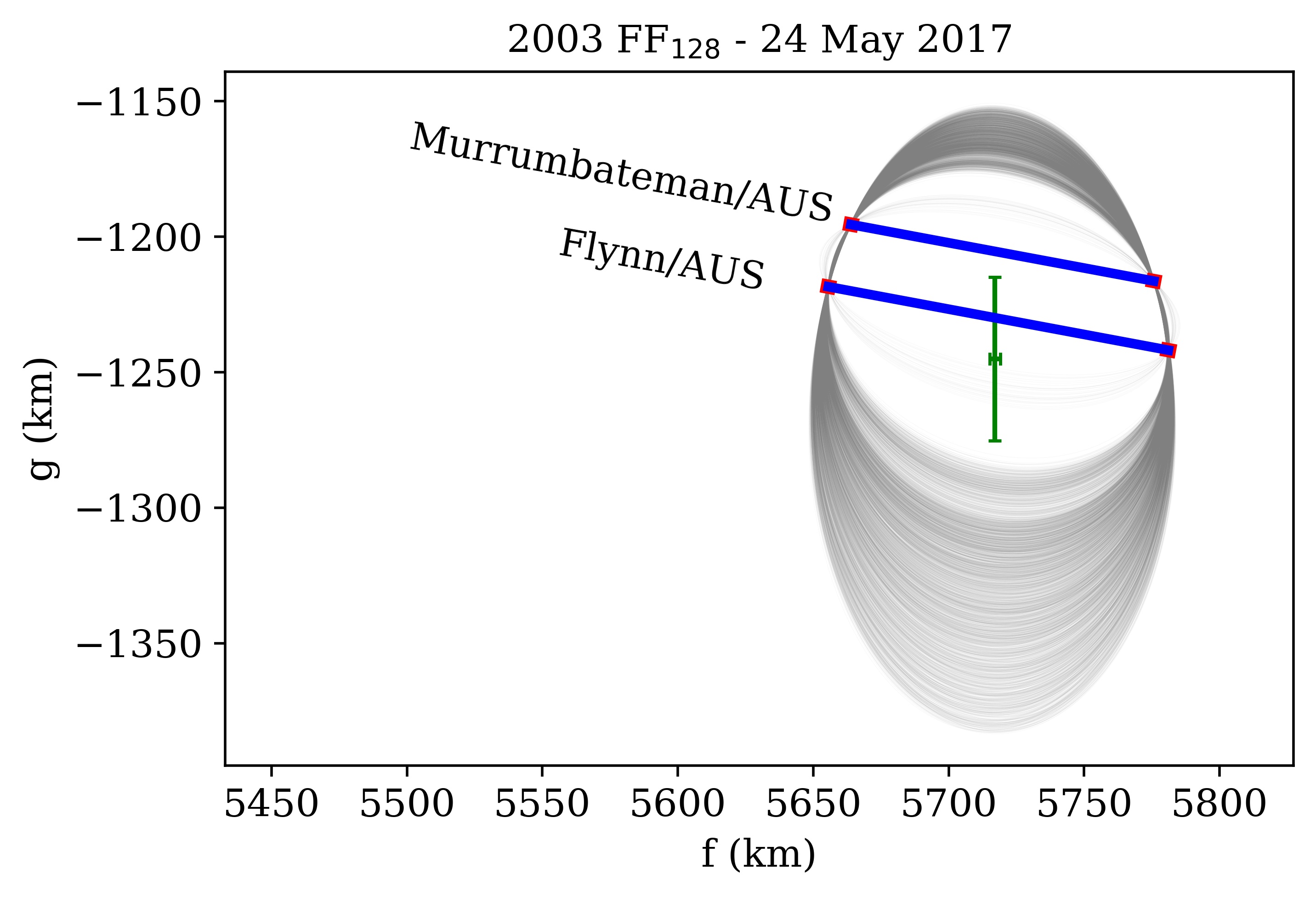}
    \caption{Same as in Fig. \ref{fig:VE95}, but for the double-chord occultation by the Plutino 2003 FF$_{128}$. Each limb solution at 1 $\sigma$ level is presented by a gray ellipse. The large error bar in declination is due to the uncertainty on the apparent oblateness of the object.}
    \label{fig:FF128}
\end{figure}

The only information available about this Plutino is that its absolute magnitude (H) equals 6.7, available at the MPC web site\footnote{$^\dag~$~\url{https://minorplanetcenter.net/iau/lists/TNOs.html}}. The first stellar occultation ever observed for this object was recorded on 24 May 2017 from two sites in Australia. The detection does not allow a circular fit, so that ellipses were fit. An upper limit for oblateness of 0.6 was imposed to avoid physically implausible solutions. At the 1$\sigma$ level, the results are ellipses with an average oblateness of 0.36 $\pm$ 0.24 and an equatorial radius of 90 $\pm$ 25 km. The astrometric position of the object is obtained from the average of the obtained centers. Fig. \ref{fig:FF128} presents  in gray all possible ellipses at the 1$\sigma$ level. The equivalent diameter ranges from 86 km to 176 km, which is smaller than the diameter obtained from its absolute magnitude, but it is not a significant disagreement, considering that the error of the latter is unknown.

% ----------

\subsection{2005 TV$_{189}$}
\label{sec:TV189}

\begin{figure}[!h]
    \centering
    \includegraphics[scale=0.63]{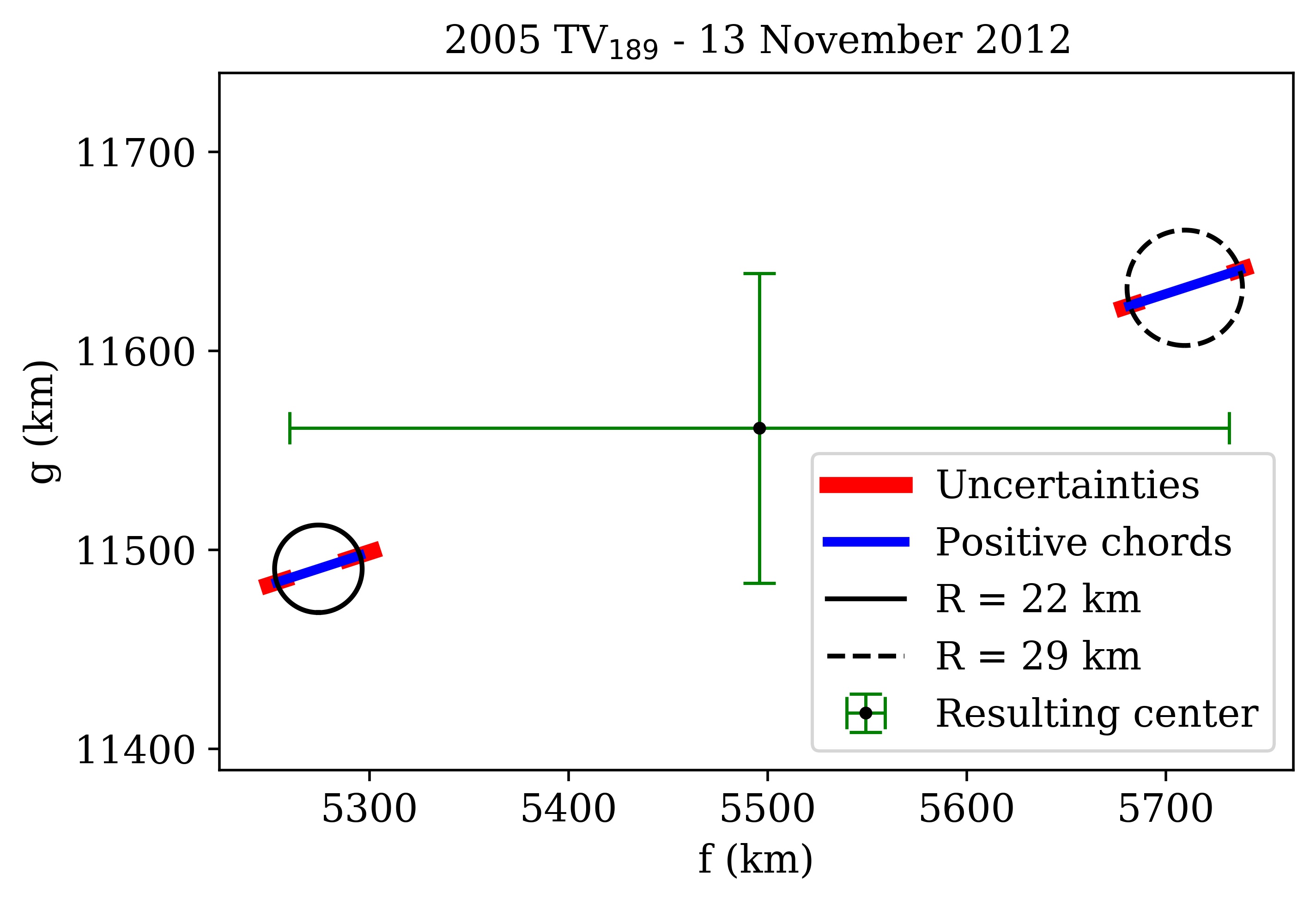}
    \caption{Occultation by the Plutino 2005 TV$_{189}$, observed visually on 13 November 2012, presented two consecutive events. The large green cross represents the astrometric position and respective uncertainty derived from this observation, obtained from the average distance between the two detections.}
    \label{fig:TV189}
\end{figure}

This TNO is also classified as a Plutino and presents the highest know inclination of Plutinos (i = $34.5\deg$) \citep{Almeida09}. According to the MPC web site$^{\dag}$ , it has H = 7.5. The only stellar occultation ever observed for this body was detected on 13 November 2012. It was a visual observation with ingress and egress times measured using a Chronograph ACH-77 instrument. Two consecutive drops in the star flux were observed. This yields two different scenarios: 1) a single object intercepting the light from a binary star, or 2) a binary object passing in front of one star. During the data processing of \textit{Gaia} DR2, all sources up to G mag 17 were considered as single stars to determine the five astrometric parameters. The star occulted by 2005 TV$_{189}$ presents G = 11.59 mag with a complete set of astrometric parameters where no duplicity or binary flags are present, as provided by Gaia DR2. We emphasize that \textit{Gaia} DR2 used "gold photometry" for this star to determine its parameters, meaning that it is probably not a binary. If the star is not double, a binary Plutino was detected upon this stellar occultation. In this case, the larger component has a diameter of 58 $\pm$ 4 km, and the secondary 44 $\pm$ 11 km (considering diametrical chords, see Fig. \ref{fig:TV189}). At the occultation instant, the two components were separated by 403 km on the sky plane and 31.26 au away from the Sun, corresponding to a separation of 18 mas.

Although an experienced observer has made this observation and the detection is not suspicous, visual observations should be taken with caution because they cannot be verified with different approaches and the information is limited. The possible duplicity of the object should therefore be verified by another observation before any discovery is claimed. For the purpose of this work, this observation does provide one useful astrometric position for this body.

%__________________________________________________________________

\section{Discussion and conclusions}
\label{Discussion}

\begin{figure}[!h]
    \centering
    \includegraphics[scale=0.32]{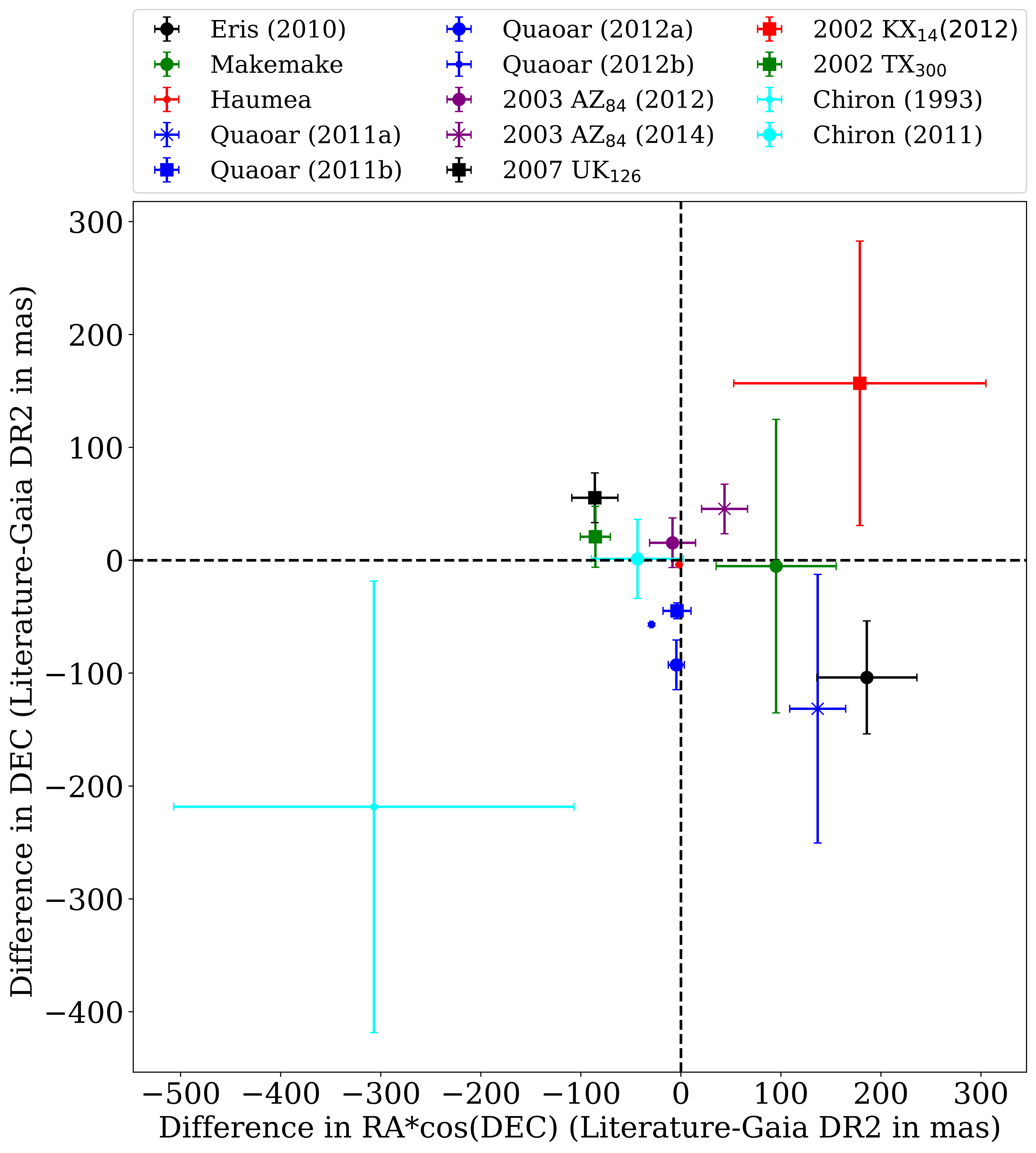}
    \caption{Difference between the star positions used to derive the object positions as published in the literature and the positions obtained from \textit{Gaia} DR2, as described in Sect. \ref{Gaia}. The x-axis presents the difference (Literature minus Gaia DR2) in Right Ascension (RA$\cdot\cos{\delta}$), and the y-axis corresponds to the difference in declination. Differences as large as $\sim$300 mas are present.}
    \label{fig:delta_star}
\end{figure}

Based on the high accuracy of \textit{Gaia} DR2, stellar occultations can now provide accurate astrometric positions at the milliarcsecond level (or better). We used stellar occultation data to obtain a set of astrometric positions for 19 TNOs and four Centaurs. We have also updated positions of objects that were observed in multichord occultation events, available in the literature. The differences between the reference star position used in the original occultation analysis (Table \ref{table:stars_published}) and the position from \textit{Gaia} DR2 propagated to the event epoch (Table \ref{table:stars_DR2}) are shown in Fig. \ref{fig:delta_star}. It is clear that large zonal errors were present, causing inaccurate offsets on the ephemeris calculated with the former positions.

No error bars for the reference stars were provided in the works about the occultation events by Haumea, Quaoar (11-02-2011), 2002 KX$_{14}$ (2012), 2002 TX$_{300}$, and Chiron (1993 and 2011), therefore we assumed them to be the errors provided by the reference catalog. For the two single-chord events of 2003 AZ$_{84}$ \citep{Dias-Oliveira17}, the stellar positions were not provided by the authors, so we were unable to compare the respective accuracy with \textit{Gaia} DR2 positions. The respective positions were obtained using the event instants that were provided.

\begin{figure*}[!h]
\resizebox{\hsize}{!}{
\includegraphics[width=0.48\linewidth]{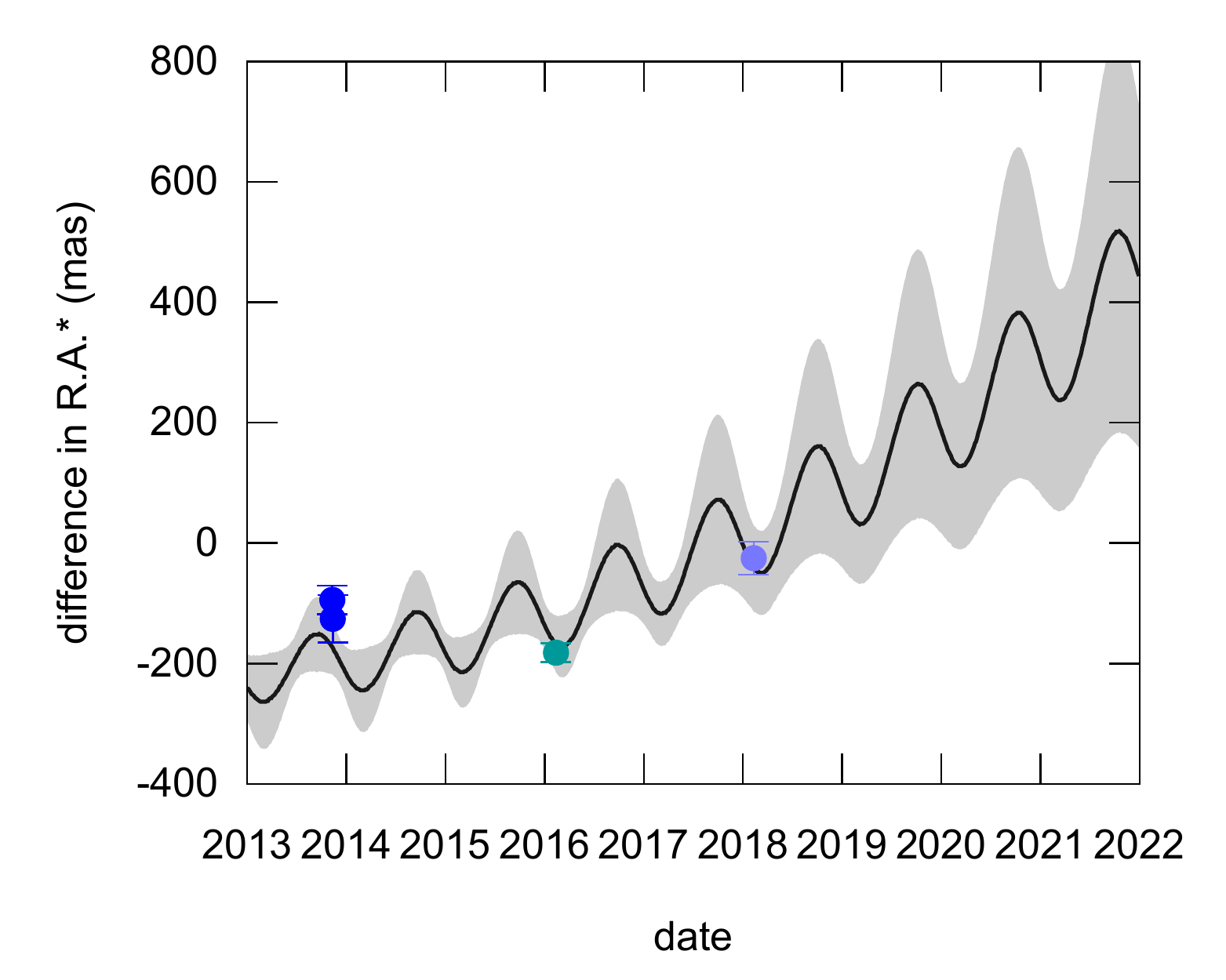}\quad
\includegraphics[width=0.48\linewidth]{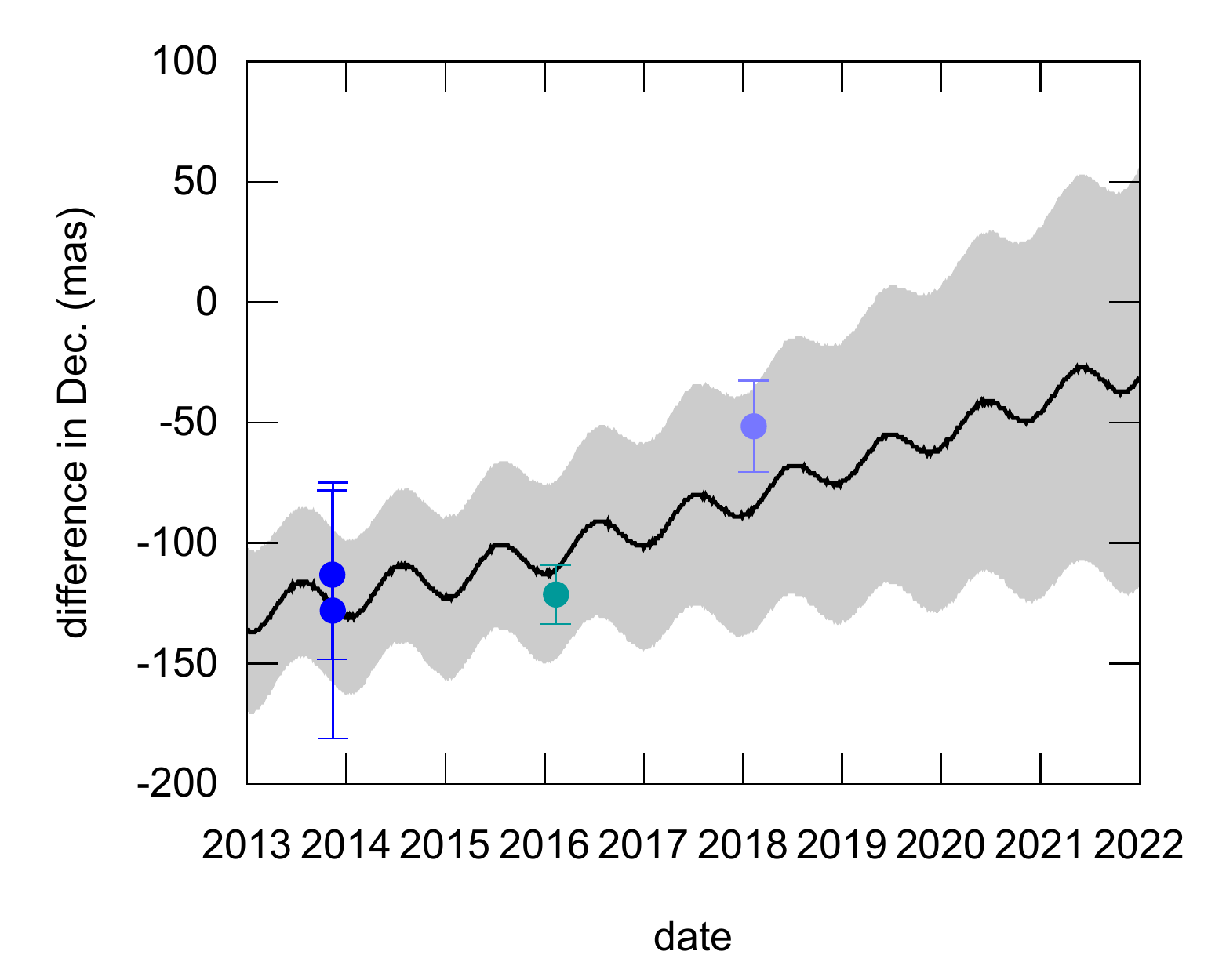}}
\resizebox{\hsize}{!}{
\includegraphics[width=0.48\linewidth]{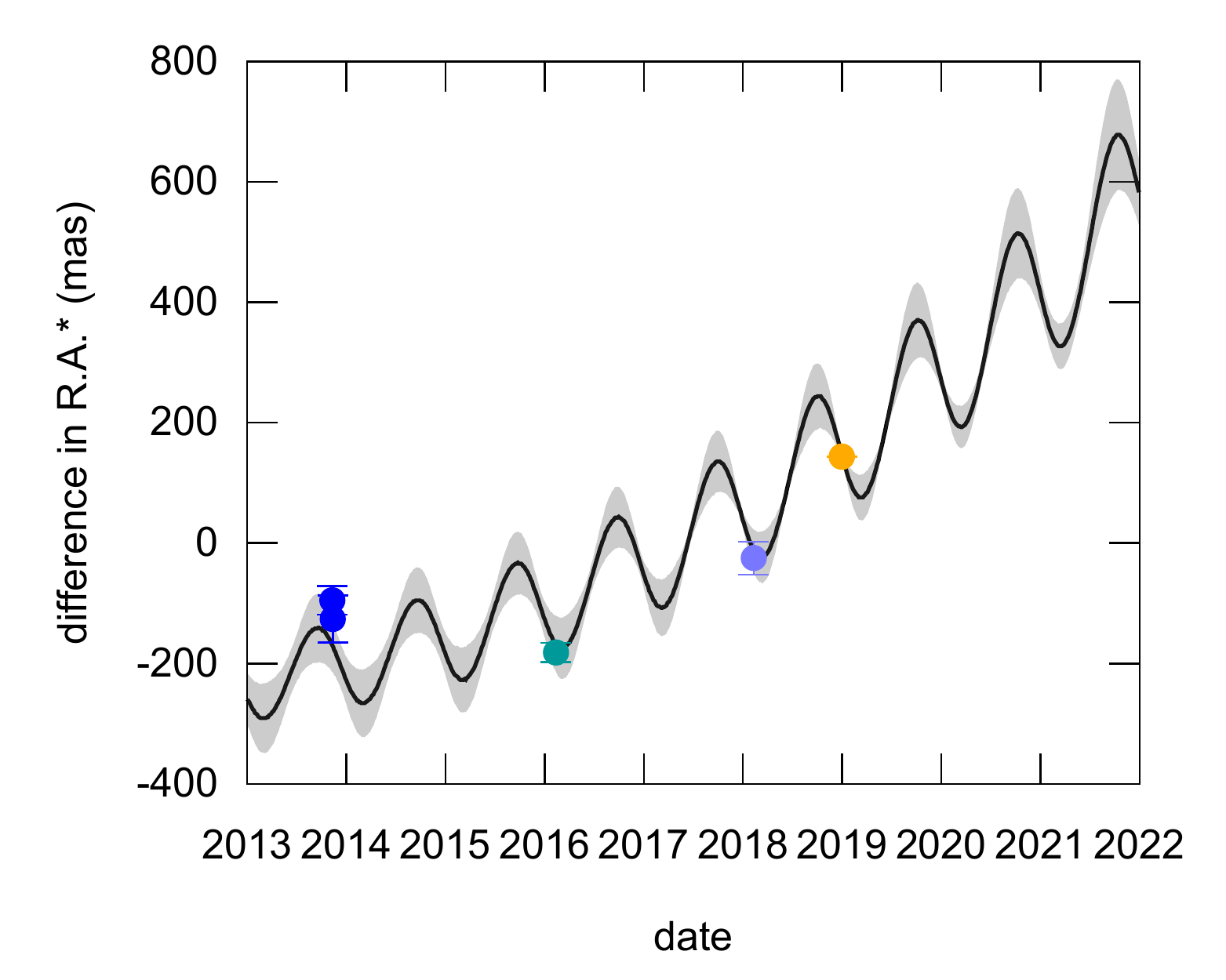}\quad
\includegraphics[width=0.48\linewidth]{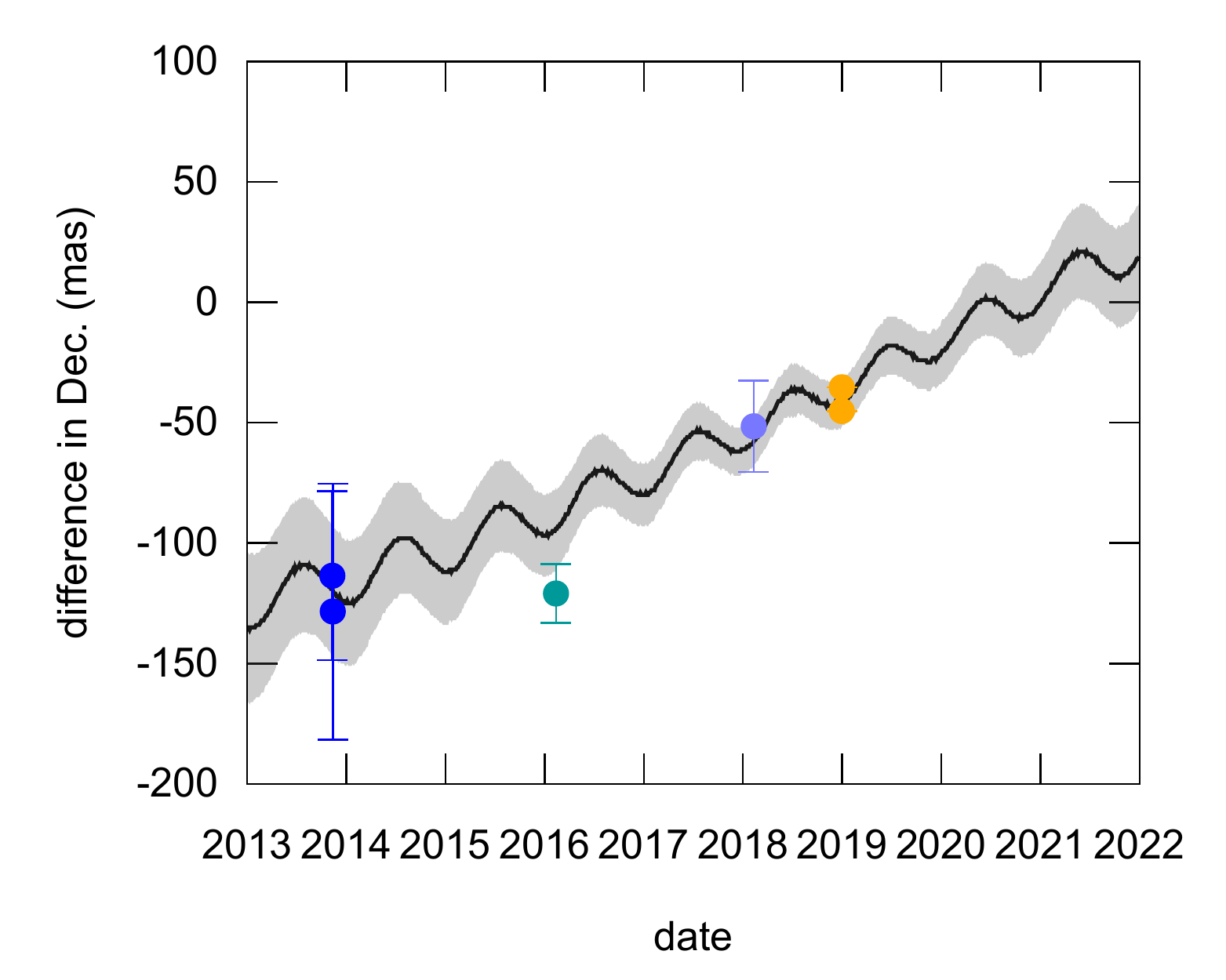}}
\caption{Upper panels: Ephemeris of 2002 WC$_{19}$ calculated with NIMA \citep{Desmars15} using positions from classic astrometry (upper panels), and including astrometric position obtained from a stellar occultation (yellow dots in the lower panels). This was a single-chord event, with two possible solutions, therefore two yellow dots are shown in the lower right panel because the northern and southern solutions were both used to calculate this orbit. The black lines represent the difference between NIMA and JPL 06 ephemerides, and the gray regions are the 1$\sigma$ uncertainties from NIMA. The uncertainty is clearly reduced when the astrometric positions derived from a stellar occultation are included.}
\label{fig:WC19}
\end{figure*}

\begin{figure*}[!h]
\resizebox{\hsize}{!}{
\includegraphics[scale=0.2]{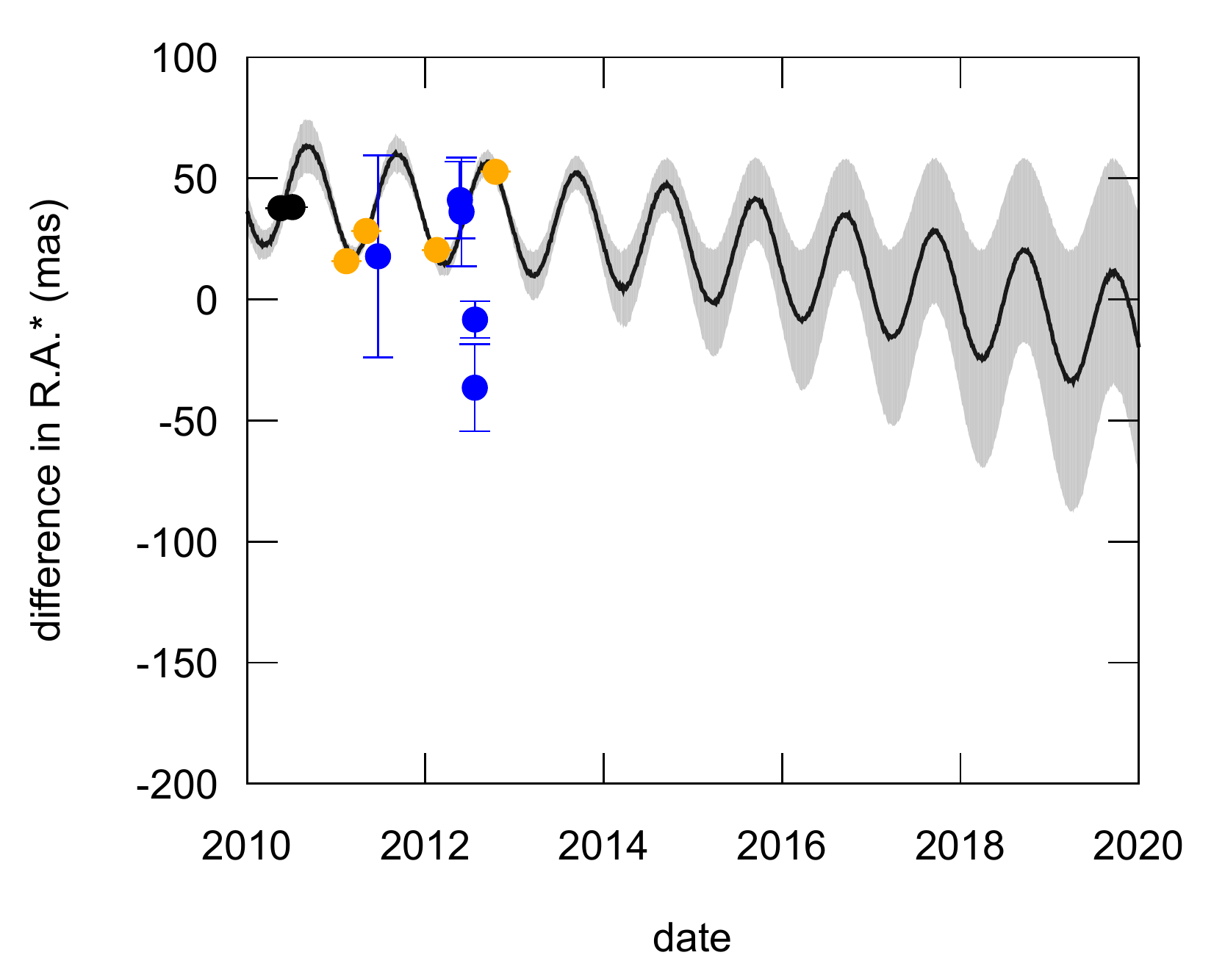}\quad
\includegraphics[scale=0.2]{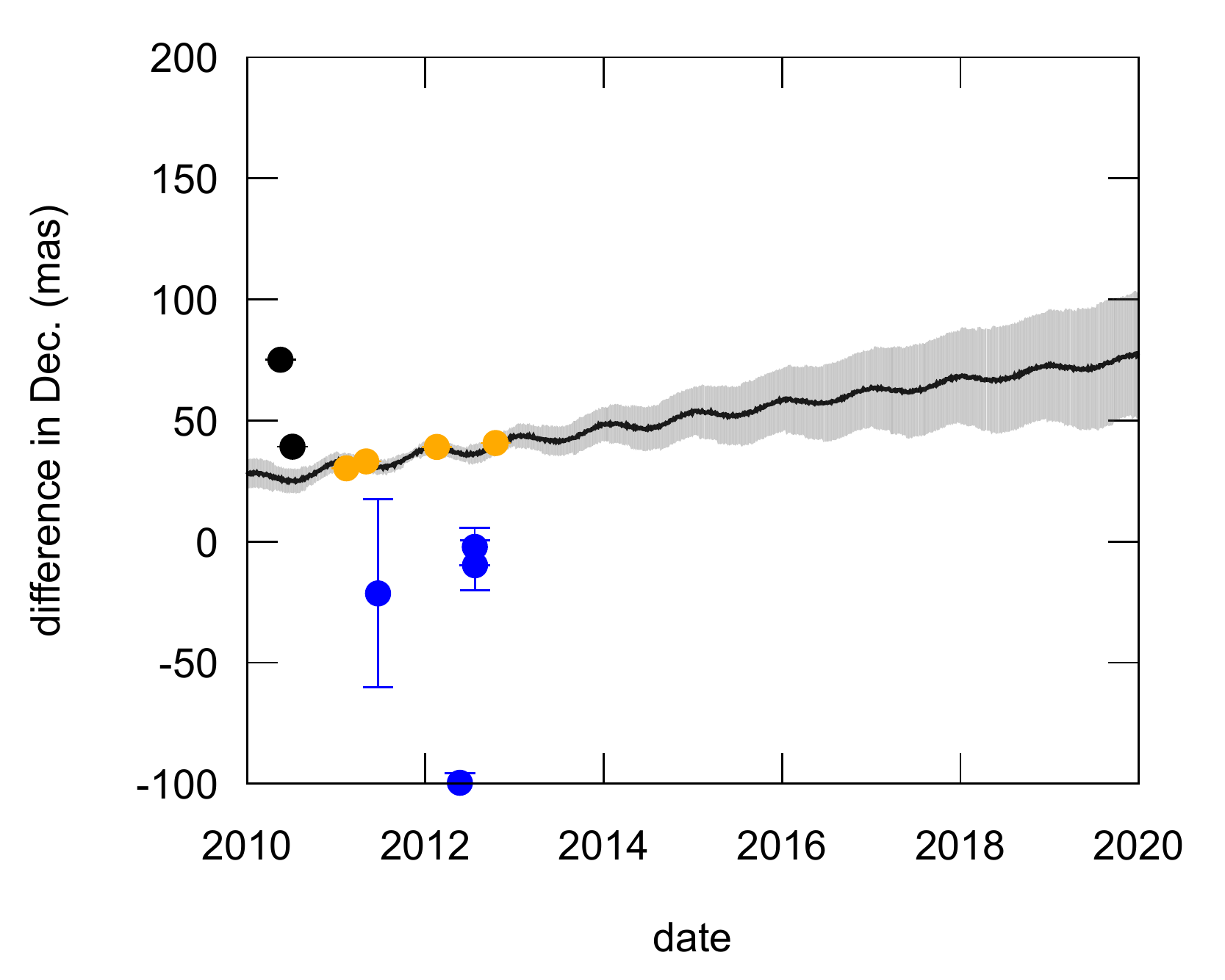}}
\resizebox{\hsize}{!}{
\includegraphics[scale=0.2]{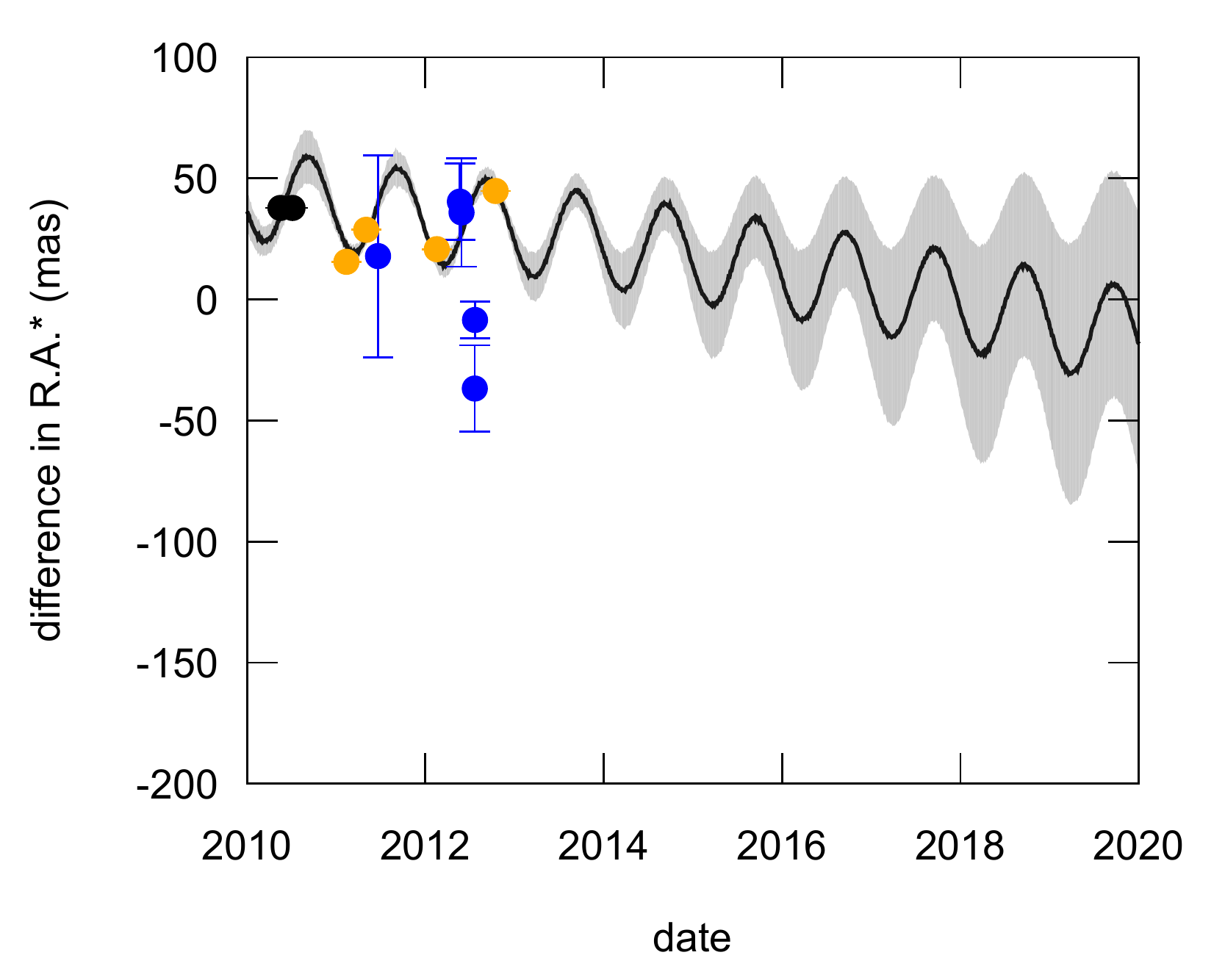}\quad
\includegraphics[scale=0.2]{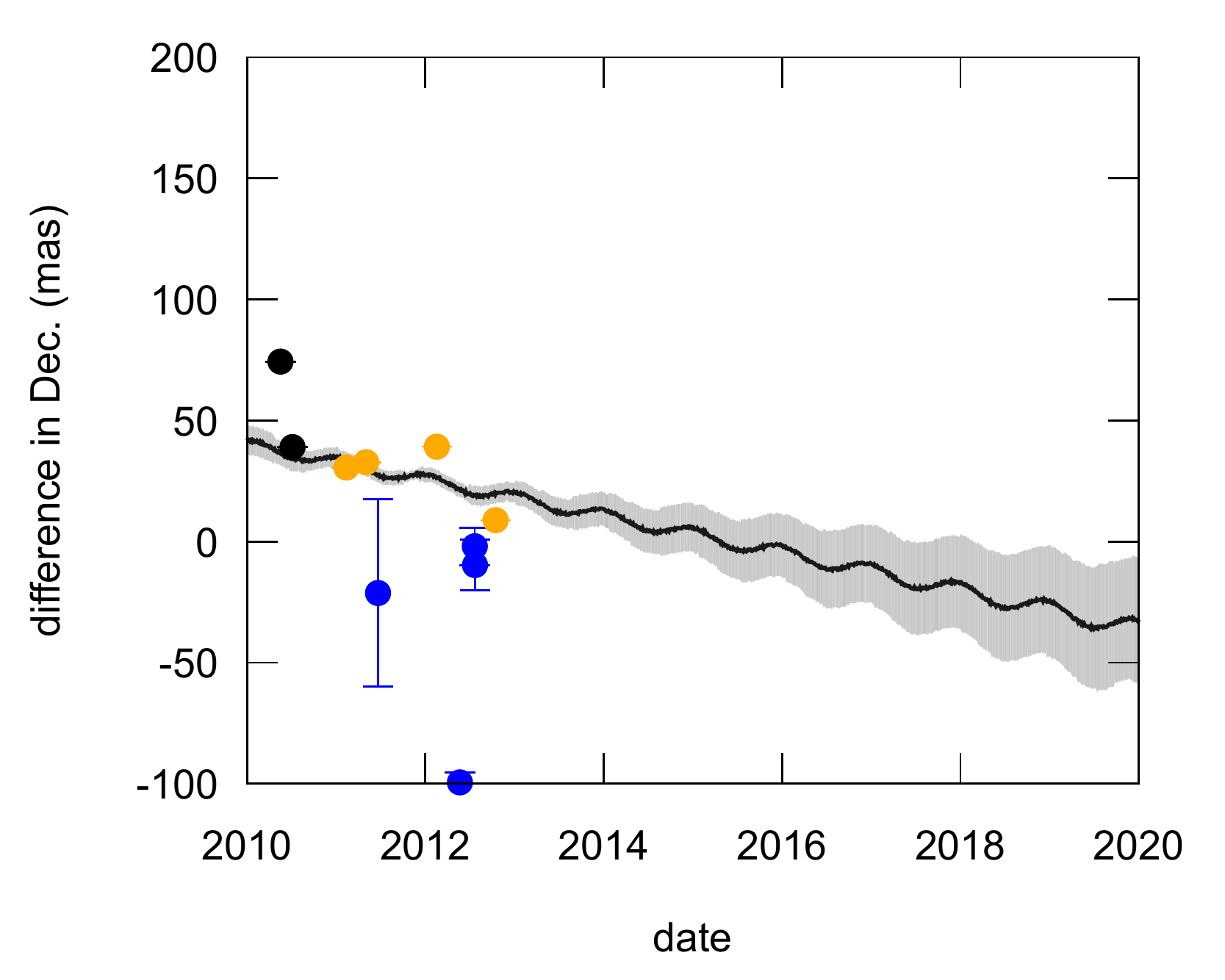}}
\caption{Ephemeris plots for Quaoar. The black lines again show the difference between NIMA and JPL. Gray regions are NIMA ephemeris uncertainties, and the yellow dots are astrometric positions from stellar occultation events. The black and blue dots are the object positions from classical astrometry. In the upper panels we show the northern astrometric solution from Table \ref{table:objects_DR2_N_S}. In the lower panels we used the southern solution. The northern portions (upper panels) fit the ephemeris better, including the other yellow dots, allowing us for choosing this one as our preferred position.}
\label{fig:Quaoar1}
\end{figure*}

The positions provided in this work significantly improve the accuracy and precision of the respective object ephemerides, at least for the coming years \citep{Desmars19}. We showed that even a single-chord detection with an ambiguous result allows for significant improvement, as shown in Fig. \ref{fig:WC19}. When a few precise astrometric positions from stellar occultations are accumulated, it is possible to break the ambiguity and choose the best solution (northern or southern) of the single-chord detections. This case is shown in Fig. \ref{fig:Quaoar1}, which presents two different solutions to Quaoar ephemerides, using the northern and southern positions for the single-chord detection in October 2012, respectively.  We noted that the northern solution gives a better fit, which was confirmed by recent occultation detections of this object, made in 2019 and 2020 (not presented in this paper).

One great benefit of precise astrometric positions derived from stellar occultations is that they allow the precise prediction of events involving distant Solar System objects in the near future, enabling focused observation campaigns to fully characterize the respective bodies. The accumulation of these precise positions will render even better predictions, to the point of enabling the selection of the best events for the study of local topographic features such as rings and satellites.

When the object ephemeris is accurate, some other factors can affect the precision of the future stellar occultation predictions. 
1) Most of the large TNOs have large satellites, indicating that some of these objects may have unknown moons. In this case, the ephemeris is given for the photocenter, while the stellar occultation position gives the center of the main body. If the secondary body is large enough, the offset between the photocenter and the center of mass can affect the prediction uncertainties.
2) For a small fraction of stars, \textit{Gaia} DR2 can flag a duplicity, which means that a second near-source was detected but discarded from data reduction. It can be a real detection of a near star or a duplicated detection of the same star. In any case, when a flag is present, it can add some uncertainties. The catalog provides this flag for any detection inside a radius of 2" from another source. However, undetected close companions can be missed by Gaia, as in the case described in \cite{Sickafoose2019}. 
3) The reference epoch for the astrometric positions provided by \textit{Gaia} DR2 is J2015.5. As we propagate the uncertainties of proper motion in time, they will reach the mas level in a few years. In our sample, the higher proper motion uncertainty is for a star with G = 18.1395 with of 0.3884 mas/year. A simple propagation in time shows that about eight years are needed to reach the one-mas level. Nevertheless, \textit{Gaia} Early Data Release 3 - EDR3\footnote{https://www.cosmos.esa.int/web/gaia/earlydr3} , which will be made available on 3 December 2020,  will improve the knowledge of these uncertainties by a factor of two. In general, the proper motion uncertainties are so small that at least two decades are needed before it becomes a significant factor in stellar occultation predictions.

%____________________________________________________________

\begin{acknowledgements}

 This study was financed in part by the Coordenação de Aperfeiçoamento de Pessoal de Nível Superior - Brasil (CAPES) - Finance Code 001 and the National Institute of Science and Technology of the e-Universe project (INCT do e-Universo, CNPq grant 465376/2014-2).The following authors acknowledge the respective CNPq grants: F.B-R 309578/2017-5; RV-M 304544/2017-5, 401903/2016-8; J.I.B.C. 308150/2016-3; M.A. 427700/2018-3, 310683/2017-3, 473002/2013-2; O.C.W. 305210/2018-1. The following authors acknowledge the respective grants: B.E.M. thanks the CAPES/Cofecub-394/2016-05 grant; G.B-R acknowledges CAPES-FAPERJ/PAPDRJ grant E26/203.173/2016; M.A. acknowledges F.A.P.E.R.J. grant E-26/111.488/2013; ARGJr acknowledges F.A.P.E.S.P. grant 2018/11239-8; O.C.W. and R.S. acknowledge F.A.P.E.S.P. grant 2016/24561-0.  D.N. acknowledges support from the French Centre National d’Etudes Spatiales (C.N.E.S.). K.H. and P.P. were supported by the project R.V.O.: 67985815. A.P. and R.S. received support from the K-125015 grant of the National Research, Development and Innovation Office (N.K.F.I.H., Hungary). Partial funding of the computational infrastructure and database servers are received from the grant KEP-7/2018 of the Hungarian Academy of Sciences. Some of the results were based on observations taken at the 1.6 m telescope on Pico dos Dias Observatory of the National Laboratory of Astrophysics (L.N.A./Brazil). Some data are based on observations collected at the Copernicus and Schmidt telescopes (Asiago, Italy) of the INAF-Astronomical Observatory of Padova. This work was carried out within the “Lucky Star” umbrella that agglomerates the efforts of the Paris, Granada and Rio teams, which is funded by the European Research Council under the European Community’s H2020 (ERC Grant Agreement No. 669416).  This work has made use of data from the European Space Agency (E.S.A.) mission {\it Gaia} (\url{https://www.cosmos.esa.int/gaia}), processed by the {\it Gaia} Data Processing and Analysis Consortium (D.P.A.C., \url{https://www.cosmos.esa.int/web/gaia/dpac/consortium}). Funding for the D.P.A.C. has been provided by national institutions, in particular the institutions participating in the {\it Gaia} Multilateral Agreement. TRAPPIST is a project funded by the Belgian F.N.R.S. under grant F.R.F.C. 2.5.594.09.F and the A.R.C. grant for Concerted Research Actions, financed by the Wallonia-Brussels Federation. E.J is a F.N.R.S. Senior Research Associate. We would like to acknowledge financial support by the Spanish grant AYA-RTI2018-098657-JI00 “LEO-SBNAF” (MCIU/AEI/FEDER, U.E.) and the financial support from the State Agency for Research of the Spanish M.C.I.U. through the “Center of Excellence Severo Ochoa” award for the Instituto de Astrofísica de Andalucía (S.E.V.- 2017-0709). Also, AYA2017-89637-R and F.E.D.E.R. funds are acknowledged. We are thankful to the following observers who participated and provided data for respective events as listed in Table \ref{table:circumstances}: Orlando A. Naranjo from Universidad de Los Andes, Mérida/VEN; Faustino Organero from La Hita Observatory - Toledo/E.S.P.

\end{acknowledgements}
\newpage

%-------------------------------------------------------------------

\begin{appendix}
%%%%%% A %%%%%%%%%%%%%%
\section{Target star positions}
Each stellar astrometric position was propagated to an arbitrary instant close to the stellar occultation mid-time. The procedure is described in detail in Section 2. The resulting stellar positions are presented in Table \ref{table:stars_DR2}.
The astrometric positions for a few target stars were published before, using information from other catalogs and are listed in table \ref{table:stars_published}.

\begin{table}[!h]
\centering
\caption{Occulted ICRS stellar positions propagated to a specified time, which is close to the predicted event time. The data are ordered by decreasing the occulting object's diameter and event date. These positions are in the ICRS as represented by the Gaia Celestial Reference Frame (Gaia-CRF2) \citep{GaiaCRF2}.}            
\label{table:stars_DR2}      
\resizebox{0.88\hsize}{!}{
\begin{tabular}{lcllc}   
\hline\hline
\multicolumn{1}{c}{\begin{tabular}[c]{@{}c@{}}\textbf{Occulting}\\ \textbf{Object}\end{tabular}}&
\multicolumn{1}{c}{\begin{tabular}[c]{@{}c@{}}\textbf{Event Date}\\ (dd-mm-yyyy) \\ \textbf{Instant - UT} \\ (hh:mm:ss)\end{tabular}}&
\multicolumn{1}{c}{\begin{tabular}[c]{@{}c@{}}\textbf{Right Ascension}\\ (hh mm ss.ss)  \\ \textbf{Declination} \\ ($^\circ$ ' '')\end{tabular}}&
\multicolumn{1}{c}{\begin{tabular}[c]{@{}c@{}} \textbf{Astrometric}\\ \textbf{Uncertainty}\tablefootmark{b} \\ (mas) \end{tabular}} &
\multicolumn{1}{c}{\begin{tabular}[c]{@{}c@{}}   \textbf{Mag. G}\\   \textbf{Gaia DR2} \\   (mag) \end{tabular}} \\  \hline 
\multirow{3}{*}{(136199)}       & 06-11-2010         &  01 39 09.92968     &  $\pm$ 1.0    &\multirow{2}{*}{  16.5271 $\pm$ 0.0013}\\	    
                                & 02:20:00.000       & -04 21 12.01520     &  $\pm$ 0.73   &\\ \cline{2-5}   
Eris                            & 29-08-2013         &  01 42 58.50363     &  $\pm$ 1.1    &\multirow{2}{*}{  18.1395 $\pm$ 0.0044}\\	    
                                & 15:38:00.000       & -03 21 17.34635     &  $\pm$ 0.69   &\\ \hline   
(136472)                        & 23-04-2011         &  12 36 11.394595    &  $\pm$ 0.98   &\multirow{2}{*}{  18.2365 $\pm$ 0.0016}\\	    
Makemake                        & 01:36:00.000       & +28 11 10.45324     &  $\pm$ 0.77   &\\ \hline 
(136108)                        & 21-01-2017         &  14 12 03.203523    &  $\pm$ 0.56   &\multirow{2}{*}{  17.8305 $\pm$ 0.0021}\\	    
Haumea                          & 03:09:00.000       & +16 33 58.64585     &  $\pm$ 0.51   &\\ \hline 
\multirow{9}{*}{(50000)}        & 11-02-2011         &  17 28 47.620587    &  $\pm$ 0.40   &\multirow{2}{*}{  15.6506 $\pm$ 0.0008}\\	    
                                & 10:04:00.000       & -15 41 59.08847     &  $\pm$ 0.33   &\\ \cline{2-5}
                                & 04-05-2011         &  17 28 50.801265    &  $\pm$ 0.40   &\multirow{2}{*}{  15.6636 $\pm$ 0.0007}\\	    
                                & 02:40:00.000       & -15 27 42.72514     &  $\pm$ 0.32   &\\ \cline{2-5}
                                & 17-02-2012         &  17 34 21.845611    &  $\pm$ 0.26   &\multirow{2}{*}{  14.8442 $\pm$ 0.0004}\\	    
 Quaoar                         & 04:30:00.000       & -15 42 10.49337     &  $\pm$ 0.21   &\\ \cline{2-5}   
                                & 15-10-2012         &  17 28 10.127719    &  $\pm$ 0.44   &\multirow{2}{*}{  16.8968 $\pm$ 0.0011}\\	    
                                & 00:45:00.000       & -15 36 23.26708     &  $\pm$ 0.39   &\\ \cline{2-5} 
                                & 09-07-2013         &  17 34 40.463695    &  $\pm$ 0.14   &\multirow{2}{*}{  14.3605 $\pm$ 0.0011}\\	    
                                & 02:40:00.000       & -15 23 37.56824     &  $\pm$ 0.11   &\\ \hline
(90377)                         & 13-01-2013         &  03 32 44.704654    &  $\pm$ 0.81   &\multirow{2}{*}{  18.1051 $\pm$ 0.0021}\\	    
Sedna                           & 12:25:00.000       & +06 47 19.02450     &  $\pm$ 0.70   &\\ \hline 
\multirow{7}{*}{(208996)}       & 08-01-2011         &  07 43 41.82694     &  $\pm$ 1.9    &\multirow{2}{*}{  18.4481 $\pm$ 0.0025}\\	    
                                & 06:29:00.000       & +11 30 23.5646      &  $\pm$ 1.3    &\\ \cline{2-5}   
                                & 03-02-2012         &  07 45 54.770163    &  $\pm$ 0.31   &\multirow{2}{*}{  15.3514 $\pm$ 0.0008}\\	    
                                & 19:45:00.000       & +11 12 43.07749     &  $\pm$ 0.24   &\\ \cline{2-5}
2003 AZ$_{84}$                  & 02-12-2013         &  07 58 56.728914    &  $\pm$ 0.89   &\multirow{2}{*}{  14.8546 $\pm$ 0.0006}\\	    
                                & 14:53:00.000       & +10 19 04.97871     &  $\pm$ 0.52   &\\ \cline{2-5}
                                & 15-11-2014         &  08 03 51.2950411   &  $\pm$ 0.079  &\multirow{2}{*}{  15.8197 $\pm$ 0.0005}\\	    
                                & 17:55:00.000       & +09 57 18.683561    &  $\pm$ 0.043  &\\ \hline
(229762)                        & 15-11-2014 	     &  04 29 30.6157327   &  $\pm$ 0.092  &\multirow{2}{*}{  15.7308 $\pm$ 0.0006}\\	    
2007 UK$_{126}$\tablefootmark{a}& 10:19:00.000       & -00 28 20.852684    &  $\pm$ 0.068  &\\ \hline 
\multirow{5}{*}{(278361)}       & 11-07-2019         &  16 46 47.305107    &  $\pm$ 0.33   &\multirow{2}{*}{  11.9109 $\pm$ 0.0003}\\	    
                                & 05:44:00.000       & -26 16 50.38512     &  $\pm$ 0.15   &\\ \cline{2-5} 
                                & 07-08-2019         &  16 45 23.13117     &  $\pm$ 1.1    &\multirow{2}{*}{  17.6713 $\pm$ 0.0010}\\		
2007 JJ$_{43}$                  & 13:06:00.000       & -26 11 07.35871     &  $\pm$ 0.51   &\\ \cline{2-5} 
                                & 24-05-2020         &  16 57 27.977189    &  $\pm$ 0.42   &\multirow{2}{*}{  15.6822 $\pm$ 0.0004}\\		
                                & 11:30:00.000       & -26 18 43.88690     &  $\pm$ 0.25   &\\ \hline
(28978)                         & 24-06-2014         &  17 17 54.229687    &  $\pm$ 0.11   &\multirow{2}{*}{  14.3729 $\pm$ 0.0004}\\	    
Ixion                           & 14:54:00.000       & -26 40 55.19406     &  $\pm$ 0.10   &\\ \hline 
\multirow{2}{*}{2017 OF$_{69}$} & 27-08-2019         &  20 13 12.06429     &  $\pm$ 1.0    &\multirow{2}{*}{  17.1538 $\pm$ 0.0013}\\	    
                                & 21:20:00.000       & -05 39 42.38066     &  $\pm$ 0.56   &\\ \hline
\multirow{3}{*}{(145451)}       & 24-12-2018         &  04 52 51.238458    &  $\pm$ 0.16   &\multirow{2}{*}{  13.4518 $\pm$ 0.0002}\\	    
                                & 02:21:00.000       & +16 45 24.165392    &  $\pm$ 0.093  &\\ \cline{2-5} 
2005 RM$_{43}$                  & 03-02-2019         &  04 50 01.048271    &  $\pm$ 0.54   &\multirow{2}{*}{  16.8597 $\pm$ 0.0007}\\	    
                                & 10:57:00.000       & +16 51 18.94980     &  $\pm$ 0.31   &\\ \hline  
(444030)                        & 16-11-2017         &  21 32 10.908885    &  $\pm$ 0.62   &\multirow{2}{*}{  17.3936 $\pm$ 0.0015}\\	    
2004 NT$_{33}$                  & 18:20:00.000       & +18 37 30.91689     &  $\pm$ 0.48   &\\ \hline
\multirow{3}{*}{(119951)}       & 26-04-2012         &  16 35 04.26616     &  $\pm$ 1.8    &\multirow{2}{*}{  18.6412 $\pm$ 0.0029}\\	    
                                & 02:35:00.00        & -22 15 22.9067      &  $\pm$ 1.3    &\\ \cline{2-5}   
2002 KX$_{14}$                  & 19-09-2018         &  17 05 27.165393    &  $\pm$ 0.37   &\multirow{2}{*}{  15.8177 $\pm$ 0.0006}\\	    
                                & 12:31:00.000       & -23 02 07.80884     &  $\pm$ 0.22   &\\ \hline
(175113)                        & 28-09-2018         &  23 01 47.577834    &  $\pm$ 0.58   &\multirow{2}{*}{  16.5718 $\pm$ 0.0010}\\	    
2004 PF$_{115}$                 & 01:36:00.000       & -20 31 01.22136     &  $\pm$ 0.40   &\\ \hline
(119979)                        & 30-12-2018         &  05 50 53.894245    &  $\pm$ 0.41   &\multirow{2}{*}{  16.4798 $\pm$ 0.0009}\\	    
2002 WC$_{19}$                  & 11:58:00.000       & +19 55 52.78043     &  $\pm$ 0.40   &\\ \hline
(55636)                         & 09-10-2009         &  00 37 13.625775    &  $\pm$ 0.21   &\multirow{2}{*}{  13.3121 $\pm$ 0.0002}\\	    
2002 TX$_{300}$                 & 10:30:00.000       & +28 22 23.00218     &  $\pm$ 0.21   &\\ \hline
(55638)                         & 03-12-2015         &  05 37 38.5279647   &  $\pm$ 0.038  &\multirow{2}{*}{  14.3702 $\pm$ 0.0004}\\	    
2002 VE$_{95}$                  & 12:30:00.000       & +08 05 09.590561    &  $\pm$ 0.027  &\\ \hline
(469506)                        & 24-05-2017         &  16 03 25.1583059   &  $\pm$ 0.097  &\multirow{2}{*}{  13.0113 $\pm$ 0.0003}\\	    
2003 FF$_{128}$                 & 13:41:00.000       & -19 45 48.983466    &  $\pm$ 0.054  &\\ \hline
\multirow{3}{*}{(54598)}        & 29-12-2017         &  02 49 15.533789    &  $\pm$ 0.11   &\multirow{2}{*}{  12.9464 $\pm$ 0.0003}\\	    
                                & 16:24:00.000       & +38 29 52.26387     &  $\pm$ 0.14   &\\ \cline{2-5}   
Bienor                          & 02-04-2018         &  03 00 51.817966    &  $\pm$ 0.13   &\multirow{2}{*}{  14.6587 $\pm$ 0.0004}\\	    
                                & 19:50:00.000       & +37 21 59.44192     &  $\pm$ 0.14   &\\ \hline
\multirow{3}{*}{(2060)}         & 07-11-1993         &  10 28 08.04049     &  $\pm$ 2.3    &\multirow{2}{*}{  13.9544 $\pm$ 0.0004}\\	    
                                & 13:15:00.000       & +03 30 35.8185      &  $\pm$ 2.0    &\\ \cline{2-5}   
Chiron                          & 29-11-2011         &  22 02 44.301833    &  $\pm$ 0.23   &\multirow{2}{*}{  14.8679 $\pm$ 0.0006}\\	    
                                & 08:16:00.000       & -05 55 12.24530     &  $\pm$ 0.26   &\\ \hline
\multirow{2}{*}{2005 TV$_{189}$}& 13-11-2012         &  05 27 31.475611    &  $\pm$ 0.24   &\multirow{2}{*}{  11.5884 $\pm$ 0.0004}\\	    
                                & 22:27:00.000       & +12 59 31.57680     &  $\pm$ 0.18   &\\ \hline 
(8405)                          & 24-11-2013         &  03 47 11.7552565   &  $\pm$ 0.085  &\multirow{2}{*}{  14.8326 $\pm$ 0.0003}\\	    
Asbolus                         & 05:23:00.000       & +36 02 10.501138    &  $\pm$ 0.064  &\\ \hline
(60558)                         & 25-06-2012         &  17 24 26.113031    &  $\pm$ 0.18   &\multirow{2}{*}{  13.6254 $\pm$ 0.0003}\\	    
Echeclus                        & 23:36:00.000       & -18 10 21.06453     &  $\pm$ 0.14   &\\ \hline \hline
\end{tabular}}
\tablefoot{\tablefoottext{a}{This object is also named as G!kún$\parallel$'hòmdímà.}\\ \tablefoottext{b}{ Uncertainties in RA$\cdot$cos(DEC) and DEC, respectively.}}
\end{table}

\begin{table}[!h]
\centering
\caption{Stellar positions obtained from the literature (see Table~\ref{table:circumstances} and references therein for details), used to obtain the updated astrometric position of the object.}
\label{table:stars_published}      
\resizebox{\hsize}{!}{        
\begin{tabular}{llll}   
\hline\hline
\multicolumn{1}{c}{\begin{tabular}[c]{@{}c@{}}\textbf{Occulting}\\ \textbf{Object}\end{tabular}}&
\multicolumn{1}{c}{\begin{tabular}[c]{@{}c@{}}\textbf{Event Date}\\ (dd-mm-yyyy) \\ \textbf{Instant - UT} \\ (hh:mm:ss)\end{tabular}}&
\multicolumn{1}{c}{\begin{tabular}[c]{@{}c@{}}\textbf{Right Ascension}\\ (hh mm ss.ss)  \\ \textbf{Declination} \\ ($^\circ$ ' '')\end{tabular}}&
\multicolumn{1}{c}{\begin{tabular}[c]{@{}c@{}} \textbf{Astrometric}\\ \textbf{Uncertainties} \\ (mas) \end{tabular}} \\  \hline 
(136199)                        & 06-11-2010         &  01 39 09.9421       &$\pm$ 50      \\            
Eris                            & 02:20:00           & -04 21 12.119        &$\pm$ 50      \\ \hline        
(136472)                        & 23-04-2011         &  12 36 11.402        &$\pm$ 130     \\            
Makemake                        & 01:36:00           & +28 11 10.448        &$\pm$ 60      \\  \hline 
(136108)                        & 21-01-2017         &  14 12 03.203400     &$\pm$ 0.60    \\            
Haumea                          & 03:09:00           & +16 33 58.64200      &$\pm$ 0.32    \\  \hline 
\multirow{7}{*}{(50000)}        & 11-02-2011         &  17 28 47.63         &$\pm$ 119     \\            
                                & 10:04:00           & -15 41 59.220        &$\pm$ 28      \\  \cline{2-4}
                                & 04-05-2011         &  17 28 50.80100      &$\pm$ 7       \\            
                                & 02:40:00           & -15 27 42.770        &$\pm$ 14      \\ \cline{2-4}
Quaoar                          & 17-02-2012         &  17 34 21.8453       &$\pm$ 22      \\            
                                & 04:30:00           & -15 42 10.5860       &$\pm$ 8       \\ \cline{2-4}        
                                & 15-10-2012         &  17 28 10.12570      &$\pm$ 1.3     \\            
                                & 00:45:00           & -15 36 23.3240       &$\pm$ 1.3     \\ \hline 
\multirow{3}{*}{(288996)}       & 03-02-2012         &  07 45 54.7696       &$\pm$ 22      \\            
                                & 19:45:00           & +11 12 43.093        &$\pm$ 23      \\ \cline{2-4}
2003 AZ$_{84}$                  & 15-11-2014         &  08 03 51.2980       &$\pm$ 22      \\            
                                & 17:55:00           & +09 57 18.729        &$\pm$ 23      \\ \hline
(229762)                        & 15-11-2014         &  04 29 30.6100       &$\pm$ 22      \\            
2007 UK$_{126}$                 & 10:19:00           & -00 28 20.908        &$\pm$ 23      \\  \hline 
(119951)                        & 26-04-2012         &  16 35 04.279        &$\pm$ 126     \\            
2002 KX$_{14}$                  & 02:35:00           & -22 15 22.75         &$\pm$ 126     \\ \hline     
(55636)                         & 09-10-2009         &  00 37 13.6193       &$\pm$ 27      \\            
2002 TX$_{300}$                 & 10:30:00           & +28 22 23.023        &$\pm$ 15      \\  \hline
\multirow{3}{*}{(2060)}         & 07-11-1993         &  10 28 08.020        &$\pm$ 200     \\          
                                & 13:15:00           & +03 30 35.60         &$\pm$ 200     \\ \cline{2-4}        
Chiron                          & 29-11-2011         &  22 02 44.2989       &$\pm$ 35      \\            
                                & 08:16:00           & -05 55 12.244        &$\pm$ 46      \\ \hline
\end{tabular}}
\end{table}

\newpage
%%%%%%%%%%%% B %%%%%%%%%%%

\section{Observational circumstances}
Table \ref{table:circumstances} presents the instruments and observational circumstances for each positive and near negative detection of the 37 stellar occultations.

\begin{table*}[!h]
\caption{Observational circumstances of the data sets. For the events available in the literature, only the corresponding reference is provided. Only the information of the negative sites that helped to constrain the object position is shown. 'Time source' refers to the source of the time used as a reference in the images: NTP (Network Time Protocol), GPS (Global Positioning System), and radio-controlled clocks (DCF77)\tablefootmark{c}.}            
\label{table:circumstances}   
\centering
\resizebox{0.9\hsize}{!}{        
\begin{tabular}{c c c c c c c }     
\hline\hline
\multirow{3}{*}{\textbf{Object}}&\multirow{3}{*}{\textbf{Event Date}}&\multirow{3}{*}{\textbf{Site/Country}}&\textbf{Latitude (\grau ' '')}&\textbf{Telescope}&\textbf{Exposure (s)}&\multirow{3}{*}{\textbf{Observers}}\\
                                     &                              &                               &\textbf{Longitude (\grau ' '')} & \textbf{Aperture (m)}&\textbf{Cycle (s)}&                  \\
                                     & \textbf{(dd-mm-yyyy)}        &                               & \textbf{Altitude (m)}          & \textbf{Instrument}  &\textbf{Time source}  &         \\ \hline
\multirow{7}{*}{(136199) Eris}       &  06-11-2010                  &\multicolumn{5}{l}{See Sicardy et al. 2011.}                                                                                 \\ \cline{2-7}
                                     &\multirow{6}{*}{29-08-2013}   &\multirow{3}{*}{\SV}           &27 22 07.0   S                  &SCT-Cass and Mark  &10.2402           &                     \\
                                     &                              &                               &152 50 53.0  E                  &0.35               &10.2487           & \JB                 \\
                                     &                              &                               &  80                            &Watec 120N+        &GPS               &                     \\ \cline{3-7}
                                     &                              &\multirow{3}{*}{\AS}           &23 42 44.274  S                 &SCT - Cass         &8.0               &                     \\
                                     &                              &                               &133 53 01.908  E                &0.36               &8.0               & \WH                 \\
                                     &                              &   \textbf{(Negative)}         &  581                           &Watec 120N+        & GPS              &                     \\ \hline
(136472) Makemake                    &  23-04-2011                  &\multicolumn{5}{l}{See Ortiz et al. 2012.}                                                                                   \\ \hline
(136108) Haumea                      &  21-01-2017                  &\multicolumn{5}{l}{See Ortiz et al. 2017.}                                                                                   \\ \hline
\multirow{7}{*}{(50000) Quaoar}      &11-02-2011                    & \multicolumn{5}{l}{See Person et al. 2011.}                                                                                 \\ \cline{2-7}
                                     &  04-05-2011                  &\multicolumn{5}{l}{See Braga-Ribas et al. 2013.}                                                                             \\ \cline{2-7}
                                     &  17-02-2012                  &\multicolumn{5}{l}{See Braga-Ribas et al. 2013.}                                                                             \\ \cline{2-7}
                                     &  15-10-2012                  &\multicolumn{5}{l}{See Braga-Ribas et al. 2013.}                                                                             \\ \cline{2-7}
                                     &\multirow{3}{*}{09-07-2013}   &\multirow{3}{*}{\Me}           &08 47 25.80  N                  &Zeiss Reflector    &30.0              &                     \\
                                     &                              &                               &70 52 21.58  W                  &1.0                &40.0              &  \ON                \\
                                     &                              &                               & 3600                           &FLI PL4240 (R filter)& GPS&              \\ \hline
\multirow{6}{*}{(90377) Sedna}       &\multirow{6}{*}{13-01-2013}   &\multirow{3}{*}{\Coral}        &16 54 26.3  S                   &RCOS               &  60.0            &                      \\
                                     &                              &                               &145 45 49.7  E                 & 0.41              &  86.0            &   \Jb                \\
                                     &                              &                               &  0.0                           &SBIG STL6K (IR filter)& GPS      &                      \\ \cline{3-7}
                                     &                              &\multirow{3}{*}{\Jewel}        &16 58 06.0  S                   &Keller Cassegrain  &  180.0           &                      \\
                                     &                              &                               &145 43 22.0  E                 & 0.50              &  244.0           &  \Tc                 \\
                                     &                              &                               &108                             &FLI Proline 16803 & NTP           &                      \\ \hline
\multirow{4}{*}{(208996) 2003 AZ\textsubscript{84}}&08-01-2011      &\multicolumn{5}{l}{See Dias-Oliveira et al. 2017.}                                                                           \\ \cline{2-7}
                                     &  03-02-2012                  &\multicolumn{5}{l}{See Dias-Oliveira et al. 2017.}                                                                           \\ \cline{2-7}
                                     &  02-12-2013                  &\multicolumn{5}{l}{See Dias-Oliveira et al. 2017.}                                                                           \\ \cline{2-7}
                                     &  15-11-2014                  &\multicolumn{5}{l}{See Dias-Oliveira et al. 2017.}                                                                           \\ \hline
(229762) 2007 UK\textsubscript{126}  &  15-11-2014                  &\multicolumn{5}{l}{See Benedetti-Rossi et al. 2016.}                                                                         \\ \hline
\multirow{18}{*}{(278361) 2007 JJ$_{43}$}&\multirow{12}{*}{11-07-2019}&\multirow{3}{*}{\Ol}          &20 42 54.27  S                  &Reflector          & 5.0              &  \CJ,              \\
                                     &                              &                               &44 47 05.82  W                  & 0.45              & 7.0              & \EP,                \\
                                     &                              &                               & 1113                           &FLI 8300           & NTP              &  \JR                \\ \cline{3-7}
                                     &                              &\multirow{3}{*}{\LS}           &29 15 16.588 S                  & TRAPPIST South    &2.5               &                     \\
                                     &                              &                               &70 44 21.818  W                 & 0.60              &4.2               &  \EJ                \\
                                     &                              & \textbf{(Negative)}           & 2315                           &FLI Proline PL3041-BB &NTP        &                     \\ \cline{3-7}
                                     &                              &\multirow{3}{*}{\LS}           &29 15 21.3  S                   & Danish            &0.1               & \CS,                \\
                                     &                              &                               &70 44 20.2  W                   &1.54               &0.1               & Jeremy Tregloan-Reed,\\
                                     &                              & \textbf{(Negative)}           & 2336                           &Lucky Imager       &GPS               & Sohrab Rahvar       \\ \cline{3-7}
                                     &                              &\multirow{3}{*}{\SPA}          &22 57 11.67  S                  & OPSPA             &5.0               & \AM,                \\
                                     &                              &                               &68 10 47.70  W                  & 0.40              &5.86              &  \JFP               \\
                                     &                              &  \textbf{(Negative)}          & 2400                           &ProLine PL16803    & NTP              &                     \\ \cline{2-7}
                                     &\multirow{3}{*}{07-08-2019}   &\multirow{3}{*}{\Mu}           &34 57 31.5  S                   &SCT - Cass         &  2.56           &                      \\
                                     &                              &                               &148 59 54.8  E                  & 0.40              & 2.56            &  \Dh                 \\
                                     &                              &                               & 594                            &Watec 910BD        &GPS              &                     \\ \cline{2-7}
                                     & \multirow{6}{*}{24-05-2020}  &\multirow{3}{*}{\Mel}          &37 59 38.16  S                  &SCT - Cass and Mark & 1.28            &                     \\
                                     &                              &                               &145 05 20.64  E                 & 0.28               & 1.28            &  \DeH               \\
                                     &                              &                               & 10                             &Watec 910HX         & GPS             &                     \\ \cline{3-7}  
                                     &                              &\multirow{3}{*}{\Ya}           &34 51 50.89  S                  &Planewave CDK20     & 1.5             &                     \\
                                     &                              &                               &148 58 35.06  E                 & 0.51               & 1.5             &  \WH                \\
                                     &                              &  \textbf{(Negative)}          & 536                            &QHY 174M            & GPS             &                     \\ \hline  
 \multirow{3}{*}{(28978) Ixion}      &\multirow{3}{*}{24-06-2014}   &\multirow{3}{*}{\AS}           &23 42 45.0  S                   &SCT - Cass          & 3.0             &                     \\
                                     &                              &                               &133 53 02.0  E                  & 0.20               & 3.1             &  \WH                \\
                                     &                              &                               & 584                            &Watec 120N+         & GPS             &                     \\ \hline
\multirow{6}{*}{2017 OF$_{69}$}&\multirow{6}{*}{27-08-2019}&\multirow{3}{*}{\On}                    &49 54 32.6  N                   & Mayer              & 8.0             &  \KH,               \\
                                     &                              &                               &14 46 53.3  E                   & 0.65               &9.7              &   \PP               \\
                                     &                              &                               & 526                            &Moravian G2-3200    &NTP              &                     \\ \cline{3-7}
                                     &                              &\multirow{3}{*}{\laHi}         &39 35 53.0  N                   &Reflector           &  8.0& Nicolás Morales,    \\
                                     &                              &                               &03 06 53.0  W                   & 0.77               &  9.02& Faustino Organero  \\
                                     &                              &\textbf{(Negative)}            & 674.9                          &  FLI Proline KAF-16803      &  NTP&                    \\ \hline
\multirow{12}{*}{(145451) 2005 RM$_{43}$}&\multirow{9}{*}{24-12-2018}&\multirow{3}{*}{\Gn}          &46 13 53.2  N                   &SCT - Celestron C11 &1.28              &                    \\
                                     &                              &                               &09 01 26.5  E                   & 0.28               &1.28              &  \Ss               \\
                                     &                              &                               &  260                           &Watec 910HX/RC      & GPS              &                  \\ \cline{3-7}
                                     &                              &\multirow{3}{*}{\Lu}           &45 59 50.6  N                   & Schmidt Cassegrain &1.28              &                    \\
                                     &                              &                               &08 55 10.5  E                   & 0.23               &1.28              &  \AO               \\
                                     &                              &                               &  350                           &Watec 910HX/RC      & GPS              &                  \\ \cline{3-7}
                                     &                              &\multirow{3}{*}{\LBJ}          &43 47 27.56  N                  & SCT - Celestron c8 &3.0               &                    \\
                                     &                              &                               &05 37 40.08  E                  & 0.20               &4.3               &  \PS               \\
                                     &                              &  \textbf{(Negative)}          &  424                           &Atik One 6.0       & NTP              &                    \\ \cline{2-7}
                                     &\multirow{3}{*}{03-02-2019}   &\multirow{3}{*}{\LT}           &43 59 06.8  S                   &Mt. John            &  1.28            &  \AG,              \\
                                     &                              &                               &170 27 50.9  E                  & 1.0                & 1.28             &  \PK               \\
                                     &                              &                               & 1029                           &Watec 120N          &GPS               &                    \\ \hline
% %%%%%%%%
\multirow{7}{*}{(444030) 2004 NT\textsubscript{33}}&\multirow{7}{*}{16-11-2017}&\multirow{7}{*}{\As}&                                &Schmidt             & 7.0              & \VN,               \\
                                     &                              &                               &                                & 0.70               & 11.7             & Domenico Nardiello \\
                                     &                              &                               & 45 50 58.00  N                 &Moravian G4-16000LC & NTP           &                    \\ \cline{5-7}
                                     &                              &                               &11 34 07.77  E                  &Copernicus          & 0.5              & Luca Zampieri,      \\
                                     &                              &                               & 1370                           & 1.82               & 0.5              &  Aleksandr Burtovoi,\\
                                     &                              &                               &                                & Aqueye+ photometer & GPS              &  Michele Fiori,     \\ 
                                     &                              &                               &                                &                    &                  &Giampiero Naletto   \\ \hline
\end{tabular}}
\end{table*}

\begin{table*}[!h]
\centering
\resizebox{0.9\hsize}{!}{        
\begin{tabular}{c c c c c c c }     
\hline
\multirow{7}{*}{(119951) 2002 KX\textsubscript{14}}&  26-04-2012    &\multicolumn{5}{l}{See Alvarez-Candal et al. 2014.}                                                                          \\ \cline{2-7}
                                     &\multirow{6}{*}{19-09-2018}   &\multirow{3}{*}{\Mu}           &34 57 31.5  S                  &SCT -Cass           & 0.64             &                    \\
                                     &                              &                               &148 59 54.8  E                 & 0.40               & 0.64             &  \Dh               \\
                                     &                              &                               &  594                           &Watec-910BD        & GPS              &                   \\ \cline{3-7}
                                     &                              &\multirow{3}{*}{\Ya}           &34 51 51.17  S                  &Planewave CDK20     & 1.5              &                    \\
                                     &                              &                               &148 58 35.14  E                 & 0.51               & 1.5              &  \WH               \\
                                     &                              &                               &  535                           &QHY 174M            & GPS              &                    \\ \hline
\multirow{3}{*}{(175113) 2004 PF\textsubscript{115}}&\multirow{3}{*}{28-09-2018}&\multirow{3}{*}{\Rec}&29 08 26.9  S                &Newtonian           & 6.0              &                    \\
                                     &                              &                               &59 38 28.1  W                   & 0.30               & 6.6              &  \as               \\
                                     &                              &                               &  50                            &ATIK -414ex       & NTP              &                    \\ \hline
\multirow{3}{*}{(119979) 2002 WC\textsubscript{19}}&\multirow{3}{*}{30-12-2018}&\multirow{3}{*}{\Ro}&23 16 10.1  S                  &SCT-Cass and Mark   & 2.56             &                    \\
                                     &                              &                               &150 30 01.6   E                 & 0.30               & 2.56             &  \SK               \\
                                     &                              &                               &  50                            &WATEC-910BD       &  GPS             &                    \\ \hline
(55636) 2002 TX\textsubscript{300}   &  09-10-2009                  &\multicolumn{5}{l}{See Elliot et al. 2010.}                                                                                  \\ \hline
 \multirow{3}{*}{(55638) 2002 VE\textsubscript{95}}&\multirow{3}{*}{03-12-2015}&\multirow{3}{*}{\SV} &27 22 07.0  S                  &SCT-Cass and Mark   & 0.32             &                    \\
                                      &                              &                               &152 50 53.0  E                 & 0.36               & 0.32             &  \JB               \\
                                      &                              &                               &  80                            &WATEC-910BD       &  GPS             &                    \\ \hline
\multirow{6}{*}{(469506) 2003 FF\textsubscript{128}}&\multirow{6}{*}{24-05-2017}&\multirow{3}{*}{\Mu}&34 57 31.5  S                 &SCT - Cass           & 0.04             &                    \\
                                     &                              &                               &148 59 54.8  E                 & 0.40               & 0.04             &  \Dh               \\
                                     &                              &                               &  594                           &Watec-910BD       &  GPS             &                    \\ \cline{3-7}
                                     &                              &\multirow{3}{*}{\Fly}           &35 11 55.34  S                 &Ritchey-Cretien    & 0.08             &                    \\
                                     &                              &                               &149 02 57.50  E                 & 0.41               & 0.08             &  \JN               \\
                                     &                              &                               &  657                           &Watec-910BD       & GPS              &                    \\ \hline
\multirow{9}{*}{(54598) Bienor}      &\multirow{3}{*}{29-12-2017}   &\multirow{3}{*}{\Yi}           &27 02 37.9  N                  &Reflector f4        & 0.267            &                    \\
                                     &                              &                               &128 26 54.0  E                 & 0.30               & 0.267            &  \YUS              \\
                                     &                              &                               &  23                            &Watec-910HX       & GPS             &         \\ \cline{2-7}
                                     &\multirow{6}{*}{02-04-2018}   &\multirow{3}{*}{\Ko}           &47 55 01.6  N                  &Reflector           &  0.3             &                    \\
                                     &                              &                               &19 53 41.5  E                  & 1.0                & 0.307            &  \AP               \\
                                     &                              &                               &  935                           &EMCCD  camera       & NTP              &                    \\ \cline{3-7}
                                     &                              &\multirow{3}{*}{\Bor}          &52 16 37.20  N                  &Reflector           & 1.5              &                    \\
                                     &                              &                               &17 04 28.56  E                  & 0.40               &  2.7             &  Anna Marciniak    \\
                                     &                              &\textbf{(Negative)}            &  123.4                         &SBIG ST-7 KAF400  & NTP              &                    \\ \hline
\multirow{2}{*}{(2060) Chiron}       &07-11-1993                    &\multicolumn{5}{l}{See Bus et al. 1996.}                                                                                     \\ \cline{2-7}
                                     &  29-11-2011                  &\multicolumn{5}{l}{See Ruprecht et al. 2015.}                                                                                \\ \hline
\multirow{3}{*}{2005 TV\textsubscript{189}}&\multirow{3}{*}{13-11-2012}&\multirow{3}{*}{\Kn}        &50 44 00.50  N                  &Newtonian           &  --              &                    \\
                                     &                              &                               &14 00 09.30  E                  & 0.25               &Chronograph ACH-77&  \TJ               \\
                                     &                              &                               &  460                           &Visual              & DCF 77&                    \\ \hline
\multirow{3}{*}{(8405) Asbolus}      &\multirow{3}{*}{24-11-2013}   &\multirow{3}{*}{\SPA}          &22 57 14.0  S                  &ASH2                & 5.0              &                    \\
                                     &                              &                               &68 10 44.0  W                  & 0.41               & 6.7              &  Nicolás Morales   \\
                                     &                              &                               &  2397                          &SBIG STL11K3      &NTP               &                    \\ \hline
\multirow{3}{*}{(60558) Echeclus}    &\multirow{3}{*}{25-06-2012}   &\multirow{3}{*}{\Ze}           &51 54 13.6  N                  &Newtonian           &  2.56            &                    \\
                                     &                              &                               &06 15 36.9  E                  & 0.31               &2.56              &  \JMW              \\
                                     &                              &                               &  66                            &Watec 120N        &GPS               &                    \\ \hline
\end{tabular}}
\tablefoot{\tablefoottext{c}{For further information, see \url{https://www.hopf.com/about-dcf77_en.php}.}}
\end{table*}

%%%%%%%%%% C %%%%%%%%%%%%
\newpage
\section{Occultation light curves}
\label{LC}
We present the occultation light curves for the events that are first published in this work. The black lines are the calibrated and normalized data, blue lines are the model, and red lines are the calculated light curve. We indicate in each figure the site and country from which the data were acquired. Each x-axis is referenced in time to the midnight of the date of the event, that is, seconds after 00:00:00.00 UTC. The y-axis presents the normalized flux of the target over the calibration stars. An integer value displaces some of the light curves for visualization purposes.

\begin{figure*}[!h]
\centering
   \includegraphics[height=4.7cm]{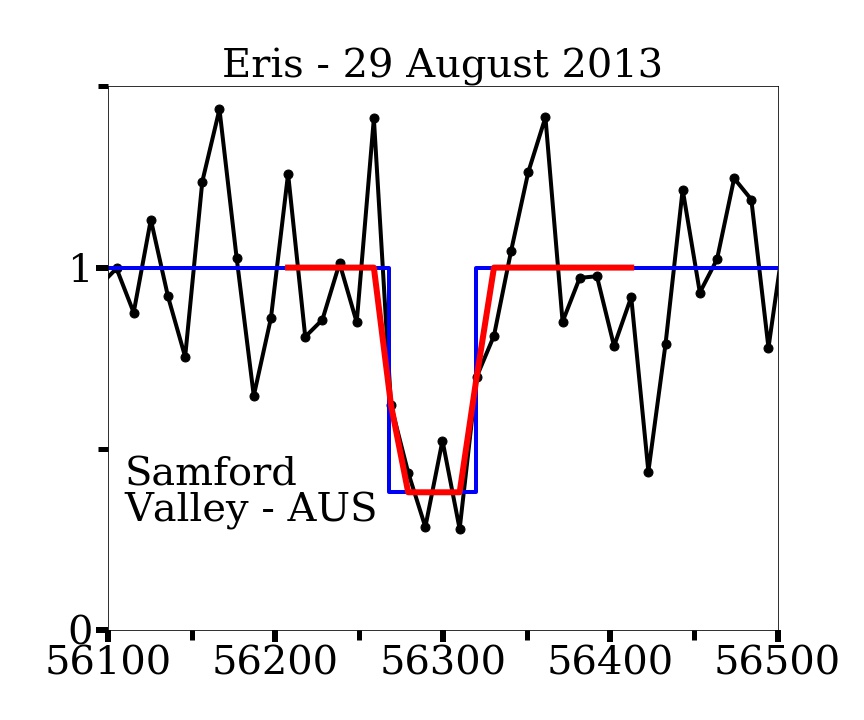}\quad
   \includegraphics[height=4.7cm]{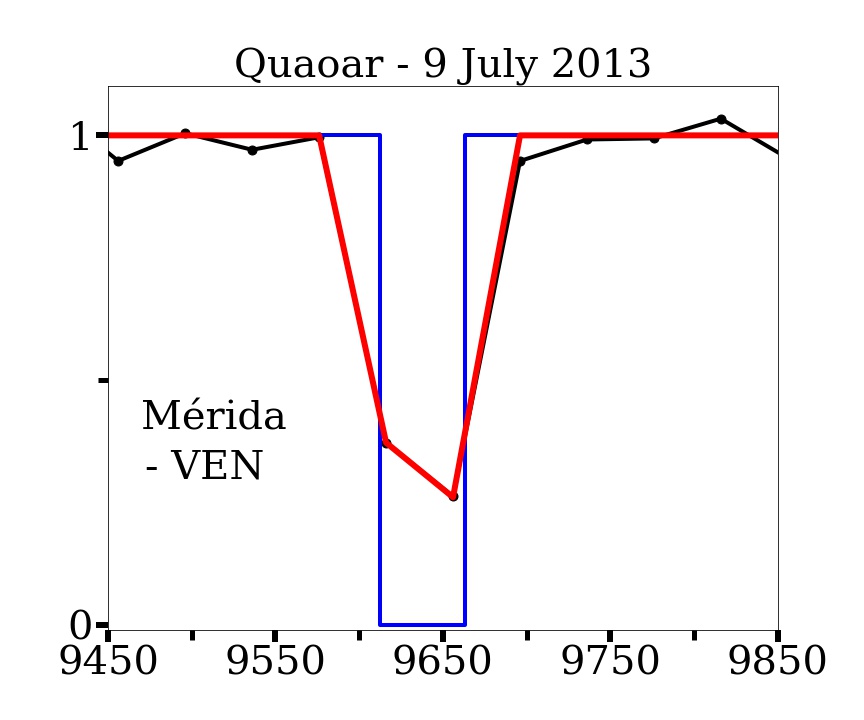}\quad
   \includegraphics[height=4.7cm]{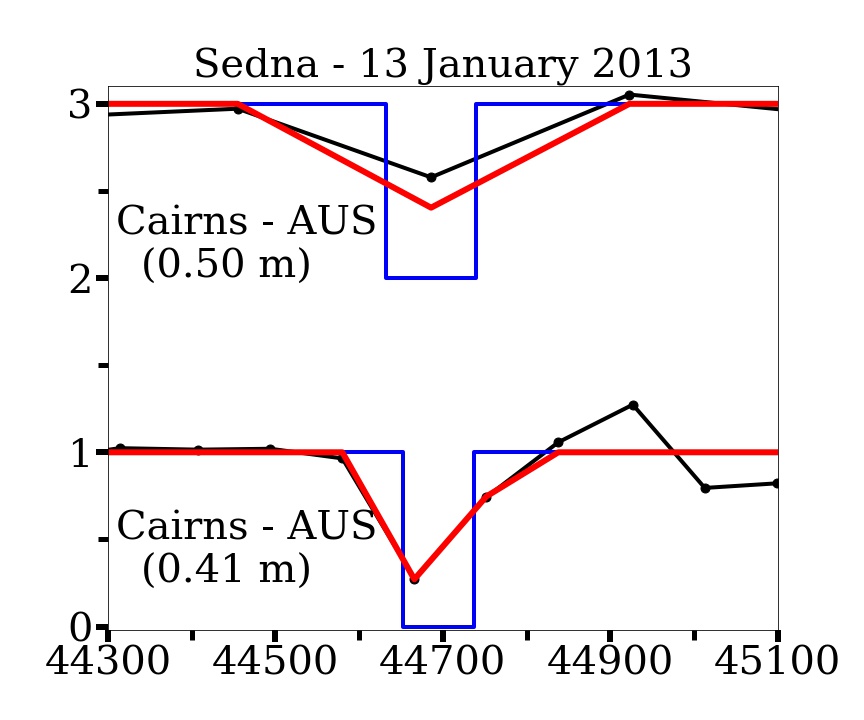}\quad
   \includegraphics[height=4.7cm]{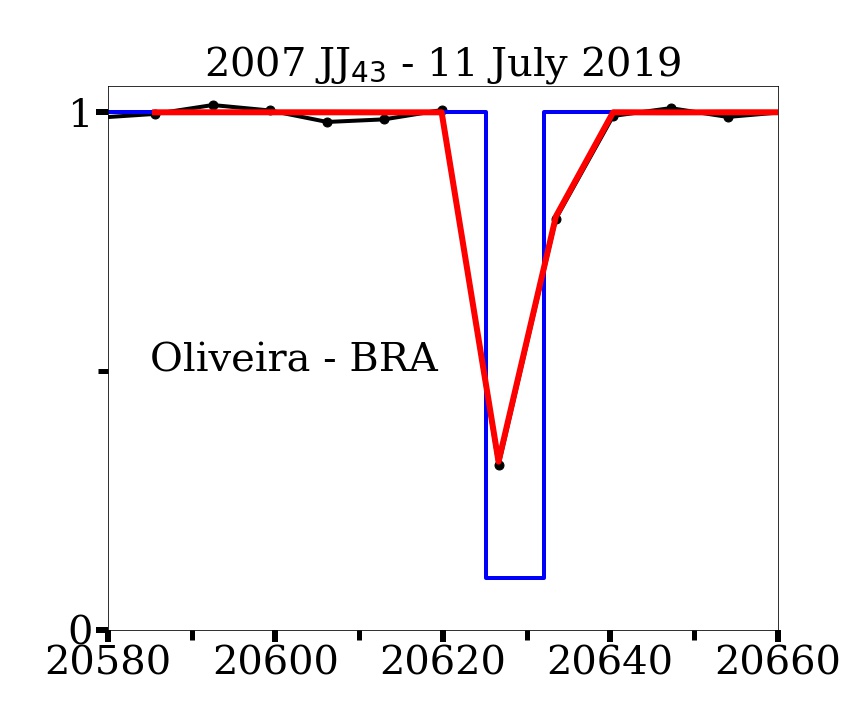}\quad
   \includegraphics[height=4.7cm]{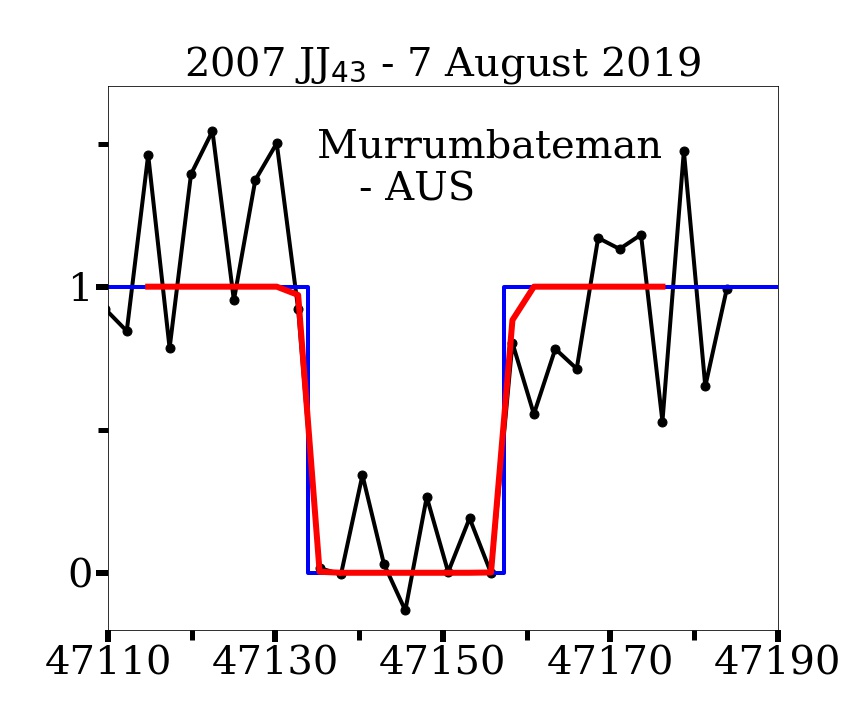}\quad
   \includegraphics[height=4.7cm]{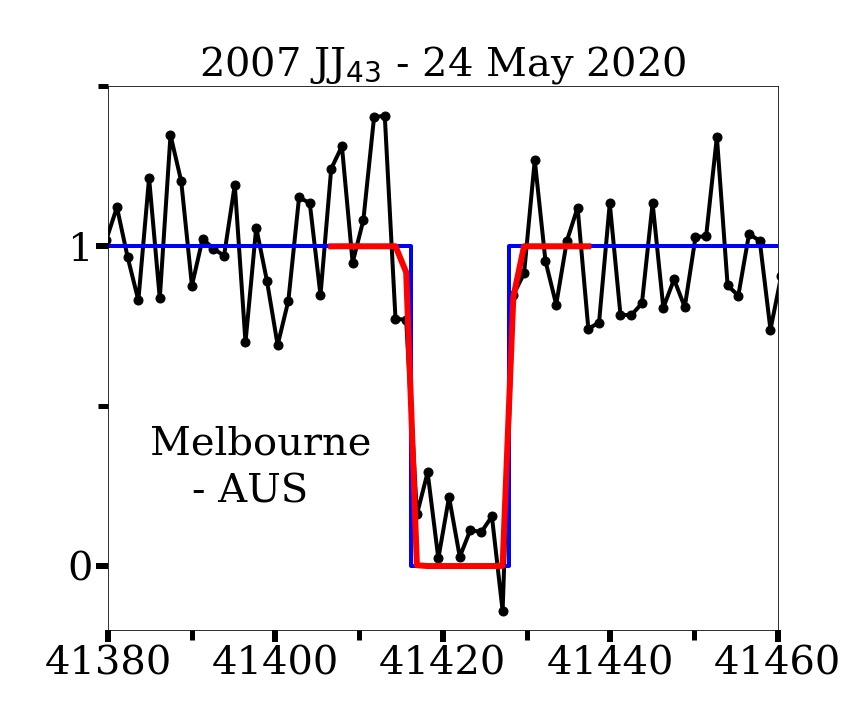}\quad
   \includegraphics[height=4.7cm]{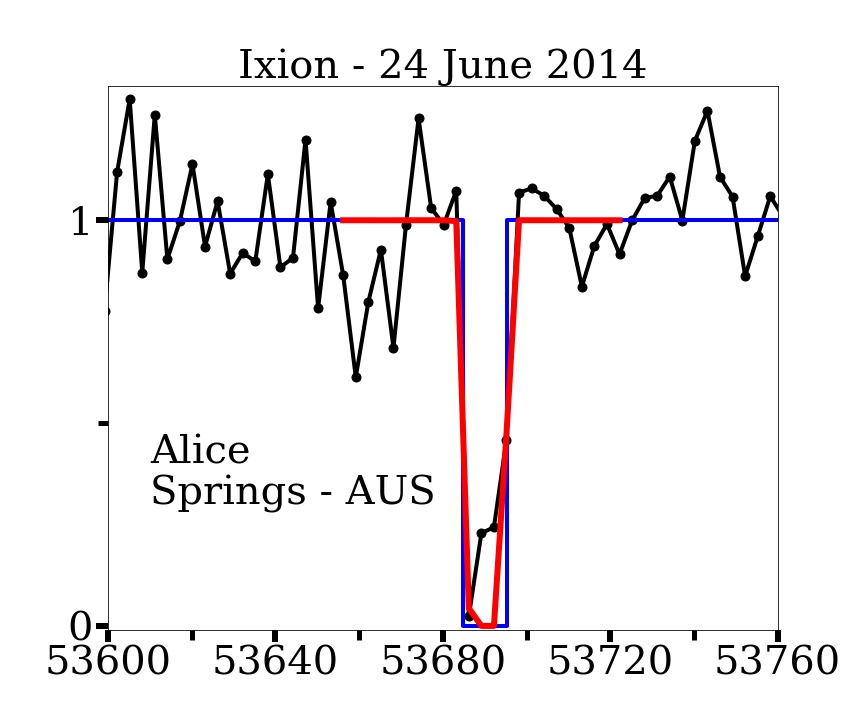}\quad
   \includegraphics[height=4.7cm]{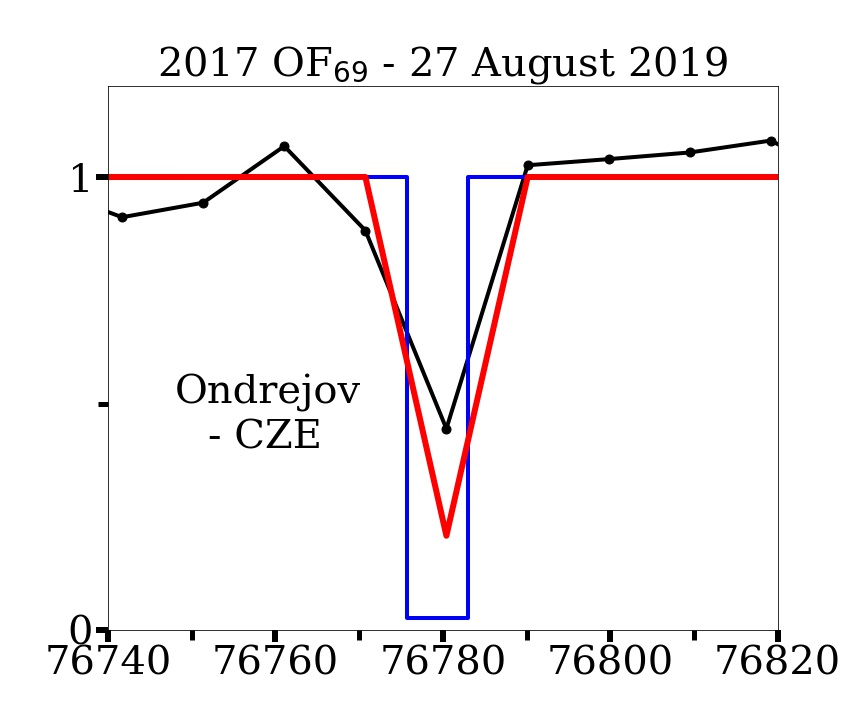}\quad
   \includegraphics[height=4.7cm]{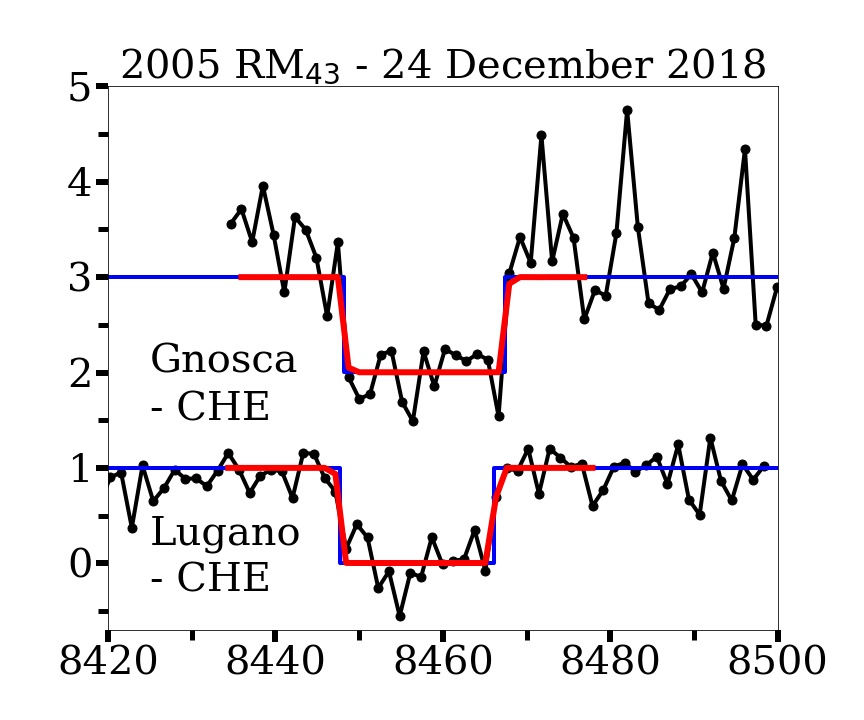}\quad
   \includegraphics[height=4.7cm]{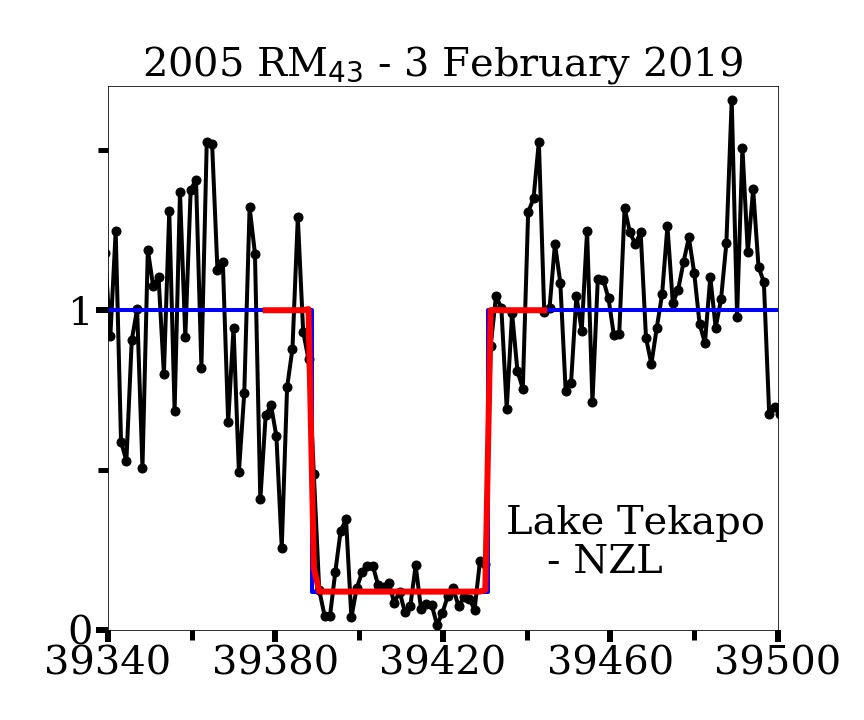}\quad
  \includegraphics[height=4.7cm]{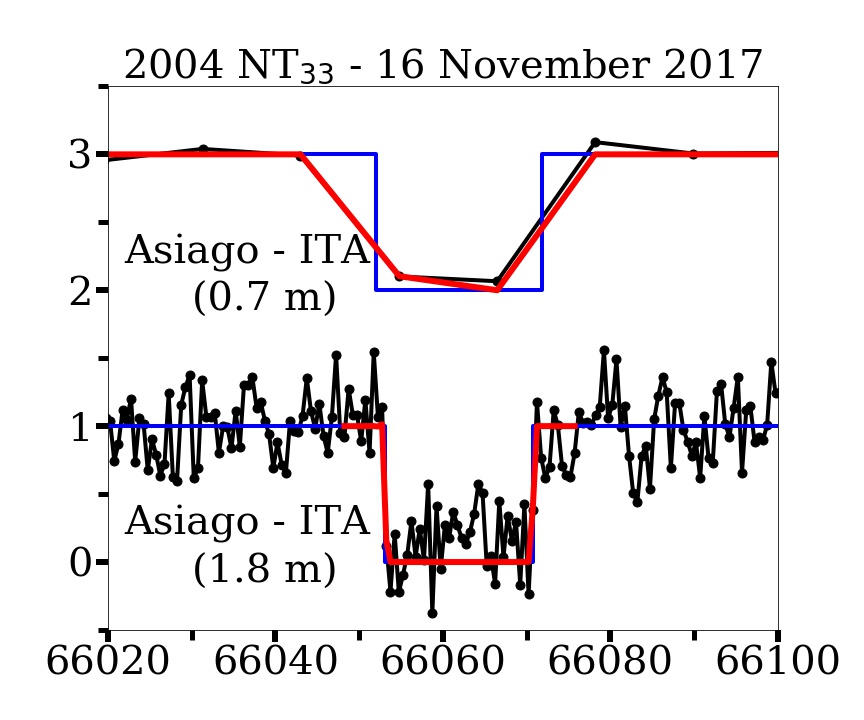}\quad
  \includegraphics[height=4.7cm]{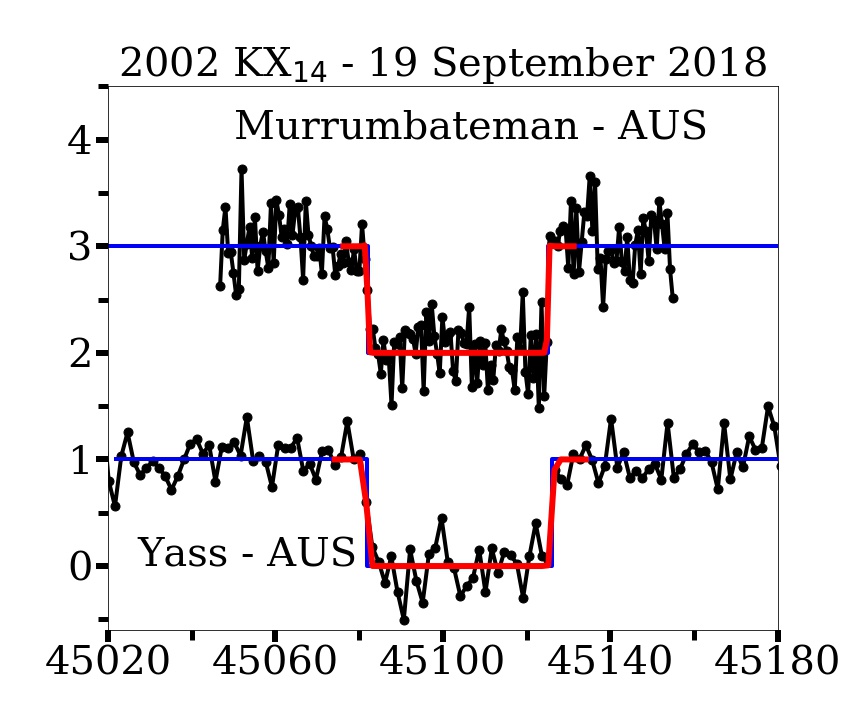}\quad
  \includegraphics[height=4.7cm]{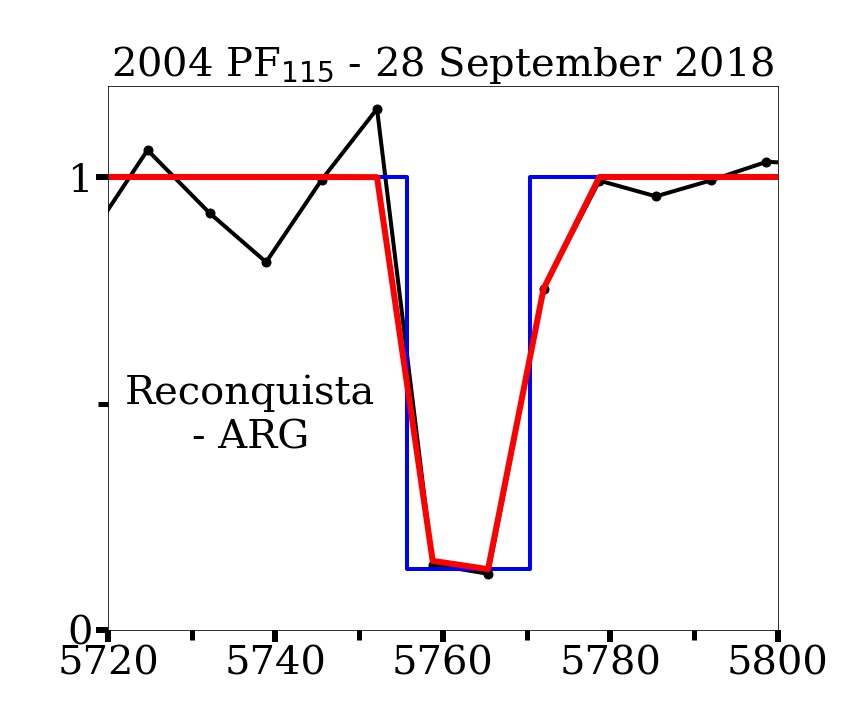}\quad
  \includegraphics[height=4.7cm]{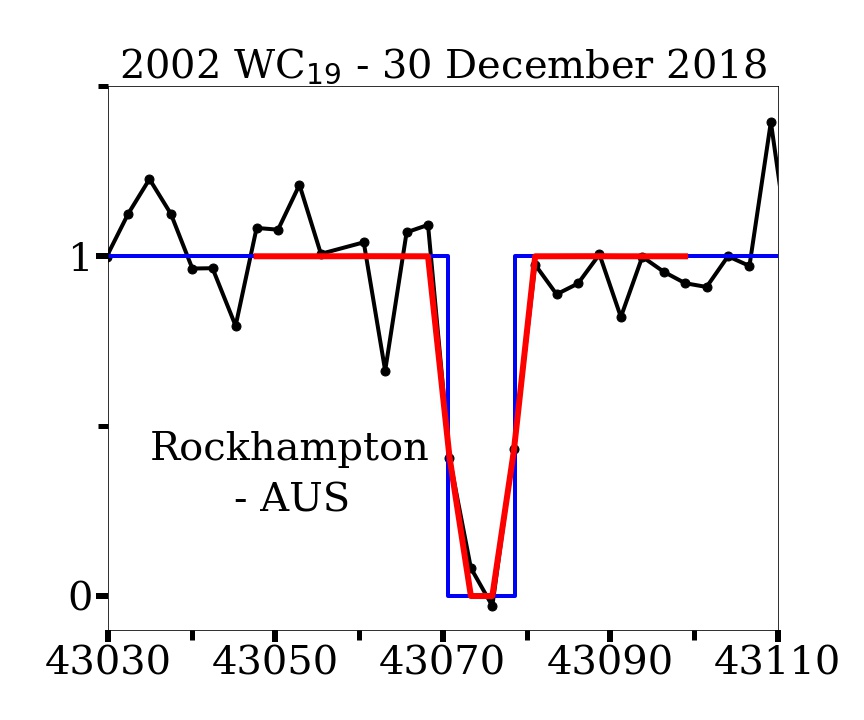}\quad
  \includegraphics[height=4.7cm]{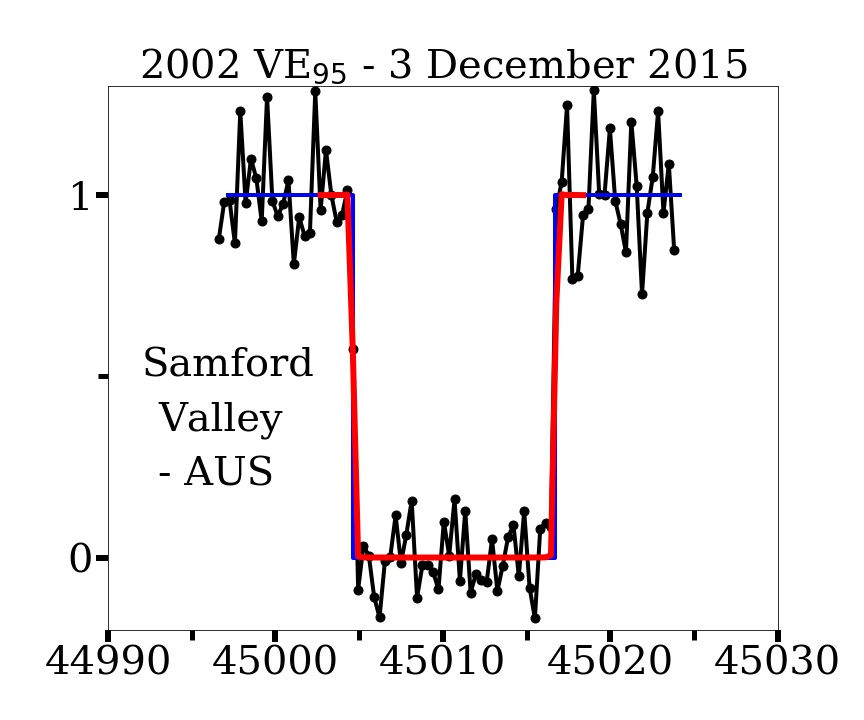}\quad
\end{figure*}

\begin{figure*}
   \includegraphics[height=4.7cm]{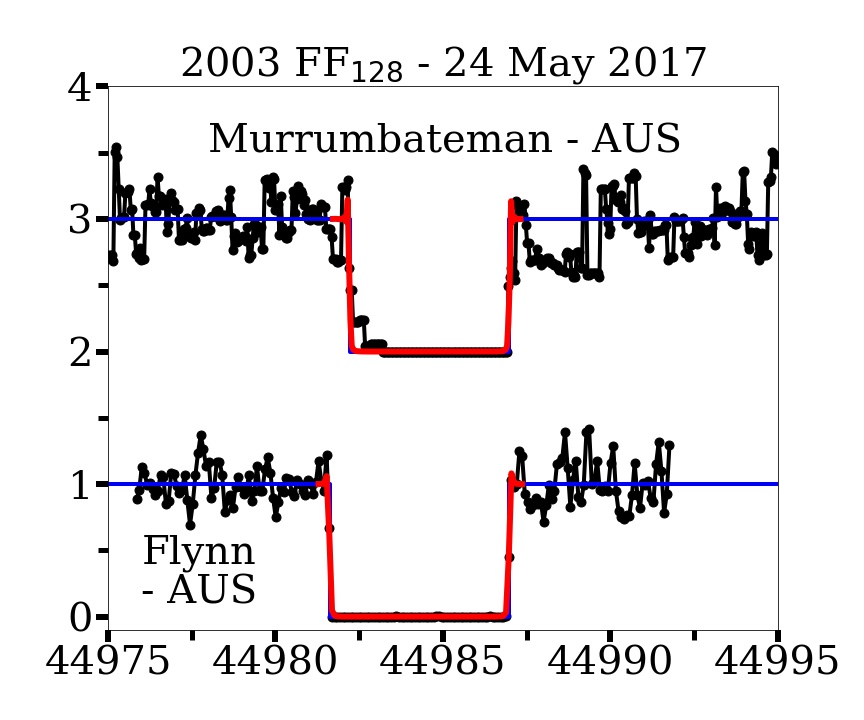}\quad
   \includegraphics[height=4.7cm]{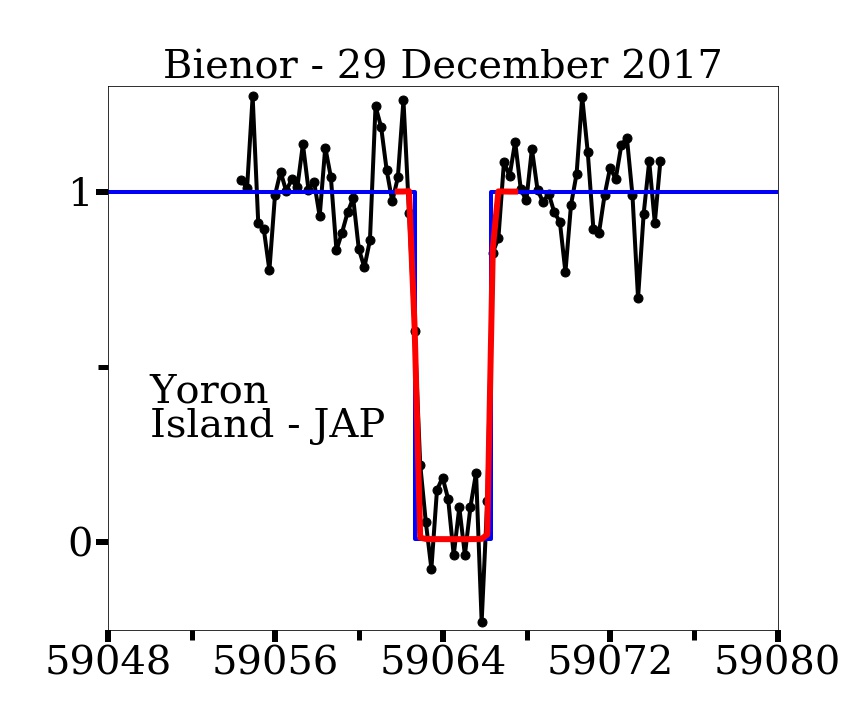}\quad
   \includegraphics[height=4.7cm]{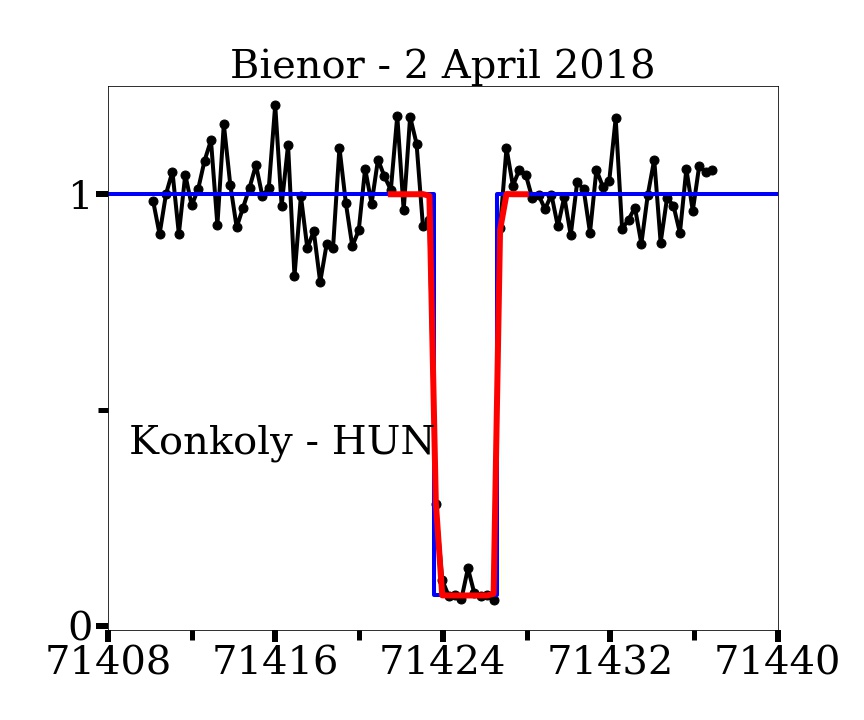}\quad
\end{figure*}

\begin{figure*}
   \includegraphics[height=4.7cm]{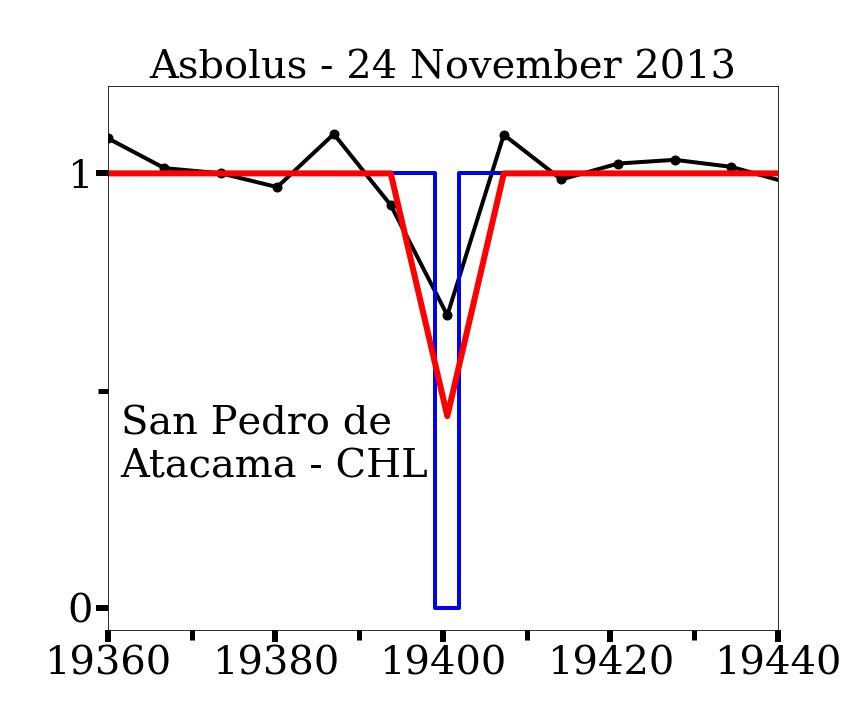}\quad
   \includegraphics[height=4.7cm]{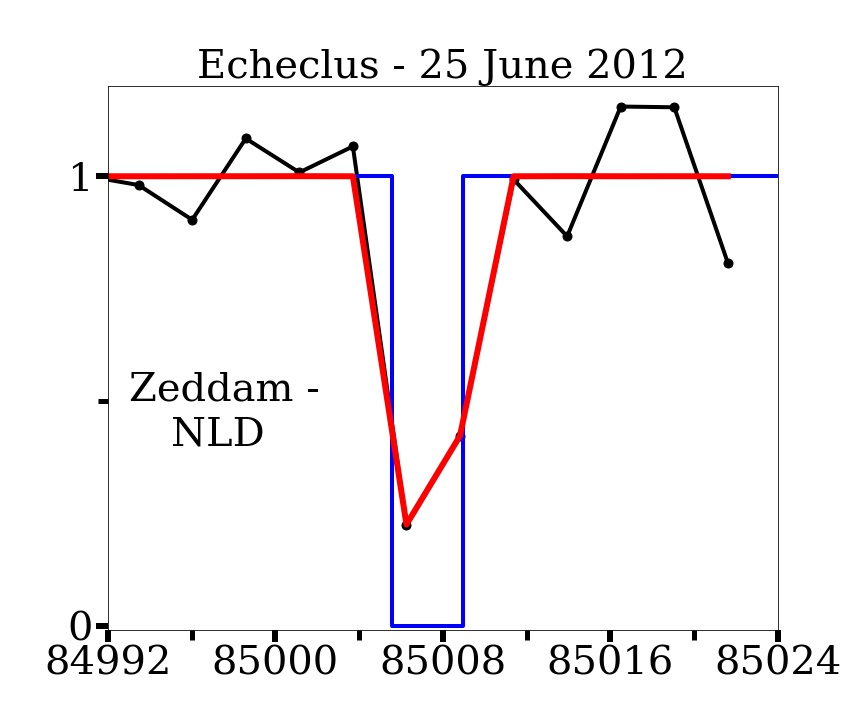}
\caption{All occultation light curves of the unpublished stellar occultations. They are ordered by decreasing diameter of the occulting object, as in Table \ref{tab:times}. The graphs present on the x-axis the time in seconds after midnight of the occultation date. On the y-axis, they show the relative flux, normalized to unity, of the occulted star divided by the reference stars. Some light curves were offset on the y-axis by an integer value for visualization purposes. Black lines represent the data, blue lines the model of a body without atmosphere occulting a punctual star at the derived times, and the red lines show the complete model where the Fresnel diffraction and the stellar diameter are convoluted to each finite exposure time.}
\label{fig:LC_figures}
\end{figure*}
\end{appendix}


\begin{thebibliography}{}
    \bibitem[Almeida et al.(2009)]{Almeida09} Almeida, A.~J.~C., Peixinho, N., Correia, A.~C.~M., 2009, \aap, 508, 2, 1021-1030.   
    
    \bibitem[Alvarez-Candal et al.(2014)]{Alvarez-Candal14} Alvarez-Candal, A., Ortiz, J.~L., Morales, N., et al.: 2014, \aap, 571, A48.
    
    \bibitem[Arias et al.(1995)]{Arias95} Arias, E.~F., Charlot, P., Feissel, M. \& Lestrade, J. -F. 1995, \aap, 303, 604. 

    \bibitem[Assafin et al.(2011)]{Assafin11} Assafin, M., Vieira Martins, R., Camargo, J.~I.~B., et al.: 2011, Gaia follow-up network for the solar system objects : Gaia FUN-SSO workshop proceedings, 85-88.  
   
    \bibitem[Assafin et al.(2012)]{Assafin12}Assafin, M., Camargo, J.I.B., Vieira Martins, et al.: 2012, \aap, 541, A142.
    
    \bibitem[Barry et al.(2015)]{Barry15}Barry, M.A.T., Gault, D., Pavlov, H., et al.: 2015, Publications of the Astronomical Society of Australia, 32, e031.

    \bibitem[Benecchi et al.(2019)]{Benecchi19}Benecchi, S.D., Porter, S.B., Buie,et al: 2019, Icarus, 334, 11.

    \bibitem[Benedetti-Rossi et al.(2016)]{Benedetti-Rossi16} Benedetti-Rossi, G., Sicardy, B., Buie, M.~W., et al.: 2016, \aj, 152, 156.
 
    \bibitem[Benedetti-Rossi et al.(2019)]{Benedetti-Rossi19} Benedetti-Rossi, G., Santos-Sanz,~P., Ortiz, J.~L., et al.: 2019, \aj, 158, 159.
    
    \bibitem[B{\'e}rard et al.(2017)]{Berard17} B{\'e}rard, D., Sicardy, B., Camargo, J.~I.~B., et al.: 2017, \aj, 154, 144.
 
    \bibitem[Braga-Ribas et al.(2013)]{Braga-Ribas13} Braga-Ribas, F., Sicardy, B., Ortiz, J.~L., et al.: 2013, \apj, 773, 1.
 
    \bibitem[Braga-Ribas et al.(2014)]{Braga-Ribas14} Braga-Ribas, F., Sicardy, B., Ortiz, J.~L., et al.: 2014, \nat, 508, 72.

    \bibitem[Braga-Ribas et al.(2019)]{Braga-Ribas19}Braga-Ribas, F., Crispim, A., Vieira-Martins, R., et al.: 2019, Journal of Physics Conference Series, 012024.

    \bibitem[Buie et al.(2020)]{Buie2020}Buie, M.W., Porter, S.B., Tamblyn, P., et al.: 2020, \aj, 159, 130.

    \bibitem[Bus et al.(1996)]{Bus96} Bus, S.~J., Buie, M.~W., Schleicher, D.~G., et al.: 1996, \icarus, 123, 2, 478-490.
    
    \bibitem[Butcher \& Stevens,(1981)]{Butcher} Butcher, H., Stevens, R. 1981, Kitt Peak National Observatory Newsletter, 16, 6.
    
    \bibitem[Camargo et al.(2014)]{Camargo14}Camargo, J.I.B., Vieira-Martins, R., Assafin, M., et al.: 2014, \aap, 561, A37.
 
    \bibitem[Dias-Oliveira et al.(2017)]{Dias-Oliveira17} Dias-Oliveira, A., Sicardy, B., Ortiz, J.~L., et al.: 2017, \aj, 154, 22.
    
    \bibitem[Desmars et al.(2015)]{Desmars15} Desmars, J., Camargo, J. I. B., Braga-Ribas, F., et al.: 2015, \aap, 584, A96.
 
    \bibitem[Desmars et al.(2019)]{Desmars19} Desmars, J., Meza, E., Sicardy, B., et al.: 2019, \aap, 625, A43.
 
    \bibitem[Duffard et al.(2014)]{Duffard14} Duffard, R., Pinilla-Alonso, N., Santos-Sanz, P., et al.: 2014, \aap, 564, A92.
    
    \bibitem[Earle et al.(2020)]{Earle20}Earle, A.M., Olkin, C., Stern, S., et al.: 2020, American Astronomical Society Meeting Abstracts , 419.03.
 
    \bibitem[Elliot et al.(1995)]{Elliot95} Elliot, J.~L., Olkin, C.~B., Dunham, E.~W., et al.: 1995, \nat, 373, 46.

    \bibitem[Elliot et al.(2010)]{Elliot10} Elliot, J.~L., Person, M.~J., Zuluaga, C.~A., et al.: 2010, \nat, 465, 897.
    
    \bibitem[Farkas-Tak{\'a}cs et al.(2020)]{Farkas20}Farkas-Tak{\'a}cs, A., Kiss, C., Vilenius, et al.: 2020, arXiv e-prints , arXiv:2002.12712.
    
    \bibitem[Fern{\'a}ndez et al.(2002)]{Fernandez02} Fern{\'a}ndez, Y.~R., Jewitt, D.~C., \& Sheppard, S.~S. 2002, \aj, 123, 1050
    
    \bibitem[Fern{\'a}ndez, J. A.(2020)]{Fernandez20}Fern{\'a}ndez, J.: 2020, The Trans-Neptunian Solar System, 1.
    
    \bibitem[Fornasier et al.(2013)]{Fornasier13} Fornasier, S., Lellouch, E., M{\"u}ller, T., et al.: 2013, \aap, 555, A15.
 
    \bibitem[Gaia collaboration et al.(2016)]{Gaia16} Gaia Collaboration, Prusti, T., de Bruijne, J. H. J., Brown, A. G. A., et al.: 2016, \aap, 595, A1.
  
    \bibitem[Gaia collaboration et al.(2018a)]{Gaia18}Gaia Collaboration; Brown, A. G. A.; Vallenari, A.; et al. 2018(a). \aap, 616, A1.
 
    \bibitem[Gaia Collaboration et al.(2018b)]{GaiaCRF2} Gaia Collaboration, Mignard, F., Klioner, S.A., Lindegren, L. et al.: 2018(b), \aap, 616, A14.
 
    \bibitem[Gladman et al.(2008)]{Gladman08} Gladman, B., Marsden, B.~G., \& Vanlaerhoven, C. 2008, The Solar System Beyond Neptune, 43.
    
    \bibitem[Horner, Evans and Bailey(2004)]{Horner04}Horner, J., Evans, N.W., and Bailey, M.E.: 2004, \mnras, 354, 798.

    \bibitem[Leiva et al.(2017)]{Leiva17} Leiva, R., Sicardy, B., Camargo, J.~I.~B., et al.: 2017, \aj, 154, 159.
 
    \bibitem[Lellouch et al.(2013)]{Lellouch13} Lellouch, E., Santos-Sanz, P., Lacerda, P., et al.: 2013, \aap, 557, A60.

    \bibitem[Lellouch et al.(2017)]{Lellouch17} Lellouch, E., Moreno, R., M\"ueller, T., et al.: 2017, \aap, 608, A45.

    \bibitem[Lim et al.(2010)]{Lim10} Lim, T.~L., Stansberry, J., M{\"u}ller, T.~G., et al.: 2010, \aap, 518, L148.

    \bibitem[Mommert et al.(2012)]{MM12} Mommert, M., Harris, A.~W., Kiss, C., et al.: 2012, \aap, 541, A93.
 
    \bibitem[M{\"u}ller, Lellouch, and Fornasier(2020)]{Mueller20}M{\"u}ller, T., Lellouch, E., and Fornasier, S.: 2020, The Trans-Neptunian Solar System, 153.
 
    \bibitem[Nesvorn{\'y} \& Morbidelli,(2012)]{Nesvorny12} Nesvorn{\'y}, D., \& Morbidelli, A. 2012, \aj, 144, 117.

    \bibitem[Ortiz et al.(2012)]{Ortiz12} Ortiz, J.~L., Sicardy, B., Braga-Ribas, F., et al.: 2012, \nat, 491, 566.
 
    \bibitem[Ortiz et al.(2015)]{Ortiz15} Ortiz, J.~L., Duffard, R., Pinilla-Alonso, N., et al.: 2015, \aap, 576, A18.
 
    \bibitem[Ortiz et al.(2017)]{Ortiz17} Ortiz, J.~L., Santos-Sanz, P., Sicardy, B., et al.: 2017, \nat, 550, 219.

    \bibitem[Ortiz et al.(2020a)]{Ortiz20_book}Ortiz, J.L., Sicardy, B., Camargo, J.I.B., Santos-Sanz, P., and Braga-Ribas, F.: 2020a, The Trans-neptunian Solar System, 413.
        
    \bibitem[Ortiz et al.(2020b)]{Ortiz20} Ortiz, J.L., Santos-Sanz, P., Sicardy, et al.: 2020b, \aap,  639, A134.
    
    \bibitem[P{\'a}l et al.(2012)]{Pal12} P{\'a}l, A., Kiss, C., M{\"u}ller, T.~G., et al.: 2012, \aap, 541, L6.
    
    \bibitem[P{\'a}l et al.(2015)]{Pal2015}P{\'a}l, A., Szab{\'o}, R., Szab{\'o}, G.M., et al.: 2015, \aj, , 804, L45.
    
    \bibitem[Person et al.(2011)]{Person2011}Person, M.J., Elliot, J.L., Bosh, et al.: 2011,  American Astronomical Society Meeting Abstracts, 218, 224.12.
    
    \bibitem[Ruprecht et al.(2015)]{Ruprecht15} Ruprecht. J. ~D., Bosh, A. ~S., Person, M. ~J., et al.: 2015, \icarus, 252, 271-276.
    
    \bibitem[Santos-Sanz et al.(2012)]{Santos-Sanz12} Santos-Sanz, P., Lellouch, E., Fornasier, S., et al.: 2012, \aap, 541, A92.
 
    \bibitem[Stern et al.(2019)]{Stern19} Stern, S. A., Weaver, H. A., Spencer, J. R., et al.: 2019, Science, 364, 6441, aaw9771.
 
    \bibitem[Schindler et al.(2017)]{Schindler17} Schindler, K., Wolf, J., Bardecker, J., et al.: 2017, \aap, 600, A12.
 
    \bibitem[Sickafoose et al.(2019)]{Sickafoose2019} Sickafoose, A.A., Bosh, A.S., Levine, S.E., et al.: 2019, Icarus, 319, 657.
 
    \bibitem[Sicardy et al.(2011)]{Sicardy11} Sicardy, B., Ortiz, J.~L., Assafin, M., et al.: 2011, \nat, 478, 493.
    
    \bibitem[Souami et al.(accepted)]{Souami} Souami, D. et al.: 2020, \aap, (accepted).
    
    \bibitem[Spencer et al.(2020)]{Spencer20}Spencer, J.R., Stern, S.A., Moore, J.M., et al.: 2020, Science, 367, aay3999.
 
    \bibitem[Umurhan et al.(2020)]{Umurhan20}Umurhan, O.M., Keane, J.T., Beyer, R.A., et al.: 2020,  American Astronomical Society Meeting Abstracts , 419.05.
 
    \bibitem[Vilenius et al.(2012)]{Vilenius12} Vilenius, E., Kiss, C., Mommert, M., et al.: 2012, \aap, 541, A94.

    \bibitem[Vilenius et al.(2014)]{Vilenius14} Vilenius, E., Kiss, C., M\"ueller, T., et al.: 2014, \aap, 564, A35.
    
    \bibitem[Vilenius et al.(2018)]{Vilenius18} Vilenius, E., Stansberry, J., M{\"u}ller, et al.: 2018, \aap, 618, A136.
    
    \end{thebibliography}
\end{document}